\documentclass[
 aip,
floatfix,
 amsmath,amssymb,
 preprint,%
]{revtex4-1}
\usepackage{caption}
\usepackage{subcaption}
\usepackage[colorlinks=true, citecolor=blue]{hyperref}
\usepackage{graphicx}
\usepackage{dcolumn}
\usepackage{bm}

\usepackage[utf8]{inputenc}
\usepackage[T1]{fontenc}
\usepackage{mathptmx}
\usepackage{etoolbox}
\usepackage[export]{adjustbox}

\makeatletter
\def\@email#1#2{%
 \endgroup
 \patchcmd{\titleblock@produce}
  {\frontmatter@RRAPformat}
  {\frontmatter@RRAPformat{\produce@RRAP{*#1\href{mailto:#2}{#2}}}\frontmatter@RRAPformat}
  {}{}
}%
\makeatother
\begin{document}

\title[Non-Linear behavior of the Electron Cyclotron Drift Instability]{Non-Linear behavior of the Electron Cyclotron Drift Instability and the Suppression of Anomalous Current}
\author{Aryan Sharma}
 \affiliation{Department of Computer Science, University of Saskatchewan, Canada.}
\author{Andrei Smolyakov}%
 \email{andrei.smolyakov@usask.ca.}
\affiliation{ 
Department of Physics and Engineering Physics, University of Saskatchewan, Saskatoon, Canada.
}%

\author{Raymond J. Spiteri}%
\affiliation{
Department of Computer Science, University of Saskatchewan, Saskatoon, Canada.
}

\date{\today}
\begin{abstract}
We present results of one-dimensional collisionless simulations of plasma turbulence and related anomalous electron current of the Electron Cyclotron Drift Instability (ECDI). Our highly resolved, long-term simulations of xenon plasma in the magnetic field performed with the WarpX particle-in-cell (PIC) code show several intermediate non-linear stages before the system enters a stationary state with significantly increased electron temperature and a finite level of energy in the electrostatic fluctuations. In early and intermediate non-linear stages, the fluctuations are driven by the electron cyclotron resonances gradually shifting from higher ($m>1$) modes to the fundamental $m=1$ resonance. Enhanced resonant growth is observed from the point when the cyclotron $m=1$ mode coincides with the most unstable ion-acoustic mode. In the final stage, the anomalous electron current existing in intermediate stages is quenched to zero. Following this quenching, our simulations reveal a transition from ECDI-driven dynamics to saturated ion-acoustic turbulence. The modification of the electron and ion distribution functions and their roles in the non-linear developments and saturation of the instability are analyzed at different non-linear stages. The non-linear development of ECDI driven by the $\mathbf{E} \times \mathbf{B}$ electron drift from the applied current and the ECDI driven by the ion beam perpendicular to the magnetic field are compared and characterized as two perspectives of the instability, observed through different Doppler-shifted frames. An extension of this work incorporating full the dynamics of  magnetized  ions  for ECDI driven by a hydrogen ion beam is shown to develop full beam inversion, with the periodic bursts of growth-saturation cycles of ECDI.

\end{abstract}

\maketitle

\section{\label{sec1:level1}Introduction}

The electron cyclotron drift instability (ECDI) belongs to a class of streaming-type instabilities,~\cite{wong1970electrostatic} where the free energy  for  the instability is provided by the relative drift between ions and electrons. Typical examples are current-driven Buneman~\cite{Buneman1962} and ion-acoustic instabilities~\cite{gary1987ion} in unmagnetized plasma. In partially magnetized plasmas, i.e., in plasmas with a moderate magnetic field that does not affect ions,  ECDI can be triggered by a beam of magnetized electrons subject to the $\mathbf{E} \times \mathbf{B}$  across the magnetic field on the background of stationary ions. Conversely, ECDI can also be initiated by a current resulting from unmagnetized ions streaming across the magnetic field.  

ECDI driven by an ion beam has been widely discussed in the context of turbulence in collisionless shocks, observed in the Earth's bow shock region,~\cite{muschietti2008electron, muschietti2013microturbulence, wilson2014quantified, wilson2010large} where it can play a significant role in particle heating and wave generation, ultimately defining the shock width. On the other hand, ECDI driven by an $\mathbf{E} \times \mathbf{B}$ electron drift is recognized as a strong candidate to explain the anomalous transport observed in regions with large electric fields, particularly in Hall thrusters,~\cite{AdamPoP2004} Penning discharge,~\cite{janhunen2017non} and other $\mathbf{E} \times \mathbf{B}$ devices.~\cite{TsikataPRL2015}

In Hall thrusters, the radial magnetic field $\mathbf{B}$ and the axial electric field $\mathbf{E}$ generate a large $\mathbf{E} \times \mathbf{B}$ electron drift in the azimuthal direction that can resonantly excite azimuthal ECDI modes at the electron cyclotron harmonics, $\omega^\prime \simeq \omega_{ce}$,~\cite{tsikata2009dispersion, tsikata2010three, BoeufJAP2017} where $\omega^\prime$ is the mode frequency in the electron reference frame. Numerous 1D azimuthal~\cite{SmolyakovPPCF2017, SmolyakovPPR2020, tavassoli2023electron,FarajiJAP2023} and 2D axial-azimuthal~\cite{AdamPoP2004, TaccognaPoP2018, TaccognaPSST2019, janhunen2017non, JanhunenPoP2018b, VillafanaPSST2021,RezaJAP2023,TaccognaRMPP2019,GarriguesJAP2021} numerical simulations of ECDI have demonstrated the generation of anomalous current during the non-linear development of the instability. Similar general features were also observed in 3D simulations.~\cite{tsikata2010three} Despite many efforts, there is no complete understanding and full consensus on the non-linear dynamics of ECDI, in particular, in applications to the problem of the anomalous transport in Hall thrusters.  
One of the important points is the relation of ECDI to the ion-acoustic modes. It has been stated in the early works~\cite{GaleevSagdeevResistivity,GaleevHandbook} that for low magnetic field, $\omega_{ce}\ll \omega_{pe}$, the current driven turbulence is not affected by the magnetic field and behaves similarly to the ion-acoustic turbulence in unmagnetized plasmas. Earlier numerical simulations~\cite{LampePF1972a,LampePRL1971,LampePF1972b,BiskampPRL1971,BiskampNF1972,BiskampIAEA1971} have shown an early saturation of the cyclotron ECDI modes and the transition to the ion-acoustic like turbulence. This analogy with the unmagnetized ion-acoustic turbulence was extensively used to interpret  the results of 1D and 2D numerical simulations in applications to anomalous transport in Hall thrusters.~\cite{LafleurPoP2016a,LafleurPoP2016b,BoeufPoP2018,CharoyPSST2019,LafleurPoP2018} It was argued, however, based on similar numerical studies\cite{ForslundPF1972a,ForslundPF1972b,ForslundPRL1970}, that many features of the non-linear growth phase of the electron-ion streaming instability are unlike those of magnetic field free ion-acoustic instability. Comparative features of ECDI and ion-acoustic instability (IAI) have been further discussed in Ref.~\onlinecite{BiskampPF1973}. In the context of the applications to turbulence in collisionless shock waves, the ECDI and IAI were compared in Ref.~\onlinecite{muschietti2013microturbulence}. Experimental measurements of turbulence in Hall thruster~\cite{BrownPRL2023} waves suggest that the effects of the magnetic field remain essential and that fluctuations are driven by the electron cyclotron resonances. Intense electron cyclotron harmonics mixed with IAI features were detected in the Earth's bow shock in several spacecraft missions.~\cite{breneman2013stereo,wilson2010large,wilson2021discrepancy} These observational results were reviewed in Ref.~\onlinecite{wilson2021discrepancy}, where concerns were also raised regarding the discrepancies between simulations and observations of electric fields in collisionless shocks.

In this paper, we perform highly resolved, long-term 1D3V simulations of ECDI along the direction of the relative drift of species for both the ECDI driven by an $\mathbf{E} \times \mathbf{B}$ electron drift (Case 1A) and ECDI driven by an ion beam (Case 1B). Our study focuses on the saturation of turbulence, the time dynamics of anomalous current, the evolution of the electron and ion distribution functions, and the coupling to the ion-acoustic modes. In 2D and 3D geometries, additional linear and non-linear effects may become essential, and other instabilities,  such as the modified two-stream instability (MTSI) and ion-ion modes, may couple to ECDI and  affect its dynamics, e.g., as observed in space studies.\cite{breneman2013stereo,wilson2010large,wilson2021discrepancy} Here, we consider a pure 1D propagation perpendicular to the magnetic field, so no MTSI nor linear transition to the IAI are possible. In previous 1D simulations, ECDI was identified as an effective mechanism of plasma heating, leading to rapid increase in temperatures above the values expected under experimental conditions. To suppress this growth, a virtual length model was used in simulations to mimic a finite residence time and thus to limit the temperature increase.~\cite{LafleurJAP2021,LafleurPoP2016b,SmolyakovPPR2020,TaccognaPSST2019} In this paper, we follow the non-linear dynamics, plasma heating, and mode saturation deeply into the non-linear stage without introducing any ad-hoc saturation mechanisms. Our long-term simulations show that after several intermediate non-linear stages, in which the electron cyclotron resonances drive the fluctuations, ECDI turbulence enters the saturated state with a finite level of ion-acoustic fluctuations but no anomalous transport.  This saturated state is characterized by a strongly flattened electron distribution function with a large electron energy, well exceeding the mean energy of the electron beam and having a much lower energy of ions and fluctuations energy. In the final stage, the electric field fluctuations are much higher than the mean electric field, and the ion distribution function has a typical quasilinear tail at high energies. 

The remainder of this paper is structured as follows. Sec.~\ref{sec2:level1} outlines the numerical and physical parameters used in this study, along with a comprehensive scaling analysis of the WarpX PIC code,~\cite{vay2018warp} which was employed to simulate the full dynamics of ECDI. Sec.~\ref{sec3:level1} presents the  analysis of the ECDI driven by the $\mathbf{E} \times \mathbf{B}$ electron drift current (Case 1A), while Sec.~\ref{sec4:level1} considers the ECDI driven by an ion beam current (Case 1B) with a comparative discussion to Case 1A. Sec.~\ref{sec5:level1} explores results from a magnetized hydrogen ion beam, demonstrating bursts of ECDI fluctuations and anomalous transport during the  complete inversion of the ion beam. Sec.~\ref{sec6:level1} discusses the energy conservation throughout the evolution of both cases 1A and 1B. Finally, Sec.~\ref{sec7:level1} concludes the paper with a summary of key findings and their implications.

\section{\label{sec2:level1}Numerical Setup and Physical Parameters}

In order to analyze the linear and non-linear evolution of ECDI, we set up a 1D3V PIC model in WarpX and analypoize the instability under three different conditions:

\begin{itemize}
\item \textbf{Case 1A}: ECDI driven by an $\mathbf{E} \times \mathbf{B}$ electron drift current (with unmagnetized xenon ions),
\item \textbf{Case 1B}: ECDI driven by an ion beam (with unmagnetized xenon ions),
\item  \textbf{Case 2}: ECDI driven by an ion beam (with magnetized hydrogen ions).
\end{itemize}

\begin{table}[htbp]
\begin{tabular}{ccc}
\hline
\hline
\textbf{ Parameters } & \textbf{ Symbol } & \textbf{ Values } \\
\hline
Simulation length      &   $ L_z $     &   $4.456$ cm \\
Magnetic field         &   $ B_0 $     &   $200$ G  \\
Electric Field         &   $ E_0 $     &   $20000$ V/m\\
Ion mass               &   $ m_{i} $ &   $131.293$ u\\
Ion temperature        &   $ T_{i} $ &   $0.2$ eV  \\
Electron mass          &   $ m_{e} $ &   $0.005$ u\\
Electron temperature   &   $ T_{e} $ &   $10.0$ eV  \\
Plasma density         &   $ n_{0} $ &   $ 10^{17} \,\, \text{m}^{-3} $\\
PIC timestep           &   $ \Delta t $ &   $0.56 \times 10^{-12}$ s \\
Spatial cell size      &   $ \Delta x $ &   $0.29 \, \lambda_{De}$ \\
\hline
Initial Electron Drift Velocity &   $ V_{de}$ &  $10^{6}$ m/s \\
Initial Ions Drift Velocity &   $ V_{di} $ &  $0$ m/s \\
\hline
Ion plasma frequency   & $ \omega_{pi} $ & $3.65 \times 10^7 \, \text{rad/s}$ \\
Electron plasma frequency & $ \omega_{pe} $ & $1.78 \times 10^{10} \, \text{rad/s}$ \\
Ion cyclotron frequency & $ \omega_{ci} $ & $1.47 \times 10^4 \, \text{rad/s}$ \\
Electron cyclotron frequency & $ \omega_{ce} $ & $3.52 \times 10^9 \, \text{rad/s}$ \\
Lower hybrid frequency & $ \omega_{LH} $ & $1.39 \times 10^6 \, \text{rad/s}$ \\
Upper hybrid frequency & $ \omega_{UH} $ & $1.82 \times 10^{10} \, \text{rad/s}$ \\
\hline
\hline
\end{tabular}
\caption{Physical, numerical, and derived parameters for the simulation setup Case 1A}
\label{tbl:initial_conditions}
\end{table}

For Case 1A, we define a constant electric field $\mathit{E_x} \, = \, E_0$ in the $x$-direction and a constant magnetic field $\mathit{B_y} \, = \, B_0$ in the $y$-direction. $E_0$ acts as an external power input to the system and drives the electron $\mathbf{E} \times \mathbf{B}$ drift current in the $z$-direction with a drift velocity given by $\mathit{V_{de}} \, = \, (E_0\,\widehat{\mathbf{e}}_x \times B_0 \widehat{\mathbf{e}}_y)/\mathit{|B_0|^{2}}$. The Poisson equation is solved in the $z$-direction using a linear solver (Multigrid solver (MLMG) class in AMReX~\cite{zhang2019amrex}) with periodic boundary conditions, to evaluate the generated electrostatic field $\mathit{E_z}$.

\begin{figure}[htbp]
\centering
\includegraphics[width=\linewidth]{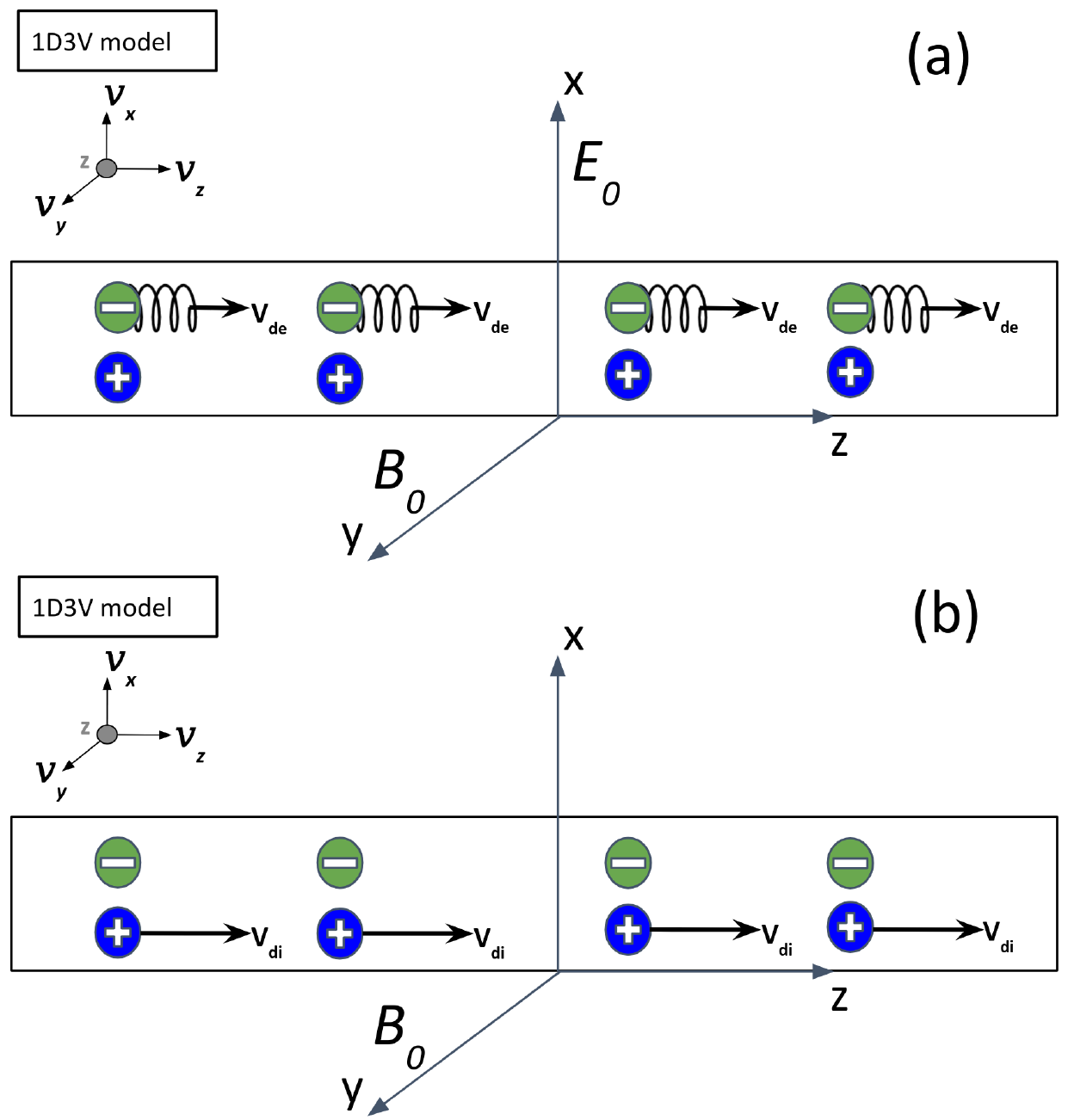}
\caption{ Setup of the 1D3V simulations  performed  in a periodic $z$ domain; a magnetic field is in the $y$-direction.  (a)  Case 1A: the instability is driven by the  electron ${\bf E} \times{\bf B} $ drift due to an applied electric field in the $x$-direction; (b)  Case 1B: the instability is driven by an ion beam the $z$-direction.}
\label{fig:1}
\end{figure}

All the physical parameters used for the 1D3V setup are given in Table~\ref{tbl:initial_conditions}, corresponding to the typical operating regime of the SPT100 Hall thruster.~\cite{janhunen2017non, tavassoli2023electron} The simulation direction is in the $z$-direction with the domain length $\mathit{L_z} = 4.456$ cm. The spatial cell size $ \Delta \mathit{z}$ is chosen to have at least 3 cells per Debye length $ \lambda_{De} \equiv \sqrt{\epsilon_0 \mathit{T_{e0}} / n_0 e^2} = 7.43 \times 10^{-3}$ cm. The ions and electrons are uniformly distributed in the spatial domain with the number of particles per cell per species $N_{\text{ppc}} = $ 1000. The electron velocities are initially sampled from a shifted Maxwellian distribution with temperature $\mathit{T_{e0}} = 10$ eV and a mean drift in the $z$-direction $V_{ze} = V_{de} = 10^{6}$ m/s. The ion velocities are sampled from a Maxwellian distribution with temperature $\mathit{T_{i0}} = 0.2$ eV and no mean drift.

The electrons are subjected to the applied fields $\mathit{E_x}, \, \mathit{B_y}$, and the generated $\mathit{E_z}$, while the ions are treated as unmagnetized and unaffected by the external electric field. Thus, ions are only subjected to $\mathit{E_z}$. The evolution of the particle trajectories under the influence of these forces is updated using a standard Boris scheme with a timestep $\Delta \mathit{t} = 0.56 \times 10^{-12}$ s.

For Case 1B, in order to simulate ECDI driven by an ion beam, we remove the external electric field $E_0$.  Now, the electron velocities are sampled from a Maxwellian distribution with temperature $\mathit{T_{e0}} = 10$ eV and no mean drift, and the ion velocities are sampled from a shifted Maxwellian distribution with temperature $\mathit{T_{i0}} = 0.2$ eV and a mean drift in the $z$-direction $V_{zi} = V_{di} = 10^{6}$ m/s. All the other physical parameters for this case are mirrored from Case 1A and taken from Table~\ref{tbl:initial_conditions}. Again, the electrons are subjected to both $\mathit{B_y}$ and $\mathit{E_z}$, while the ions are only subjected to $\mathit{E_z}$.

For Case 2, we use the Case 1B setup and replace xenon with hydrogen, and introduce the effect of magnetic field $\mathit{B_y}$ for ions as well.

We use the WarpX~\cite{vay2018warp} code for our simulations.  WarpX is a highly optimized, time-based electromagnetic and electrostatic PIC code, capable of running on multi-core CPUs and multi-GPUs. It has been recognized with 2022 ACM Gordon Bell prize for its exascale computing capabilities.~\cite{Exascale2022} In the PIC method employed in WarpX, the fields are defined on a spatially discretized grid, and the particle trajectories are tracked under the influence of these fields, while periodically updating these fields based on the particles' collective interactions. For further implementation details, we refer to Ref.~\onlinecite{vay2018warp}.

We have analyzed the scalability of WarpX for ECDI studies. First, we benchmark  WarpX  results against previous Vlasov and PIC simulations~\cite{tavassoli2023electron} and assess the scalability of the code. Then, we use this setup to simulate a xenon ion beam system, compare the results with an $\mathbf{E} \times \mathbf{B}$ electron drift setup, and extend the model to simulate the ECDI instability driven by the hydrogen beam, including full ion dynamics.

The strong scaling performance for the setup mentioned in Section \ref{sec2:level1} is shown in  Table \ref{tbl:cores_vs_time}. The results demonstrate excellent scaling up to 512 cores, utilizing up to 8 nodes equipped with AMD 7532 processors, each with 64 cores per node, 2 sockets, and 8 memory channels per socket. All simulations in this study were conducted on the Narval supercomputing cluster, supported by the Digital Research Alliance of Canada. We note that EDIPIC code used in previous studies~\cite{tavassoli2023electron} and WarpX code perform similarly for 32 processors.

\begin{table}[h]
\renewcommand{\arraystretch}{1.2}
\begin{tabular}{c|c}
\hline
\hline
\textbf{Number of CPU Cores Used} & \textbf{Time Taken (in days)} \\
\hline
16  & $\sim 14$ days \\
32  & $\sim 7$ days \\
64  & $\sim 3.5$ days \\
128 & $\sim 1.7$ days \\
256 & $\sim 0.9$ days \\
512 & $\sim 0.5$ days \\
\hline
\hline
\end{tabular}
\caption{{Strong scaling analysis results for ECDI parameters defined in Table 1. The numbers are reported for simulation times of $t = 2000$ ns.}}
\label{tbl:cores_vs_time}
\end{table}


We note that the numbers in Table~\ref{tbl:initial_conditions} are computed when the data output is limited to simple integral quantities. To derive a highly resolved dispersion relation, the frequency of the electric field $E_z$ data output was increased to occur at every 100 timesteps for a simulation with a timestep of $\Delta t = 0.56 \times 10^{-12}$ s (to resolve frequencies as high as $\sim 8 \omega_{ce}$). As the number of cores (128, 256, 512) increased, the number of processes and associated data files also rose, leading to a proportional increase in the number of file writes. Consequently, the overall simulation time scaled approximately by a factor of two, largely due to the additional file handling required in these highly parallel simulations.
\section{\label{sec3:level1}Results: ECDI driven by external $\mathbf{E} \times \mathbf{B}$ current in xenon plasma (Case 1A)}

\subsection{\label{sec3:level2}Overview of the long-term  dynamics}

The non-linear dynamics of ECDI shows complex behaviour and several transitions between different stages. These stages (phases) are characterized by specific features in the growth of the wave energy, the characteristic wavelengths and the frequency of turbulent fluctuations, and the evolution and phase space patterns in the electron and ion distribution functions. Some of these features are clearly specific to distinct phases, whereas the others are common and persist for several phases.

\subcaptionsetup{font=large}

\begin{figure}[htbp]
\centering
\begin{subfigure}[b]{\linewidth}
\centering
\caption{ }
\includegraphics[width=\linewidth]{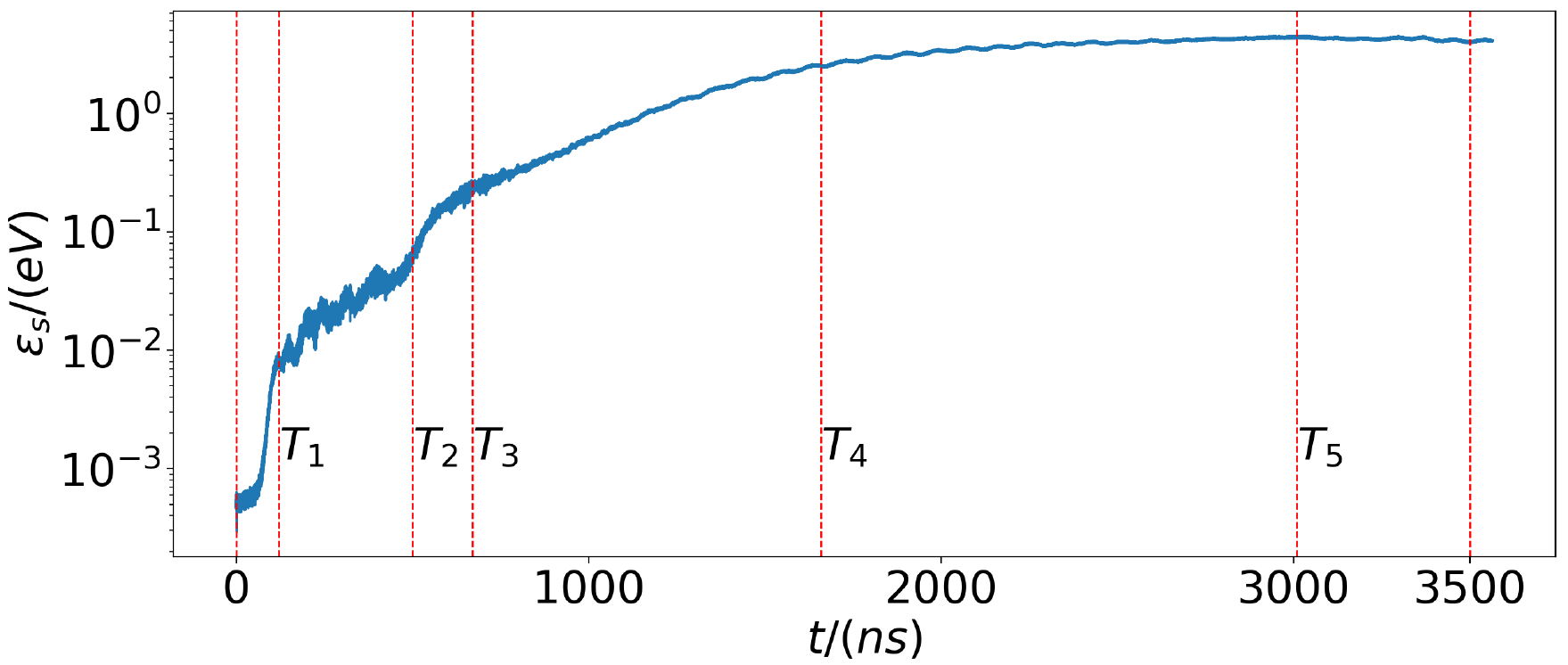}
\label{fig:2a}
\end{subfigure}
\hfill
\begin{subfigure}[b]{\linewidth}
\centering
\caption{ }
\includegraphics[width=\textwidth]{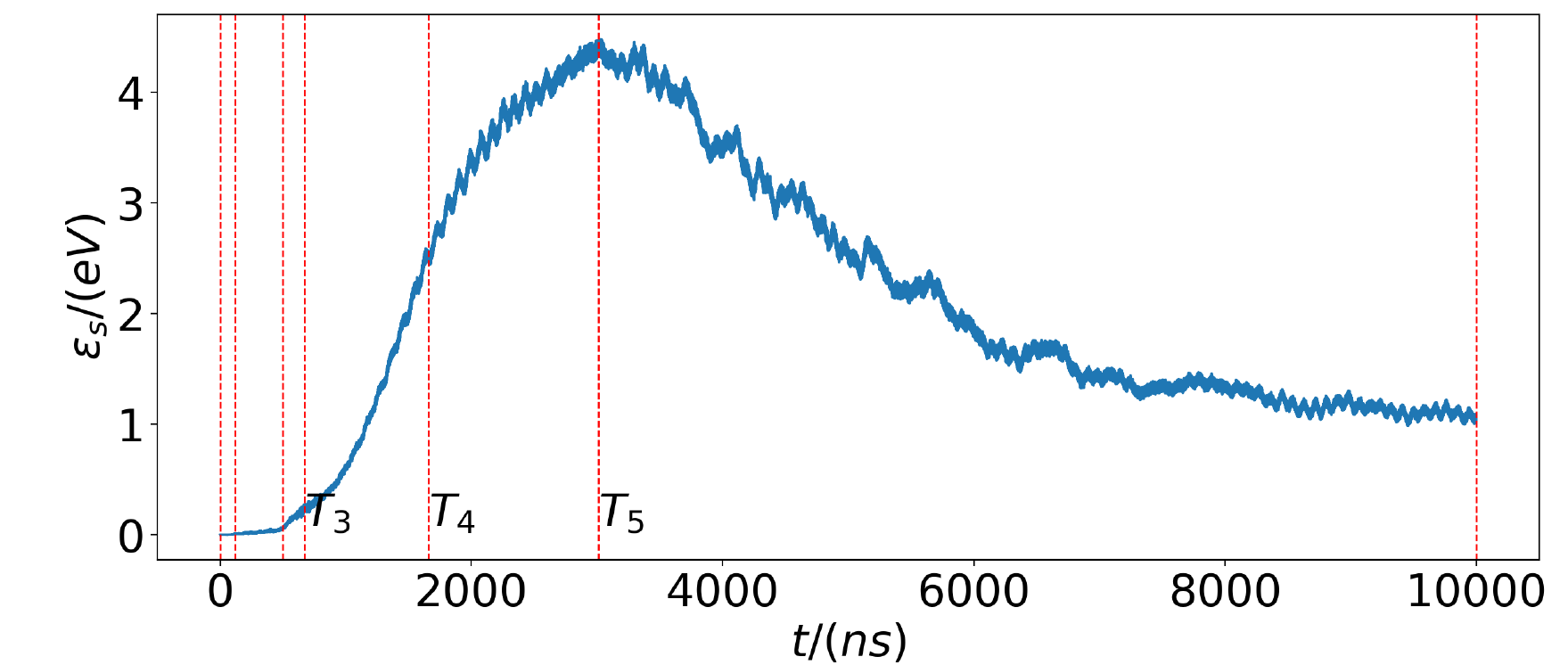}
\label{fig:2b}
\end{subfigure}
\caption{Time evolution of the electrostatic field energy ${\epsilon_s}$ (Eq.~\eqref{eq:es}) for Case 1A. Vertical red dotted lines shows time moments ($\mathit{T_1}, T_2, \dots, \mathit{T_5}$) for the transitions between different phases as described in the text. (a) $\epsilon_s$ in logscale for 0 < t < 3500 ns; (b) Evolution of ${\epsilon_s}$ for the entire simulation time 0 < t < 10000 ns (depicts complete evolution of $\epsilon_s$ from Phase 1 -- Phase 5).}
\label{fig:2}
\end{figure}

The full Case 1A is run up to $t=10000$ ns. The growth and saturation of the normalized electrostatic field energy per electron,
\begin{equation}
\label{eq:es}
\epsilon_s =E_s/(n_0L_z)= \frac{\epsilon_0}{2 n_0 L_z} \int_{-L_z/2}^{L_z/2} E_z^{2} \, dz,
\end{equation}
is shown in Figs.~\ref{fig:2a} and \ref{fig:2b}. Fig.~\ref{fig:2a} shows the evolution of $\epsilon_s$ (log-scale) up to $t = 3500$ ns, and the evolution up $t = 10000$ ns is shown in Fig.~\ref{fig:2b}. 
Overall, we distinguish five different stages (phases), with timings for each phase given in Table~\ref{tbl:ecdi_phases}, which are characterized by the following main features:


\begin{itemize}
\item Phase \textbf{1}: $0 < t < T_1$: Linear stage of ECDI;

\item Phase \textbf{2}: $T_1 < t < T_2$: Quasi-linear flattening of the electron distribution function (modified by the magnetic field) and resonance broadening of Doppler-shifted electron Bernstein modes;

\item Phase \textbf{3}: $T_2 < t < T_3$: Resonant interaction of the electron cyclotron $m=1$ mode with the ion-acoustic mode at $k_0=k_s$, where $k_s=1/\sqrt{2}\lambda_{De}$;

\item Phase \textbf{4}: $T_3 < t < T_5$: Enhanced growth of the ion-acoustic wave (at $\omega_{pi} = 1$), driven by the  $m=1$ ($k/k_0=1$) cyclotron resonance, and secondary ion-acoustic waves (at $\omega_{pi} = 2,3,\dots$), driven by higher mode cyclotron resonances $k/k_0 = 2,3,\dots $. This stage is characterized by intense  electron heating, and ion wave-particle interactions become visible in the ion distribution function leading to non-linear ion-trapping;

\item Phase \textbf{5}: $t \gg T_5$:  At $t=T_5$, the total electrostatic energy in fluctuations is at its maximum. The intensity of fluctuations concentrated at the $m=1$ cyclotron resonance $\omega/\omega_{pi} = 1$ and  $k/k_0 = 1$ starts to decrease spreading along the ion-acoustic like dispersion curve at $\omega_{pi}$ and its non-linear harmonics (at $\omega/\omega_{pi} = 2,3,\dots$). For $t \gg T_5$, a quasi-equilibrium is reached with fully suppressed anomalous current.
\end{itemize}

\begin{table}[htbp]
\renewcommand{\arraystretch}{1.2}
\centering
\begin{tabular}{|c|c|c|}
\hline
\textbf{Phase} & \textbf{Time Interval (ns)} & \textbf{End Time (ns)} \\ \hline
\textbf{1}     & $0 < t < T_1$   & $T_1 = 110$ ns    \\ \hline
\textbf{2}     & $T_1 < t < T_2$ & $T_2 = 460$ ns    \\ \hline
\textbf{3}     & $T_2 < t < T_3$ & $T_3 = 650$ ns    \\ \hline
\textbf{4}     & $T_3 < t < T_5$ & $T_5 = 3010$ ns   \\ \hline
\textbf{5}     & $t \gg T_5$     &  $\sim 10000$ ns    \\ \hline
\end{tabular}
\caption{Timings for five different stages characterized for the complete evolution of ECDI, for Case 1A. Note that Phase 4 is defined from $T_3 < t < T_5$; at $T_4 \, (=1660 \, \text{ns})$ the anomalous current $J_{xe}$ reaches its maximum and is included in the discussion for Phase 4.}
\label{tbl:ecdi_phases}
\end{table}

\begin{figure*}[htbp]
\centering
\begin{subfigure}[b]{0.8\textwidth}
\centering
\caption{ }
\includegraphics[width=\textwidth]{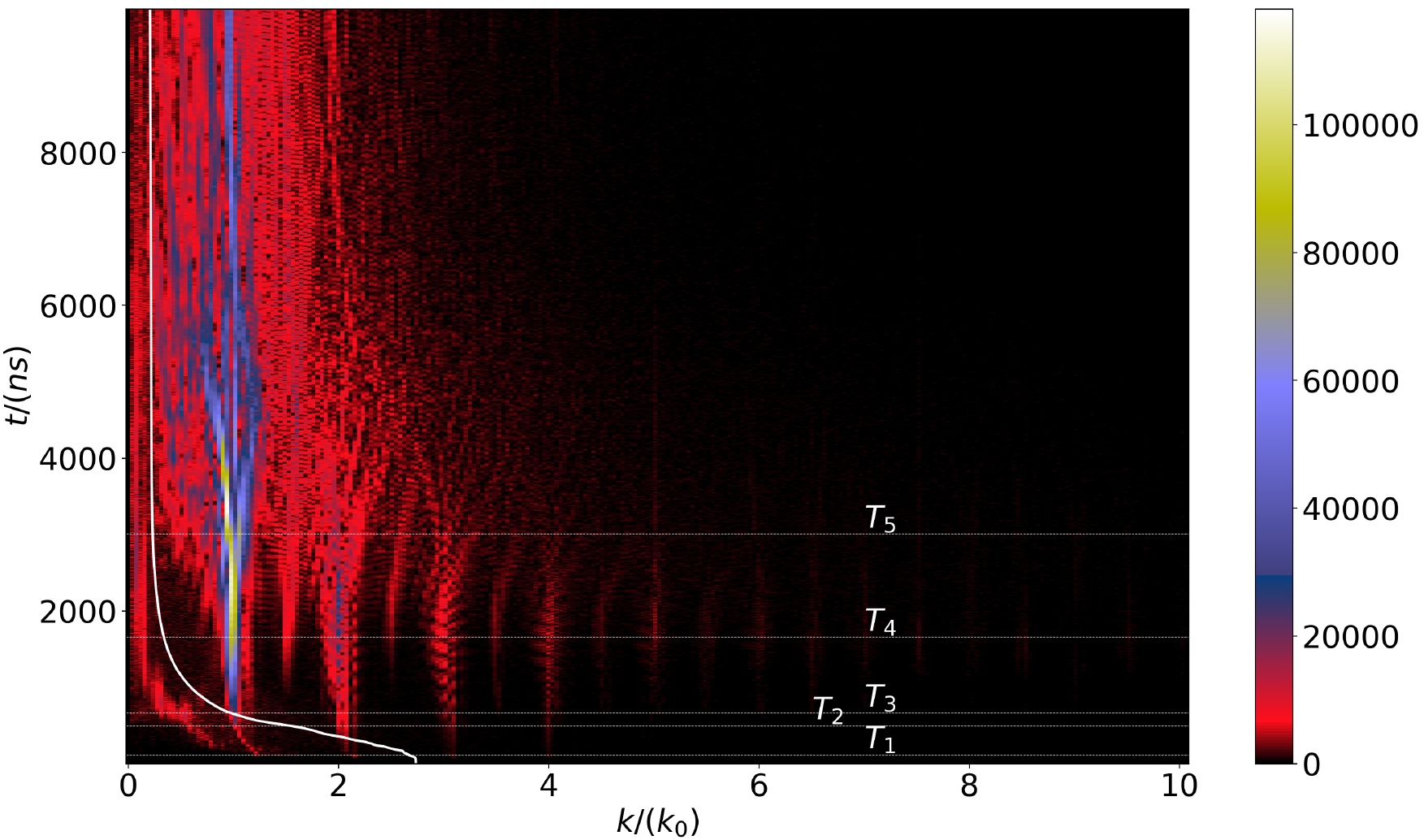}
\label{fig:3a}
\end{subfigure}
\hfill
\begin{subfigure}[b]{0.48\textwidth}
\centering
\caption{ }
\includegraphics[width=\textwidth]{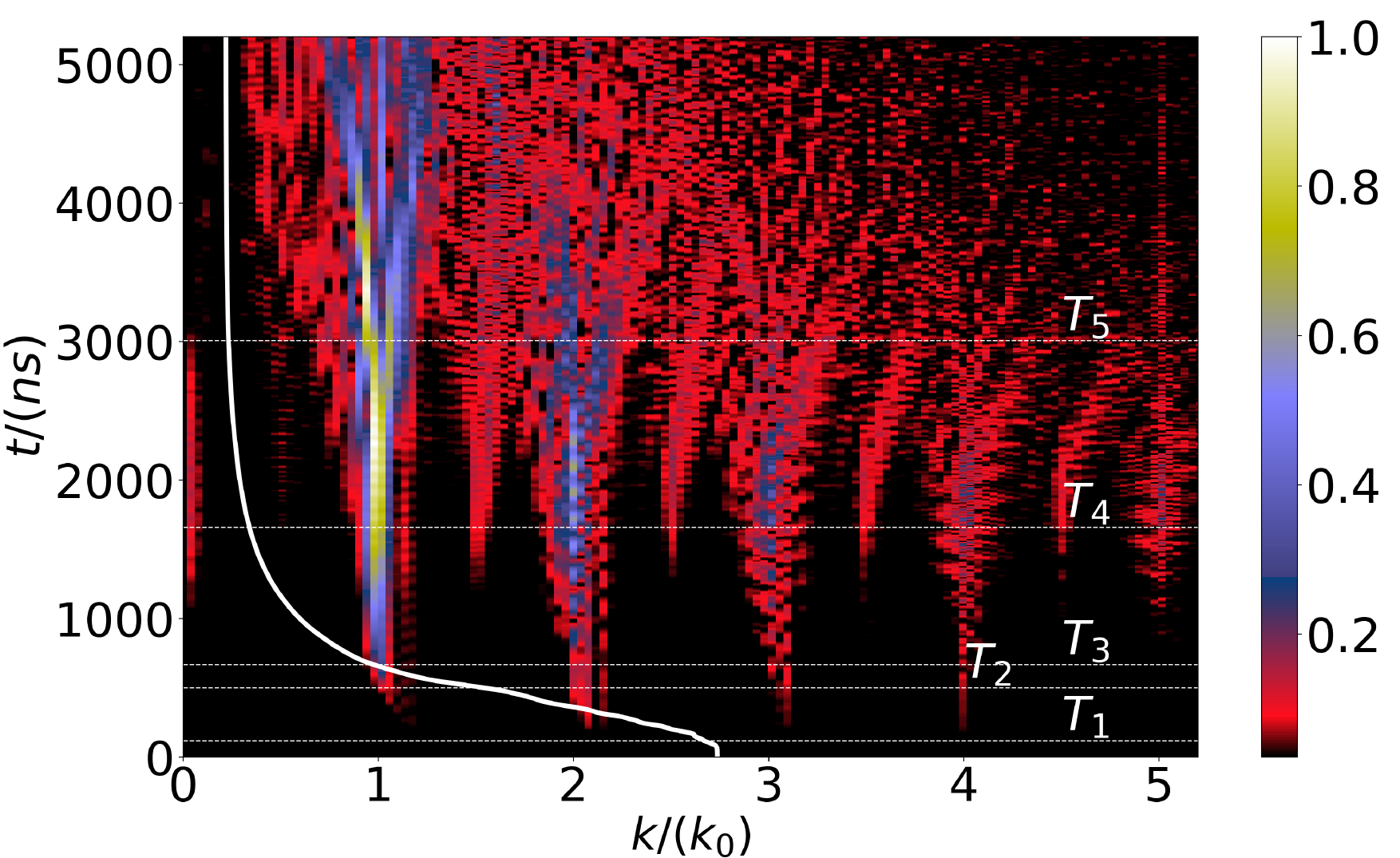}
\label{fig:3b}
\end{subfigure}
\hfill
\begin{subfigure}[b]{0.48\textwidth}
\centering
\caption{ }
\includegraphics[width=\textwidth]{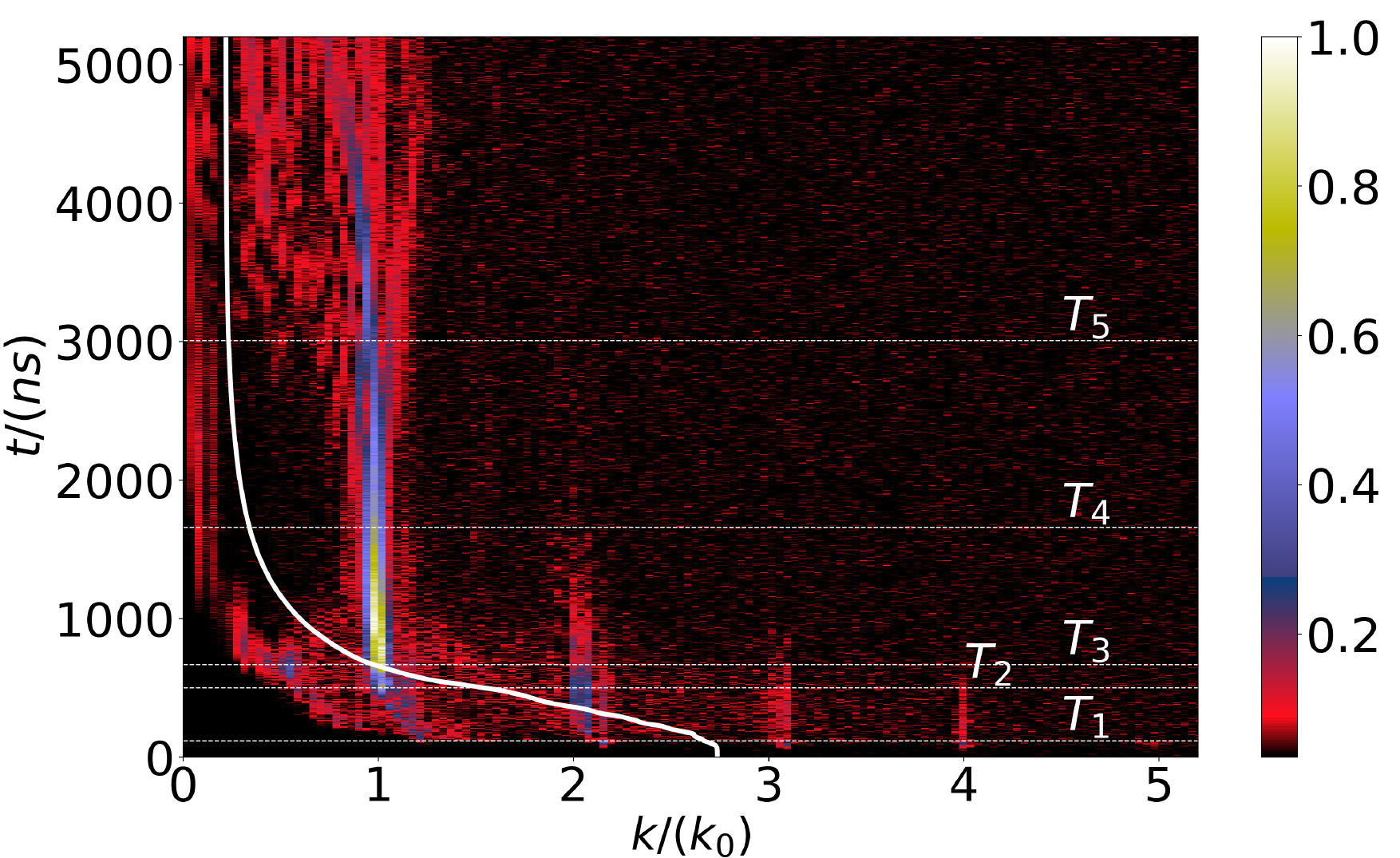}
\label{fig:3c}
\end{subfigure}
\caption{One-dimensional Fourier transform of (a) Electric field in the $z$-direction $\tilde{E}_z (k ,t)$ as given by Eq.~\eqref{eq:1dfft}, (b) Ion density in the $z$-direction $\tilde{n}_i (k ,t)$, and the (c) Electron density in the $z$-direction $\tilde{n}_e (k ,t)$. Horizontal dotted lines show the time moments ($\mathit{T_1}, T_2, \dots, \mathit{T_5}$) for the transitions between different phases as described in the text.}
\label{fig:3}
\end{figure*}

\subcaptionsetup{font=small}
\begin{figure*}[htbp]
\centering
\begin{subfigure}[b]{\textwidth}
\centering
\caption{ $\sim$ End of Phase 1}
\includegraphics[width=\textwidth]{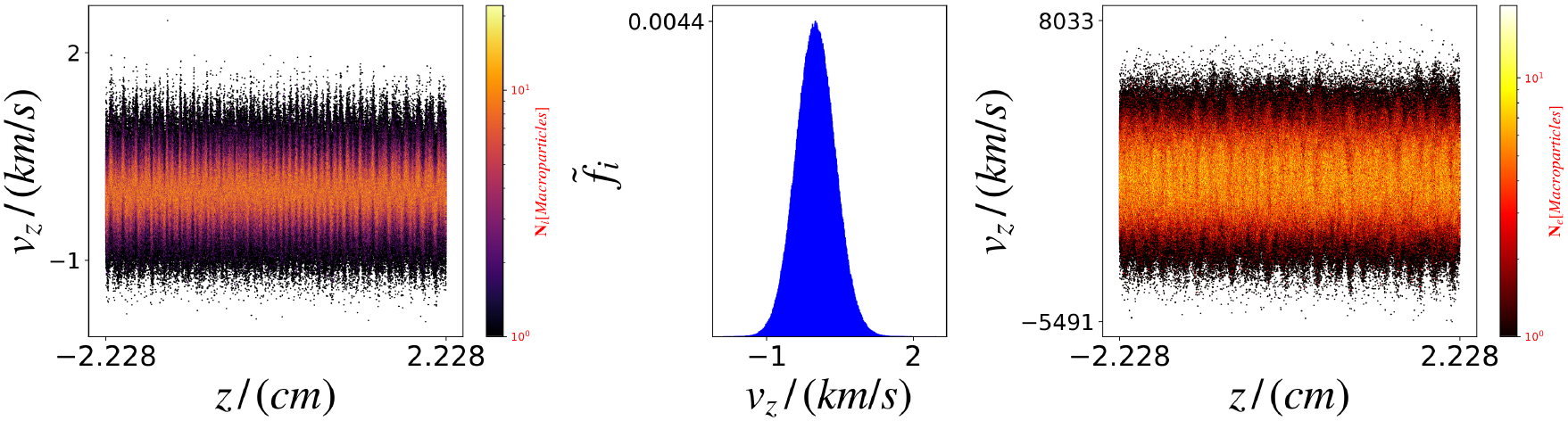}%
\label{fig:4a}
\end{subfigure}
\begin{subfigure}[b]{\textwidth}
\centering
\caption{ $\sim$ End of Phase 2}
\includegraphics[width=\textwidth]{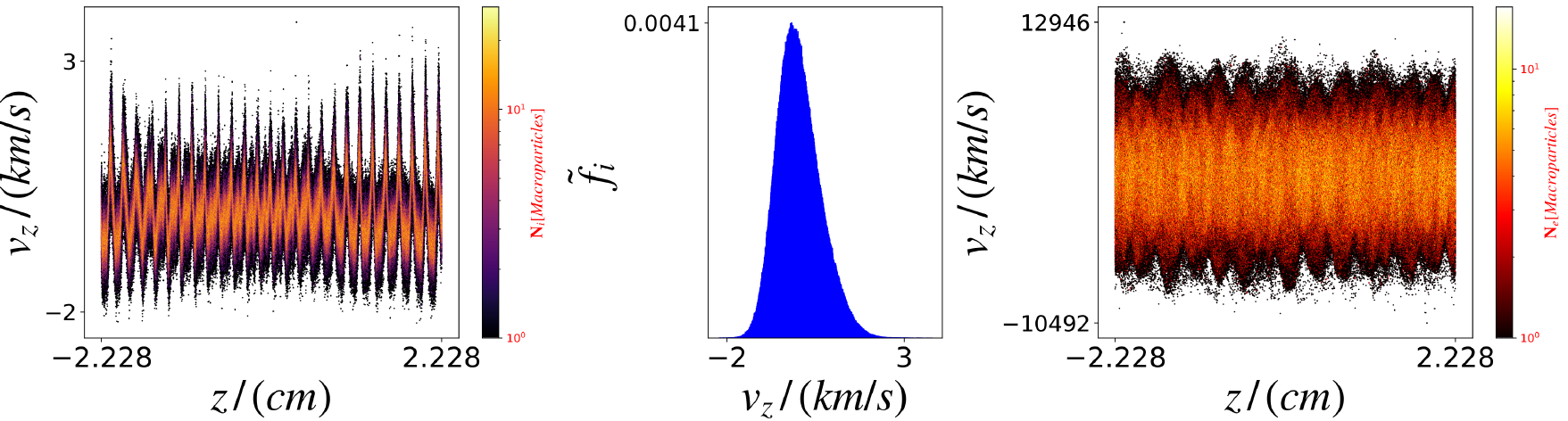}
\label{fig:4b}
\end{subfigure}
\begin{subfigure}[b]{\textwidth}
\centering
\caption{ $\sim$ End of Phase 3}
\includegraphics[width=\textwidth]{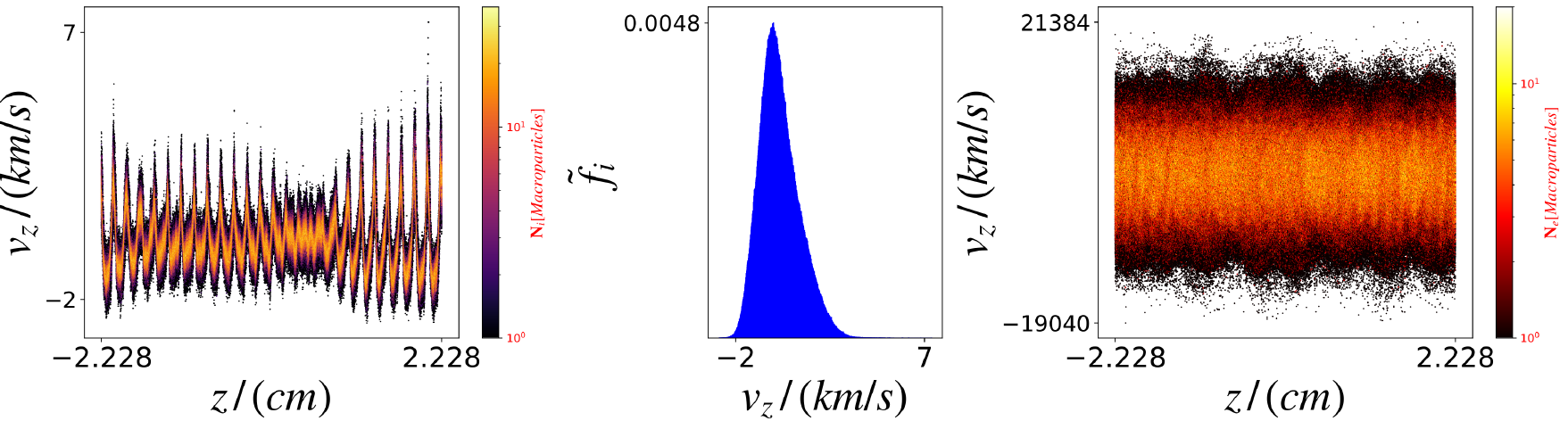}
\label{fig:4c}
\end{subfigure}
\begin{subfigure}[b]{\textwidth}
\centering
\caption{ $\sim$ Middle of Phase 4}
\includegraphics[width=\textwidth]{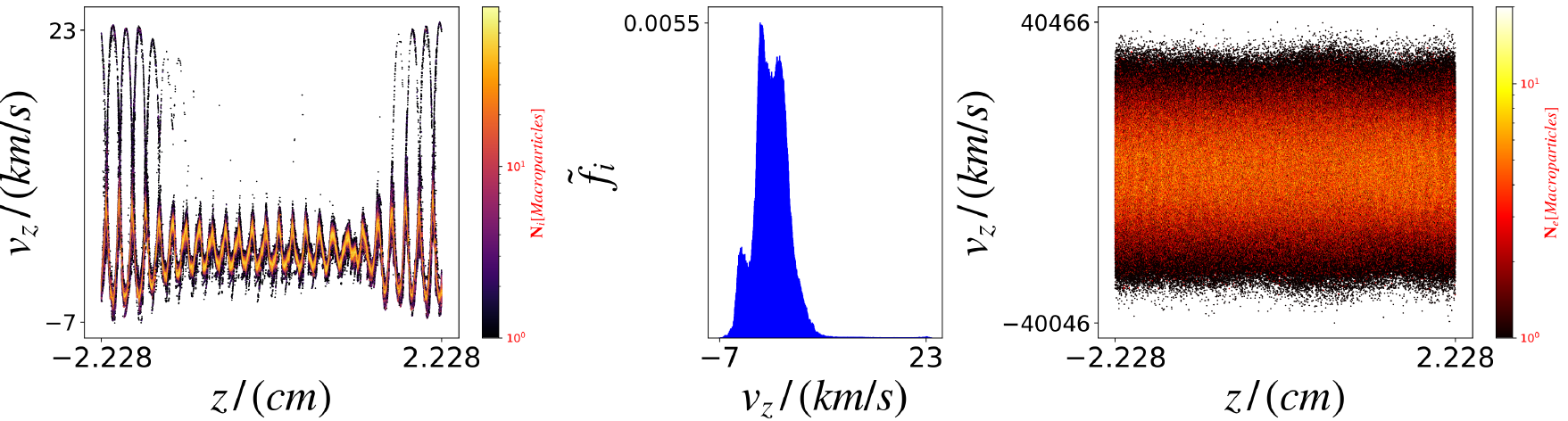}
\label{fig:4d}
\end{subfigure}
\end{figure*}

\begin{figure*}[htbp]\ContinuedFloat
\begin{subfigure}[b]{\textwidth}
\centering
\caption{ $\sim $ End of Phase 4}
\includegraphics[width=\textwidth]{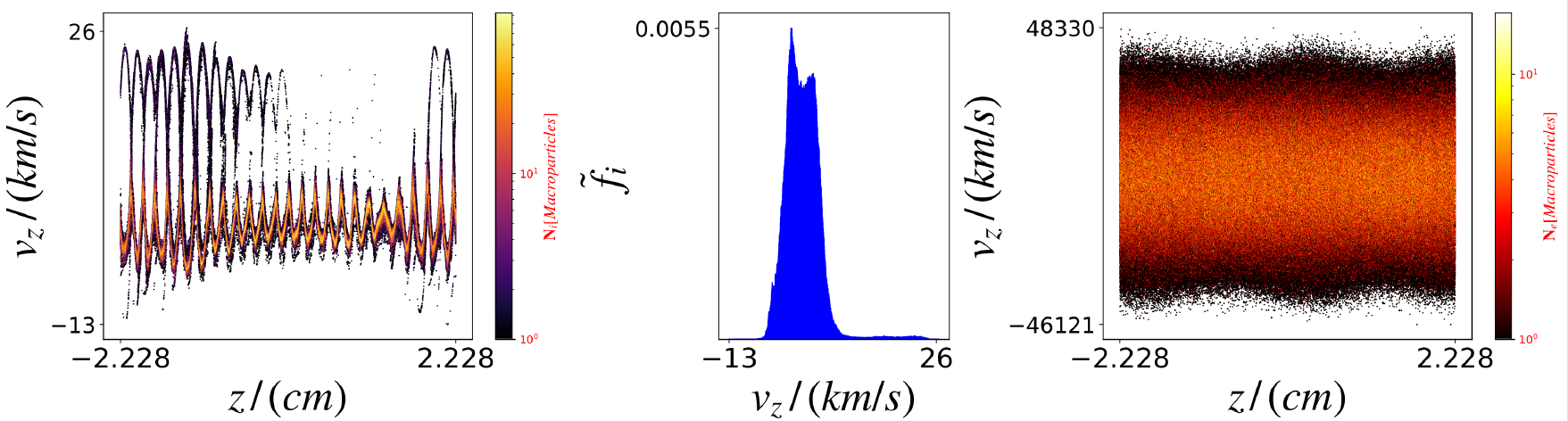}
\label{fig:4e}
\end{subfigure}
\begin{subfigure}[b]{\textwidth}
\centering
\caption{ $\sim$ End of Phase 5}
\includegraphics[width=\textwidth]{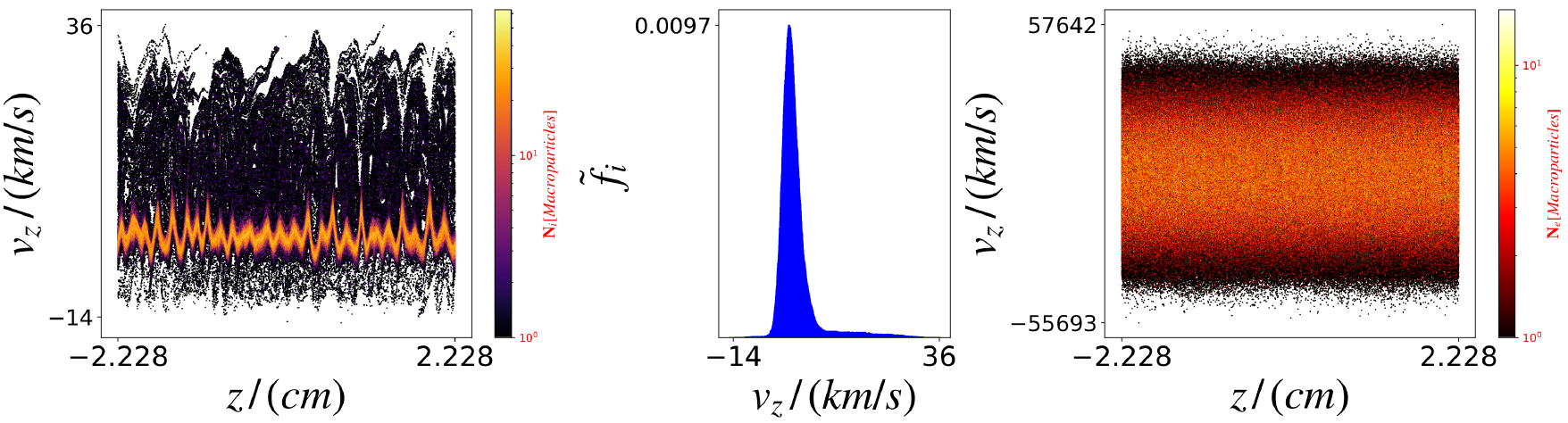}
\label{fig:4f}
\end{subfigure}
\caption{Snapshots of the evolution of three quantities shown for six different timeframes from Phase 1 to Phase 5. Column 1: Ion phase space in the $z$-direction, Column 2: Ion distribution function $\mathit{\tilde{f_i}}$ as given by Eq.~\eqref{eq:fe}, and Column 3: Electron phase space in the $z$-direction.}
\label{fig:4}
\end{figure*}

\begin{figure*}[htbp]
\centering
\begin{subfigure}[b]{0.48\textwidth}
\centering
\caption{ $\sim$ Middle of Phase 1}
\includegraphics[width=\textwidth, height=5.5cm]{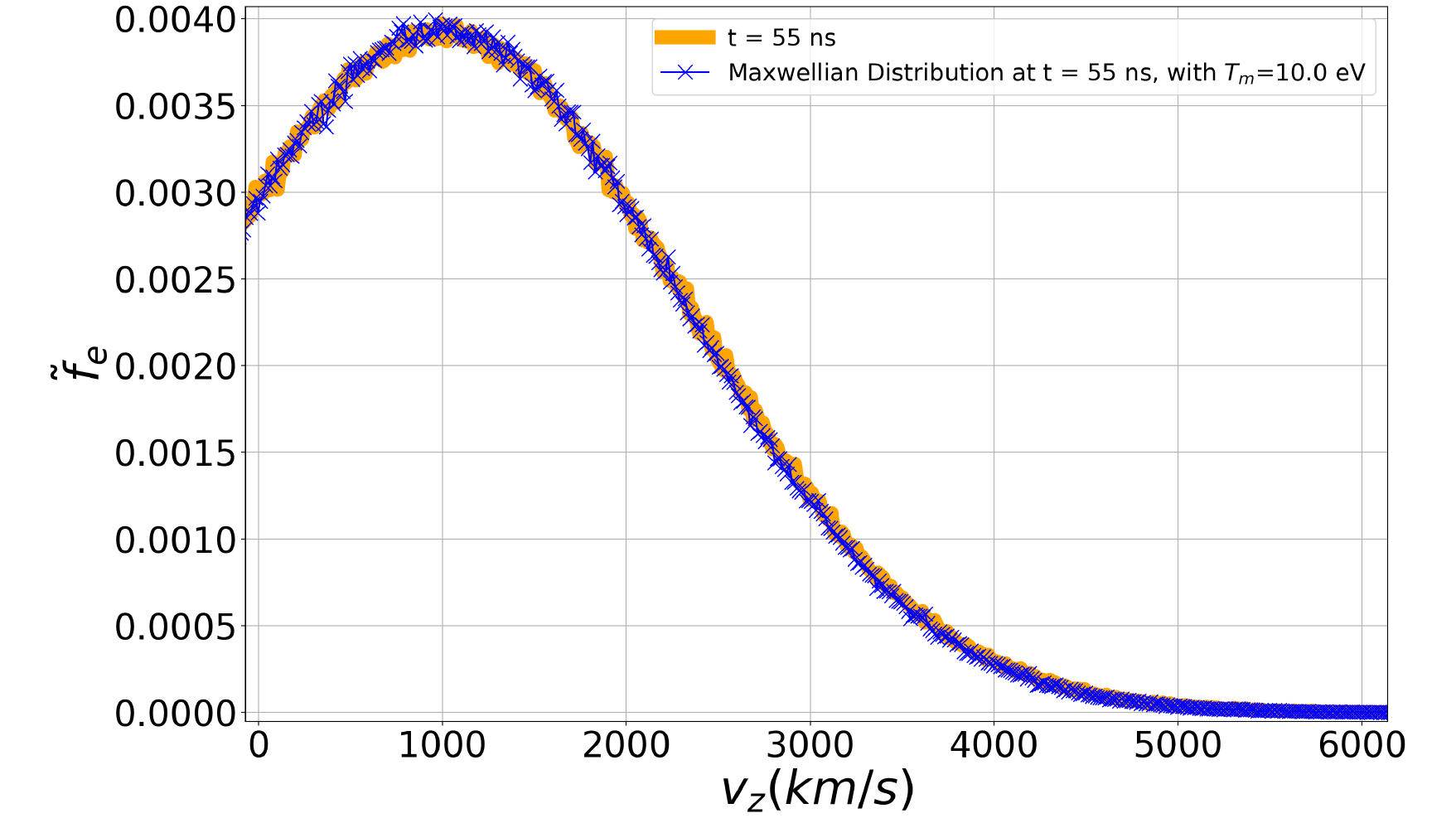}
\label{fig:5a}
\end{subfigure}
\hfill
\begin{subfigure}[b]{0.48\textwidth}
\centering
\caption{ $\sim$ Middle of Phase 2}
\includegraphics[width=\textwidth, height=5.5cm]{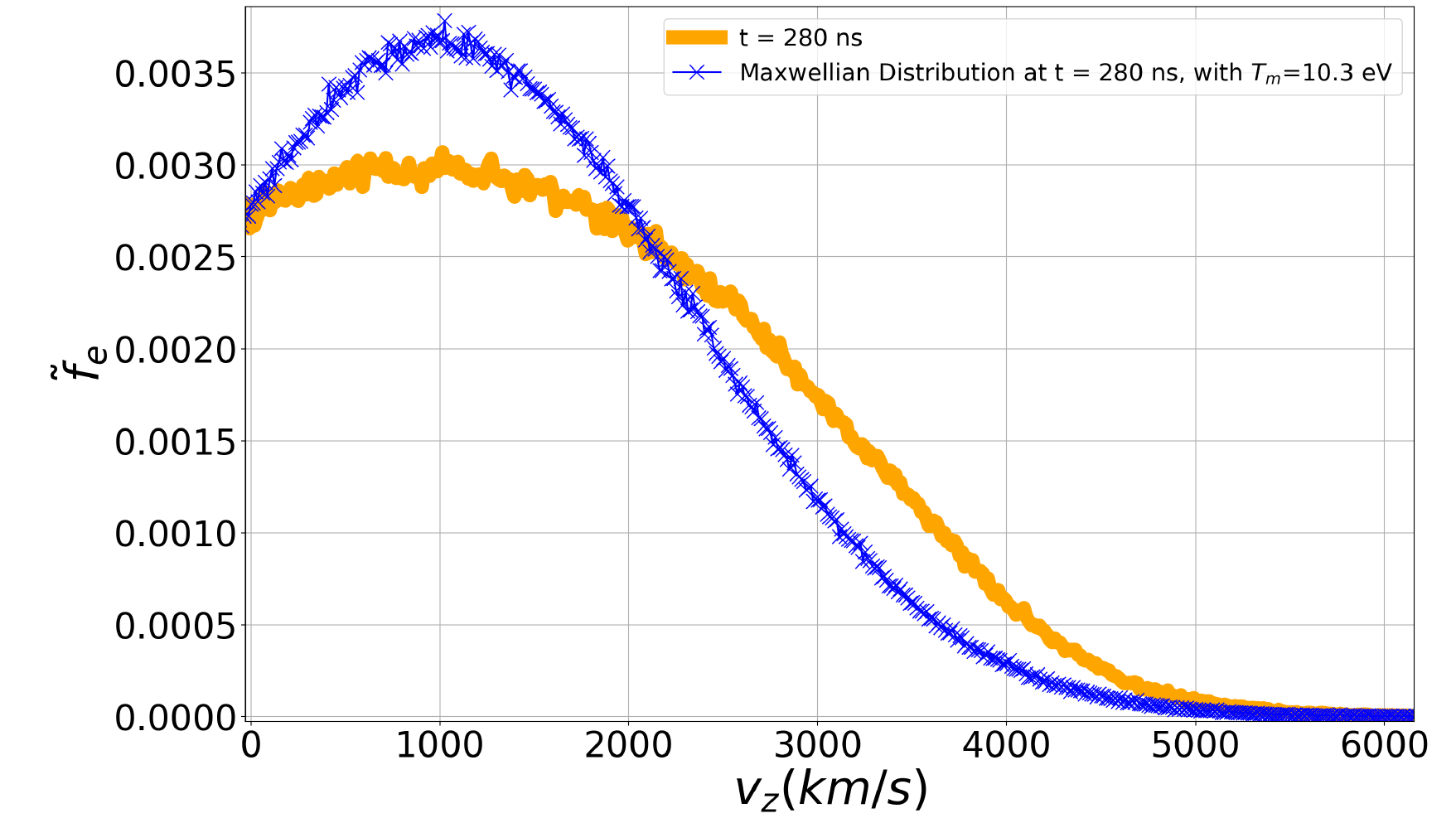}
\label{fig:5b}
\end{subfigure}
\begin{subfigure}[b]{0.48\textwidth}
\centering
\caption{ $\sim$ Middle of Phase 3}
\includegraphics[width=\textwidth, height=5.5cm]{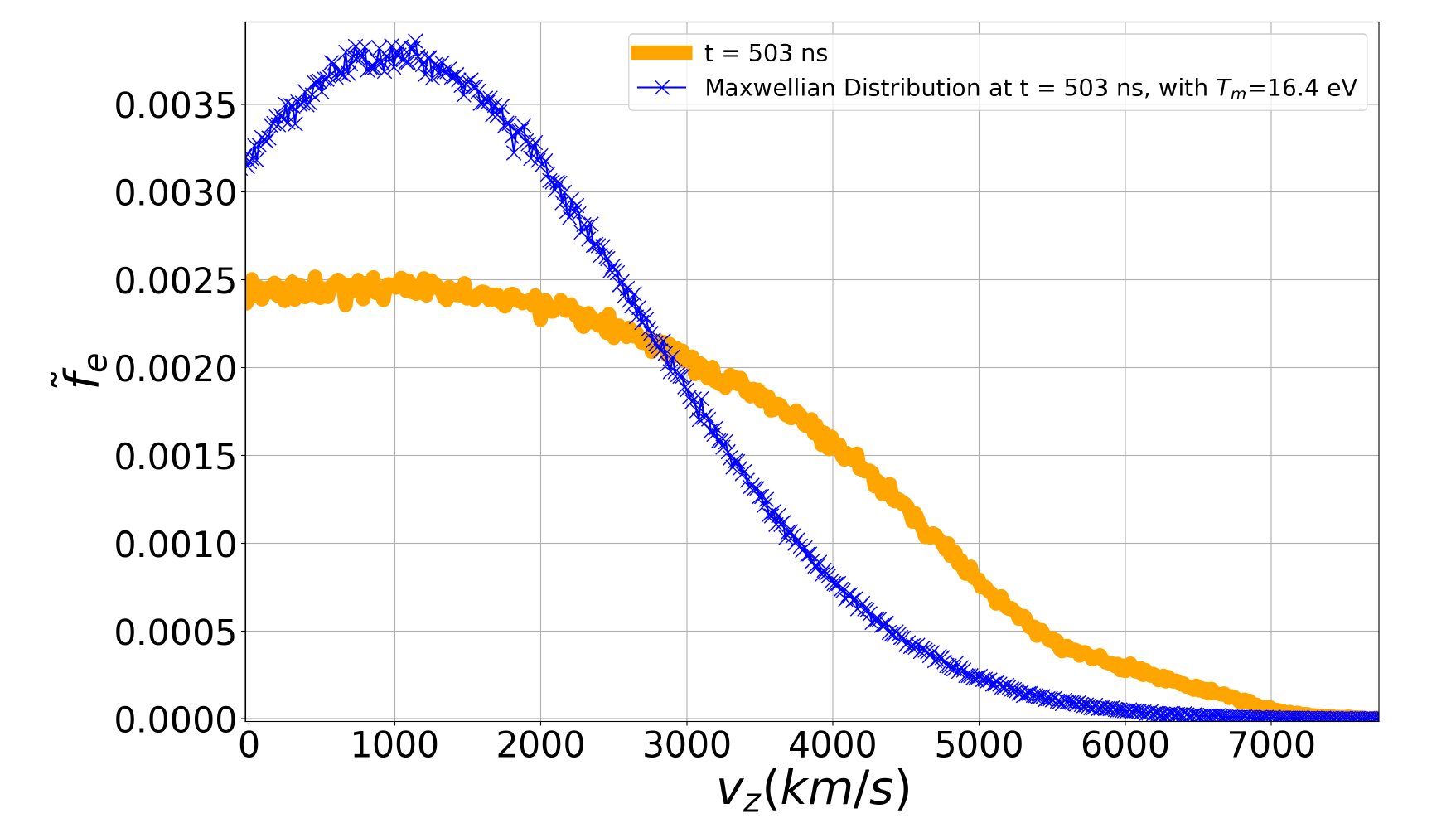}
\label{fig:5c}
\end{subfigure}
\hfill
\begin{subfigure}[b]{0.48\textwidth}
\centering
\caption{ $\sim$ End of Phase 3}
\includegraphics[width=\textwidth, height=5.5cm]{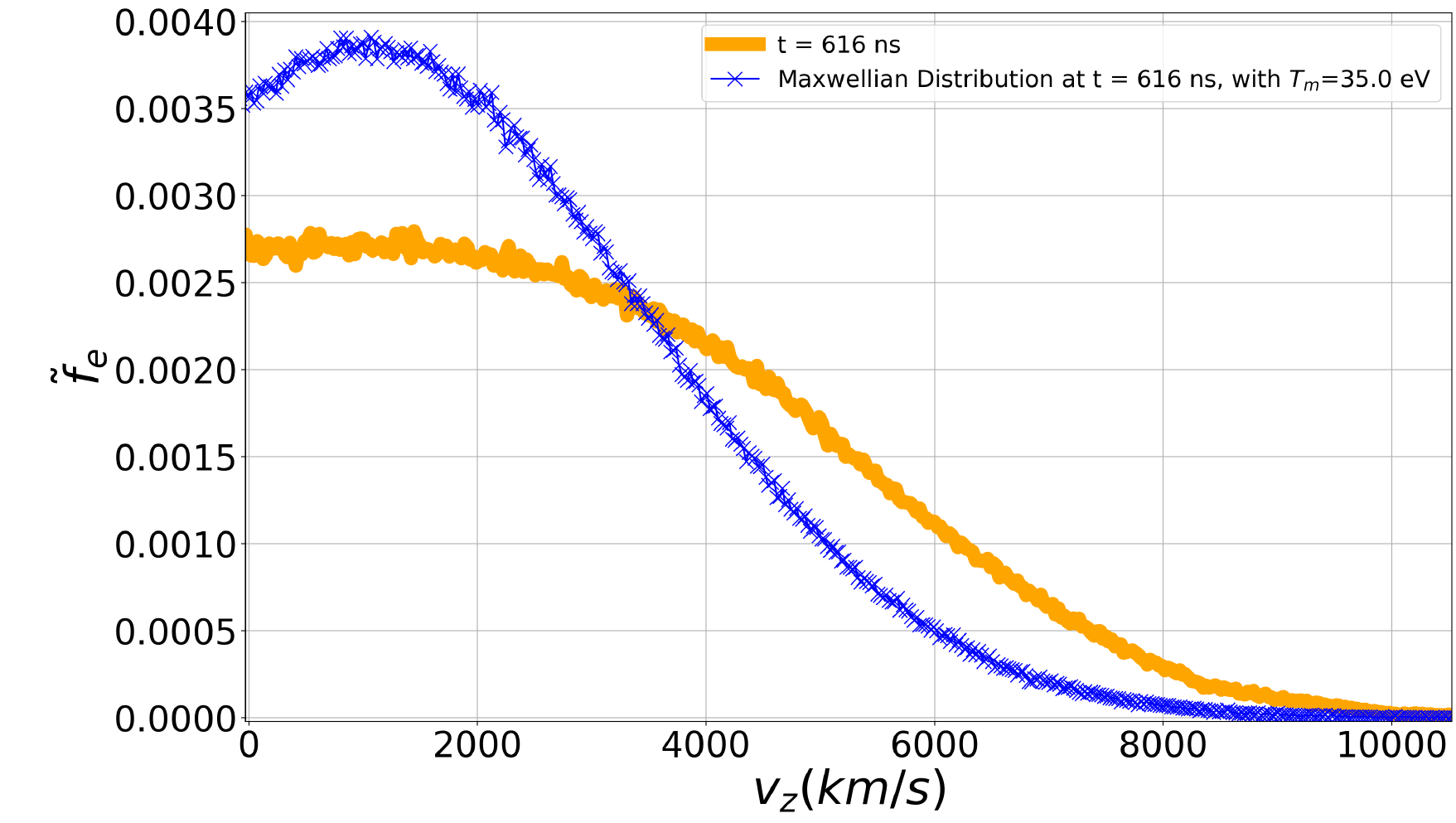}
\label{fig:5d}
\end{subfigure}
\begin{subfigure}[b]{0.48\textwidth}
\centering
\caption{ $\sim$ During Phase 4}
\includegraphics[width=\textwidth, height=5.5cm]{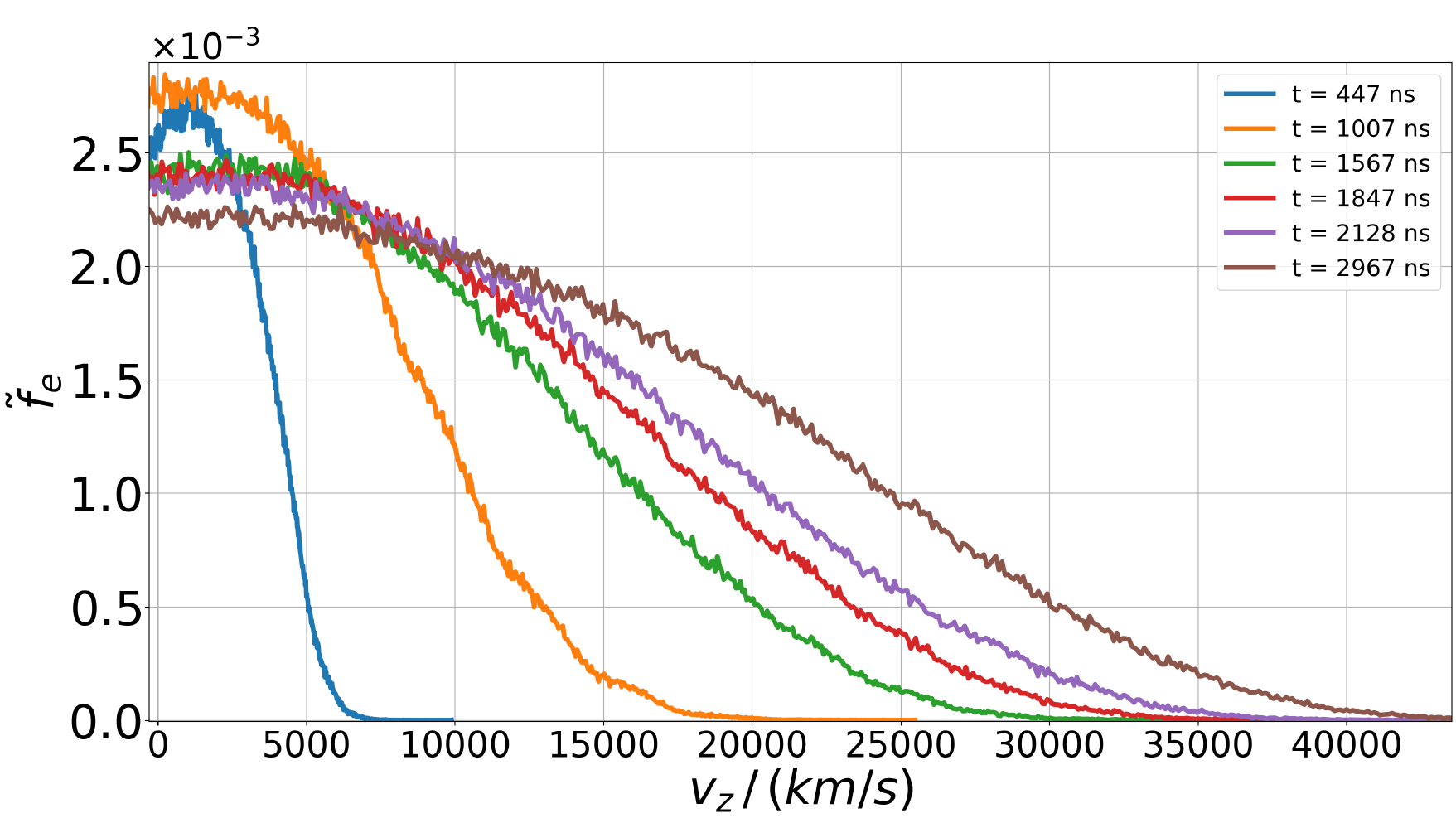}
\label{fig:5e}
\end{subfigure}
\caption{Time evolution of the electron distribution function $\tilde{f}_e$ as given by Eq.~\eqref{eq:fe} at different times, (a)--(d) from Phase 1 to Phase 3, and (e) during Phase 4.}
\label{fig:5}
\end{figure*}

\begin{figure*}[htbp]
\centering
\begin{subfigure}[b]{0.48\textwidth}
\centering
\caption{ Phase 1}
\includegraphics[width=\textwidth]{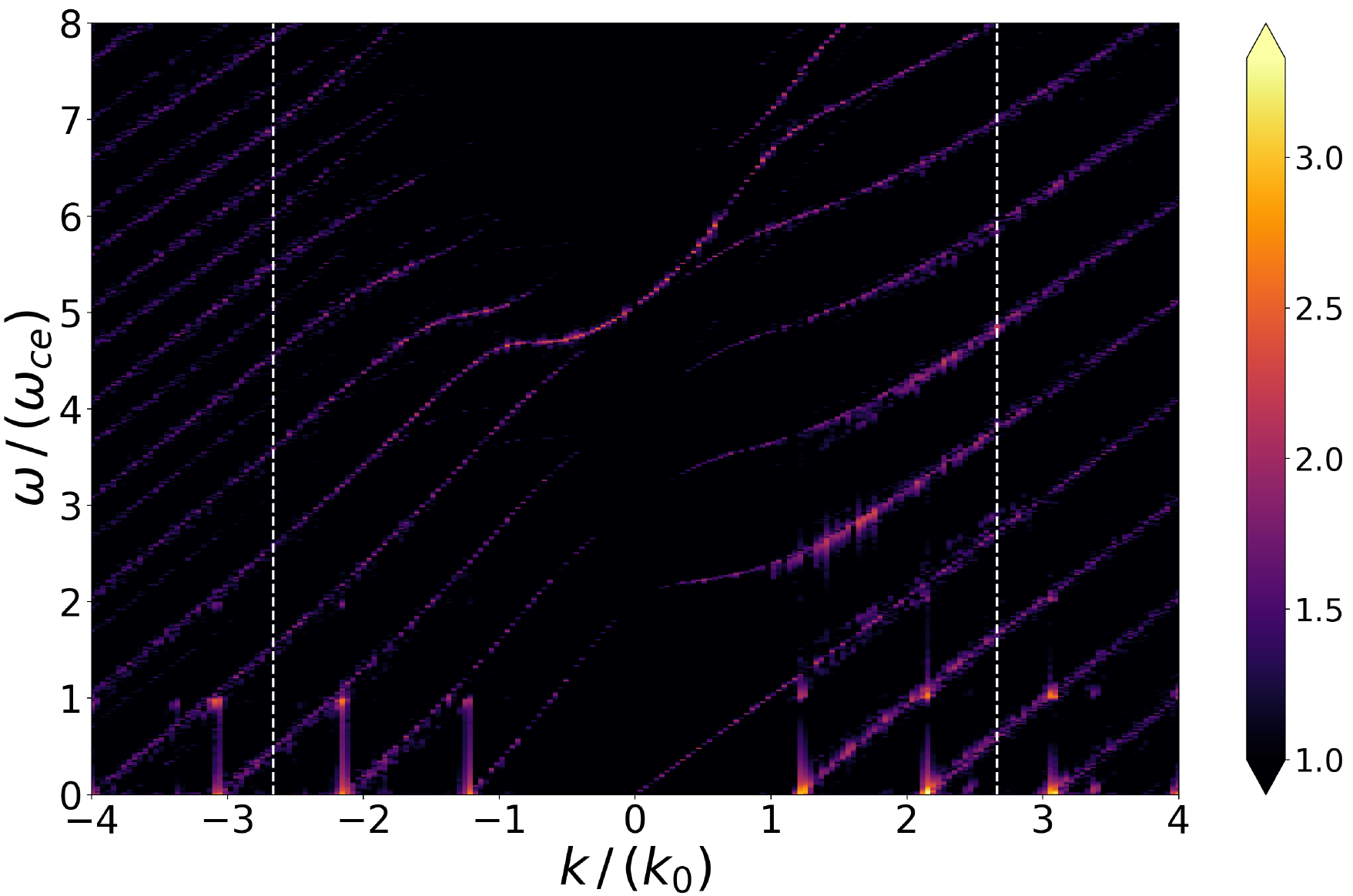}
\label{fig:6a}
\end{subfigure}
\begin{subfigure}[b]{0.48\textwidth}
\centering
\caption{ Phase 2}
\includegraphics[width=\textwidth]{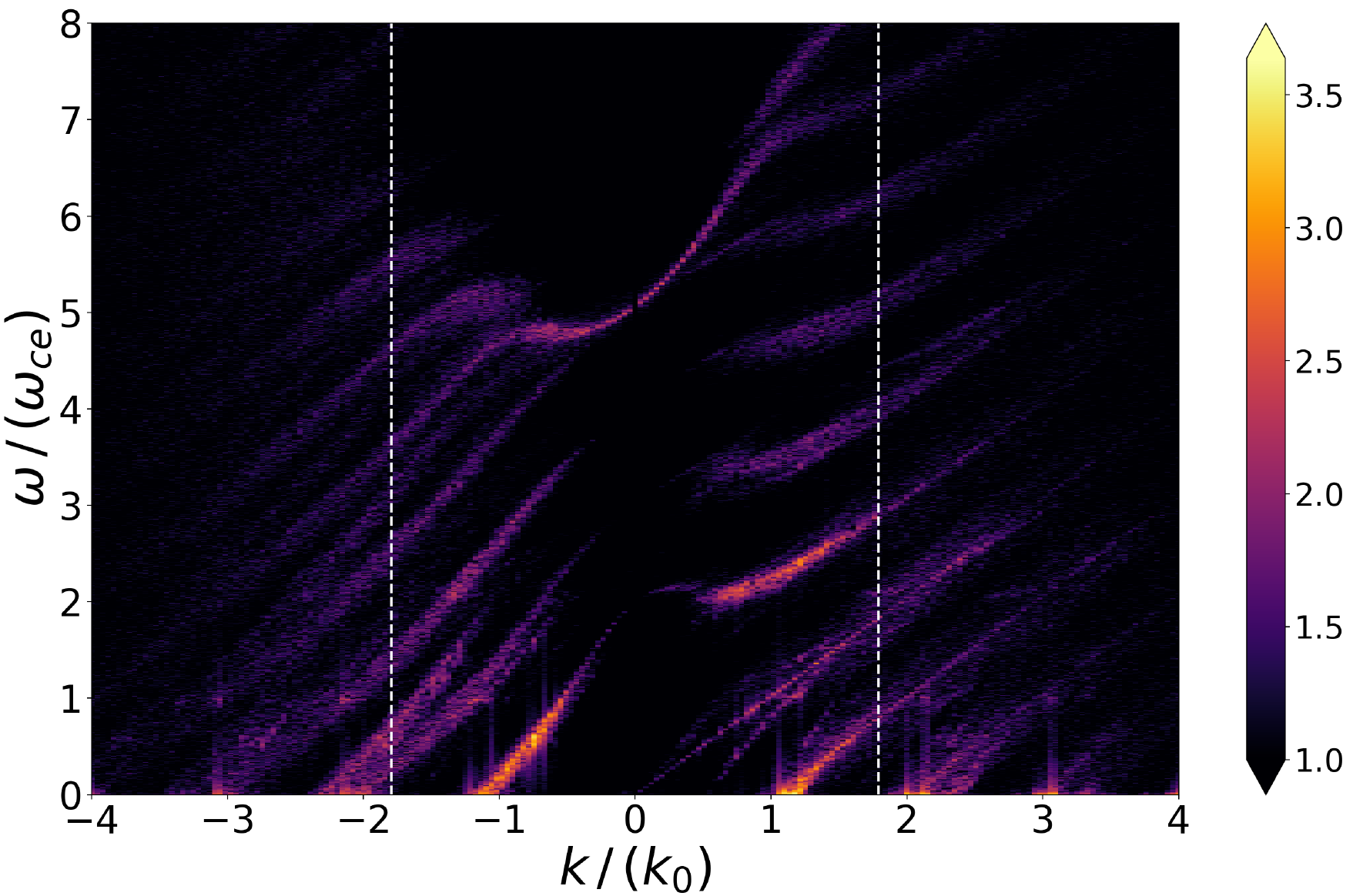}
\label{fig:6b}
\end{subfigure}
\begin{subfigure}[b]{0.48\textwidth}
\centering
\caption{ Phase 3}
\includegraphics[width=\textwidth]{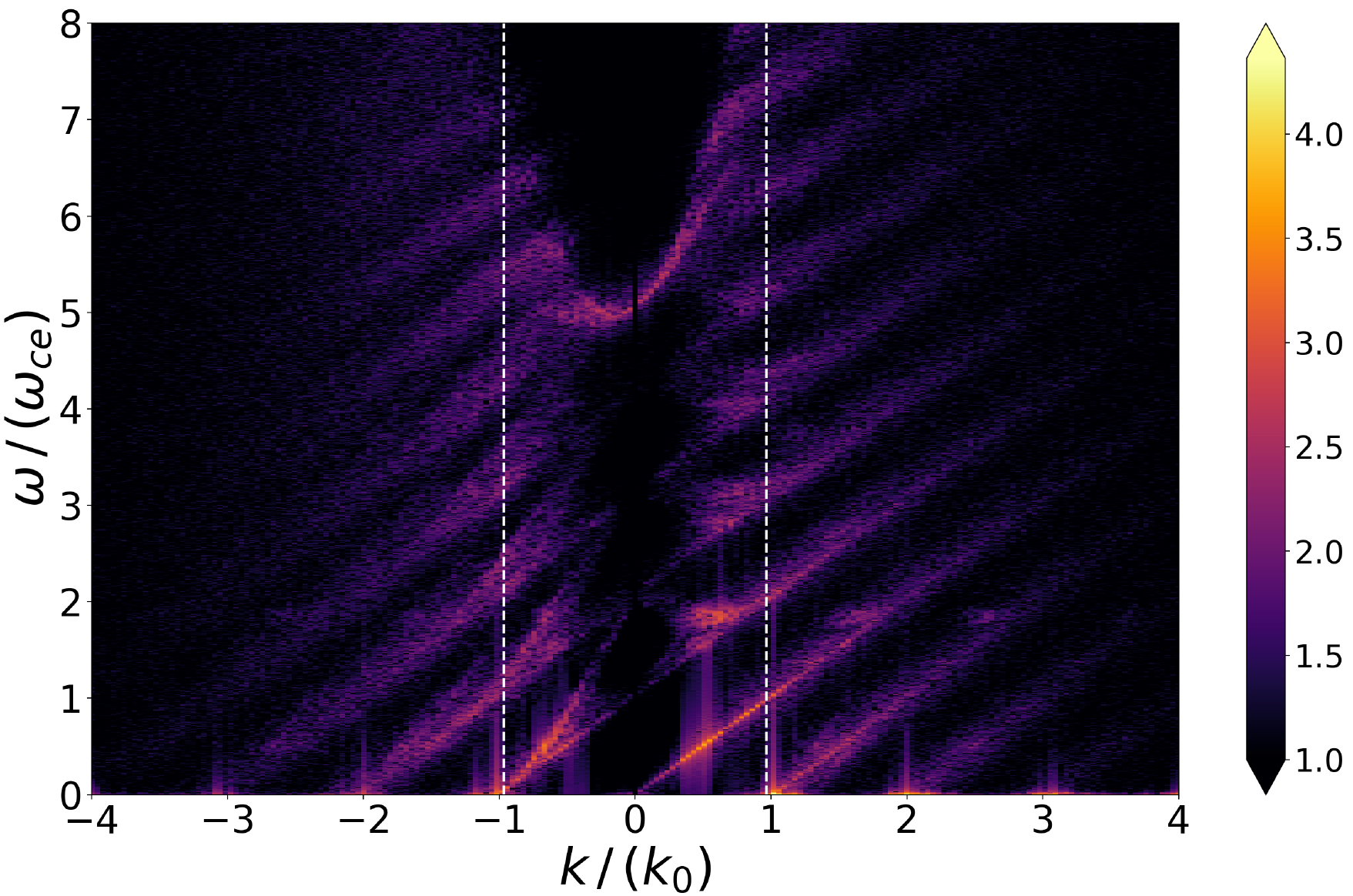}
\label{fig:6c}
\end{subfigure}
\begin{subfigure}[b]{0.48\textwidth}
\centering
\caption{ Phase 4}
\includegraphics[width=\textwidth]{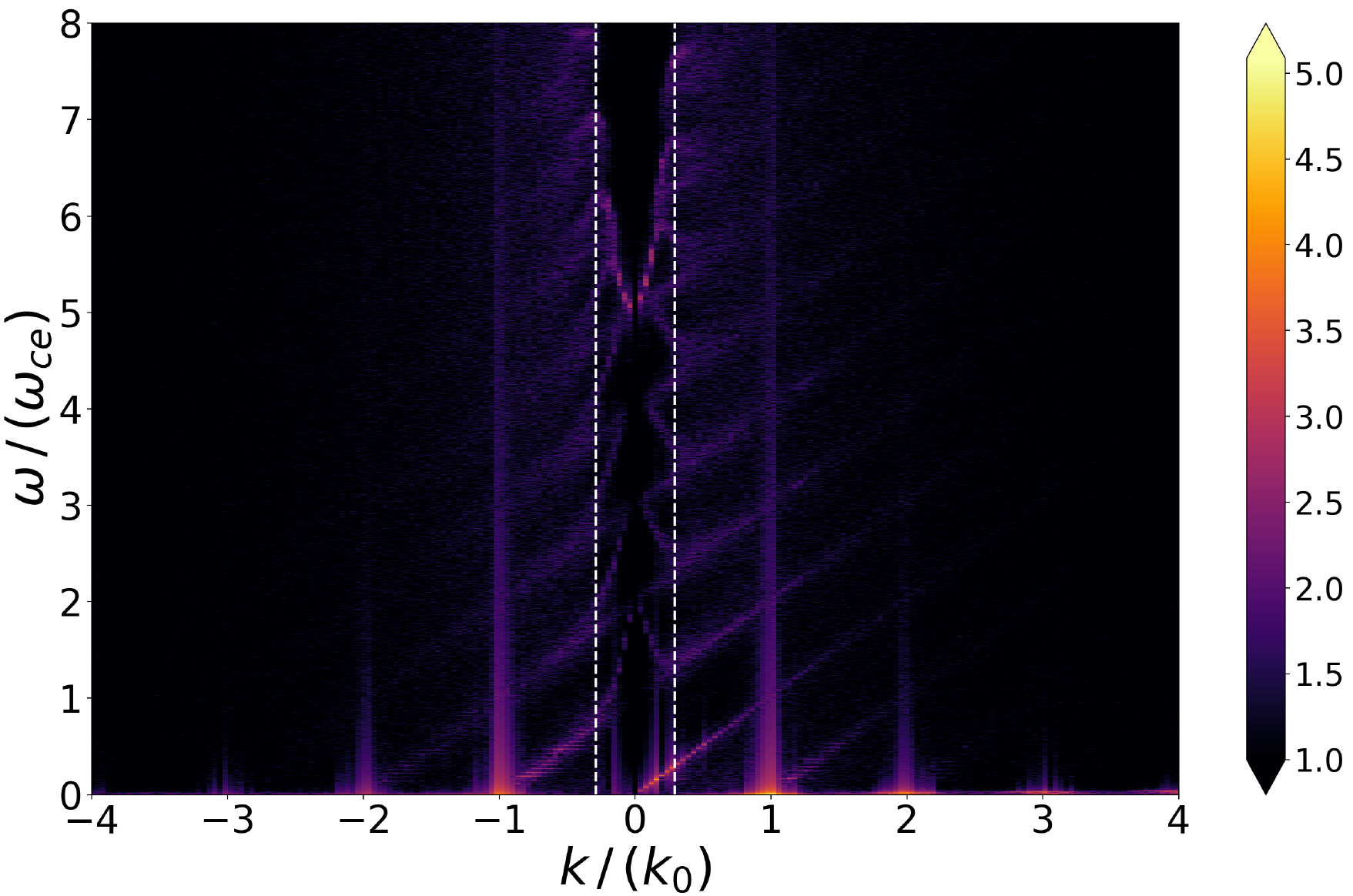}
\label{fig:6d}
\end{subfigure}
\begin{subfigure}[b]{0.48\textwidth}
\centering
\caption{ Phase 5}
\includegraphics[width=\textwidth]{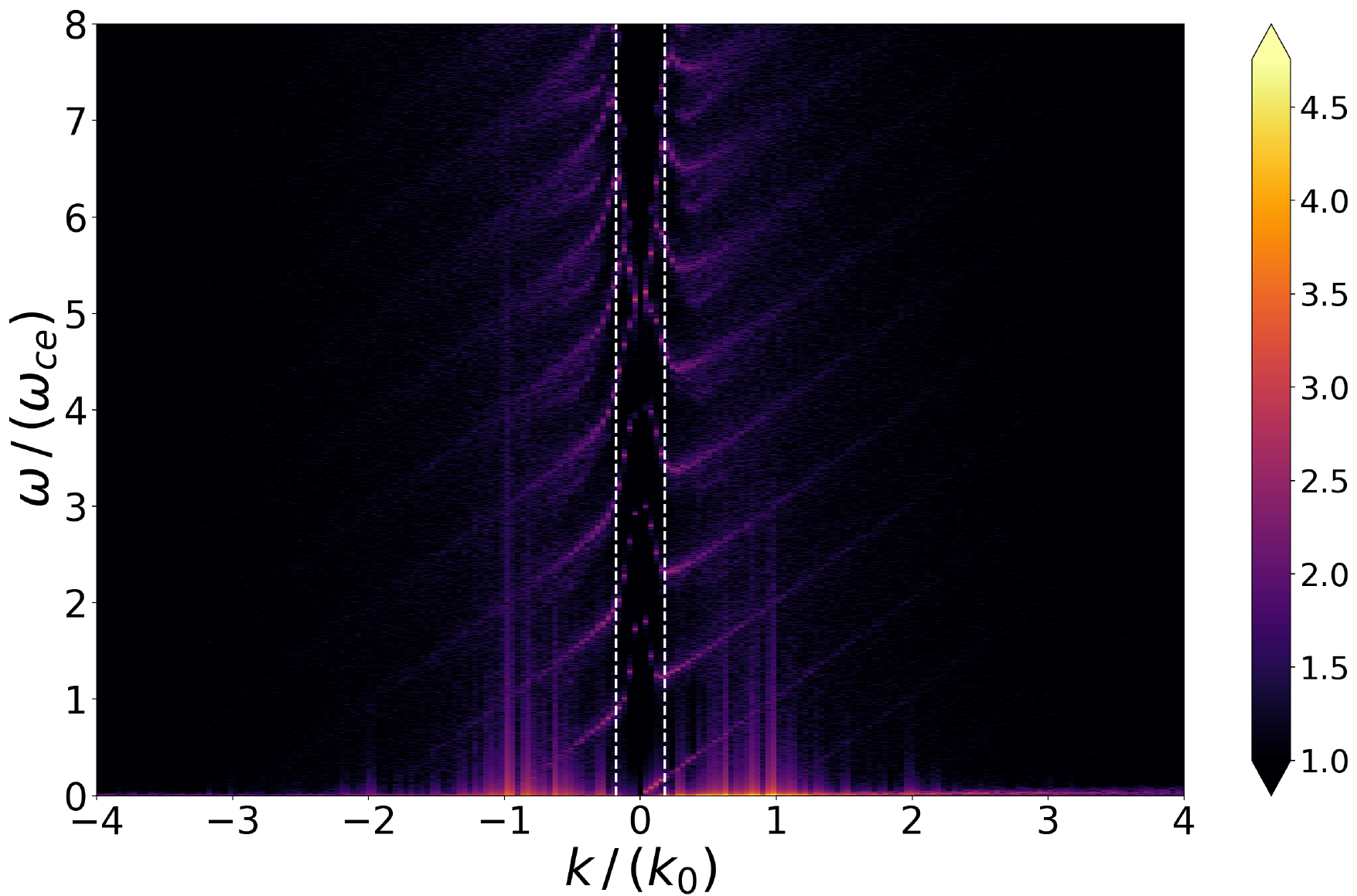}
\label{fig:6e}
\end{subfigure}
\caption{The logarithm of the 2D Fourier transform of the electric field $\log(\tilde{E}_z (\omega, k))$ given by Eq.~\eqref{eq:2dfft} for five different time ranges. (a) Phase 1 (0 < t < $\mathit{T_1}$), (b) Phase 2 ($\mathit{T_1} < t < \mathit{T_2}$), (c) Phase 3 ($\mathit{T_2}$ < t < $\mathit{T_3}$), (d) Phase 4 ($\mathit{T_3}$ < t < $\mathit{T_5}$), and (e) Phase 5 ($t \gg \mathit{T_5}$).}
\label{fig:6}
\end{figure*}

\begin{figure*}[htbp]
\begin{subfigure}[b]{0.48\textwidth}
\centering
\caption{ Phase 4}
\includegraphics[width=\textwidth]{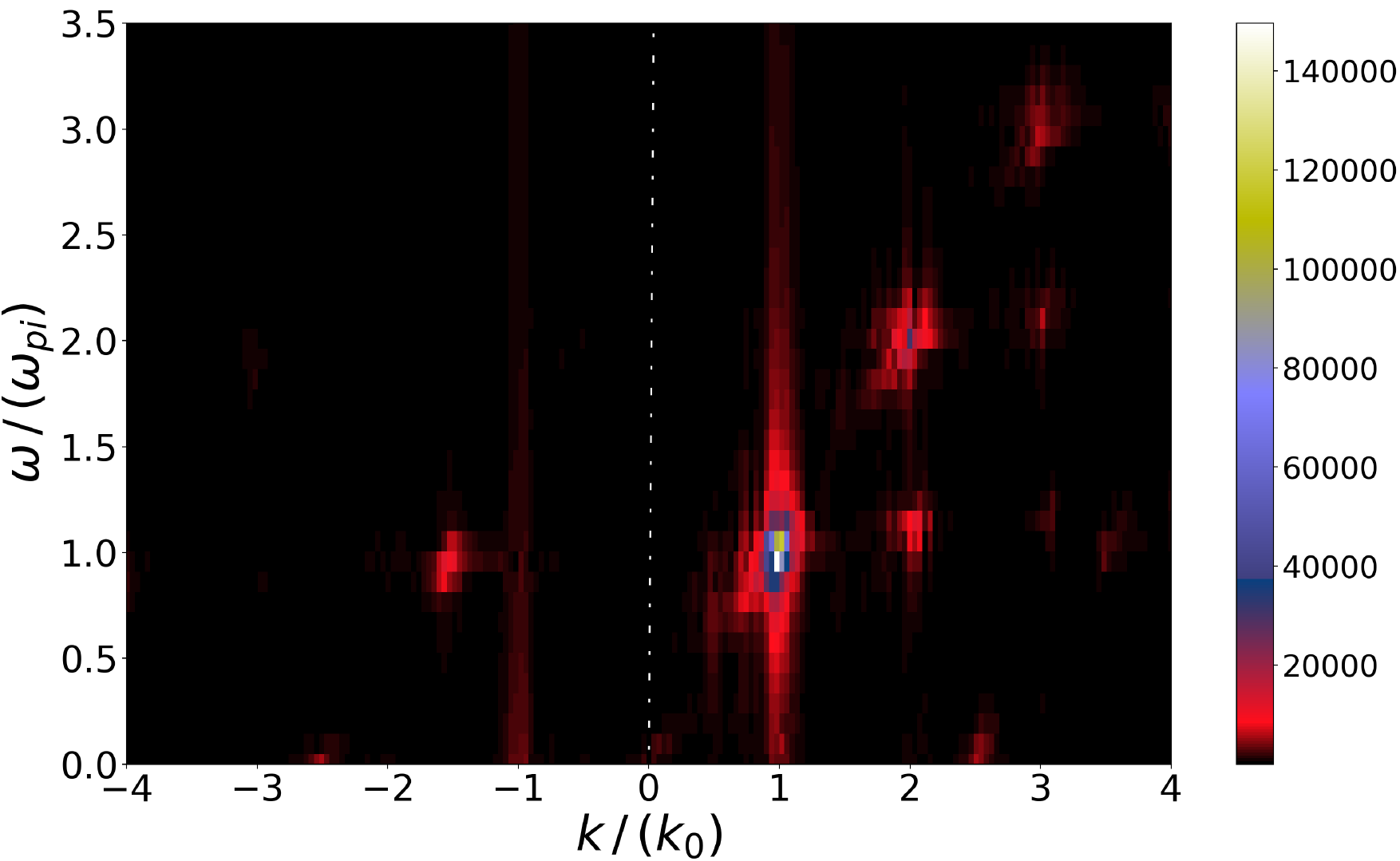}
\label{fig:7a}
\end{subfigure}
\begin{subfigure}[b]{0.48\textwidth}
\centering
\caption{ Phase 5}
\includegraphics[width=\textwidth]{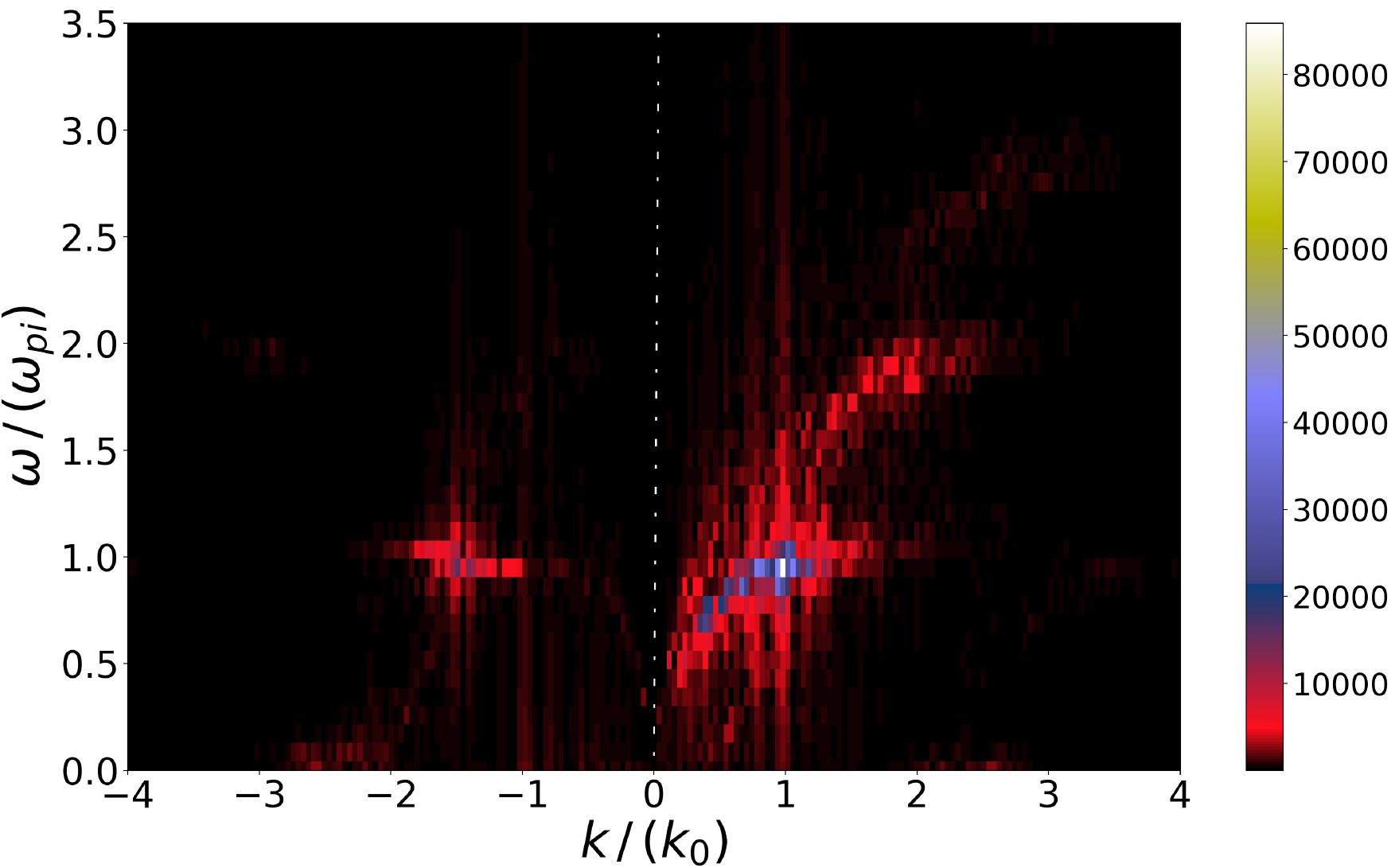}
\label{fig:7b}
\end{subfigure}
\caption{The 2D Fourier transform of the electric field $\tilde{E}_z (\omega, k)$, given by Eq.~\eqref{eq:2dfft} for (a) Phase 4 ($\mathit{T_3}$ < t < $\mathit{T_5}$), and (b) Phase 5 ($t \gg \mathit{T_5}$). $\tilde{E}_z (\omega, k)$ is plotted on a normal scale to highlight the non-linear harmonics at $\omega_{pi}$ in Phase 4 and Phase 5.}
\label{fig:7}
\end{figure*}
\begin{figure*}[htbp]
\centering
\begin{subfigure}[b]{0.44\textwidth}
\centering
\caption{ Phase 1}
\includegraphics[width=\textwidth]{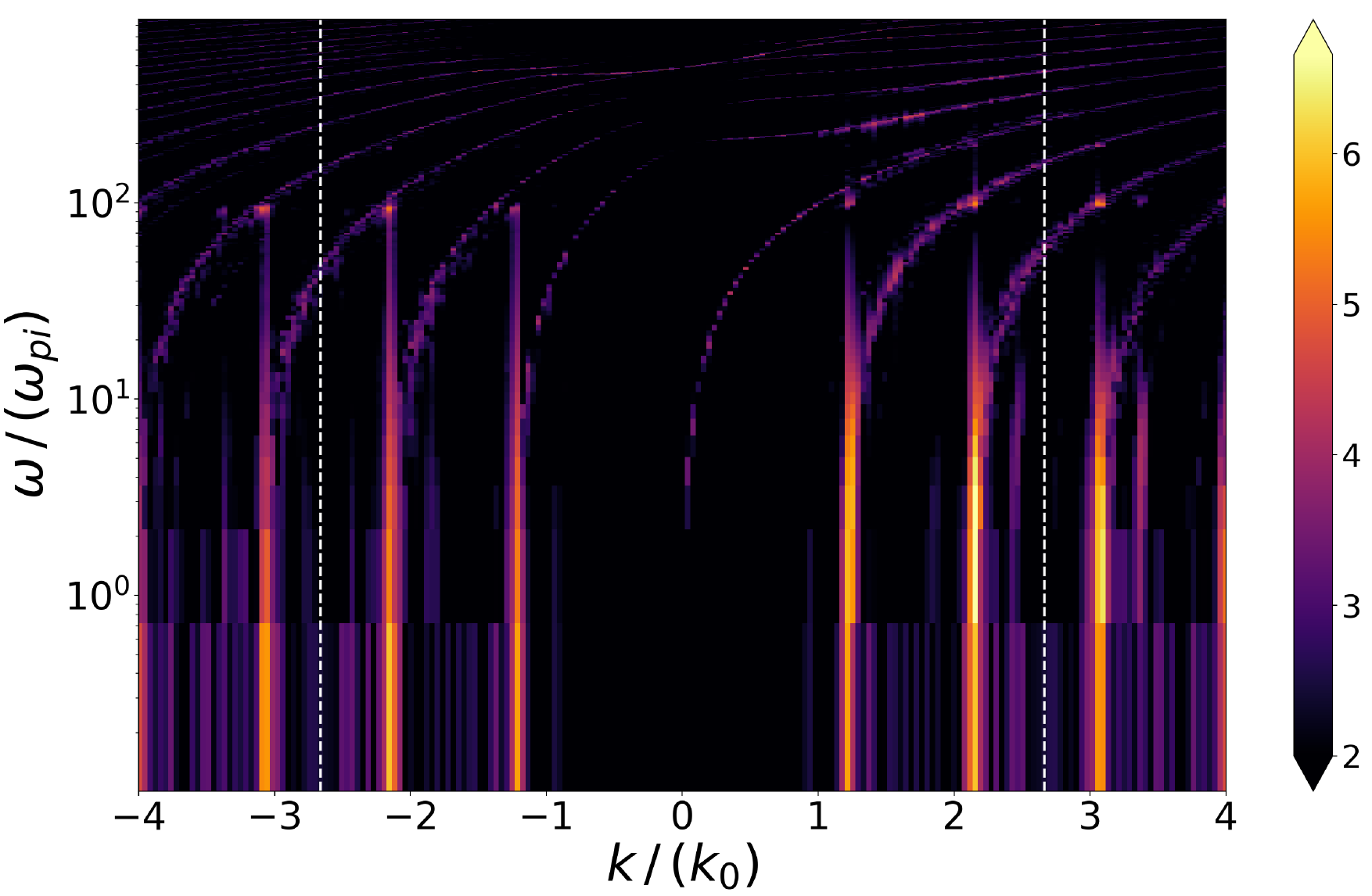}
\label{fig:8a}
\end{subfigure}
\begin{subfigure}[b]{0.44\textwidth}
\centering
\caption{ Phase 2}
\includegraphics[width=\textwidth]{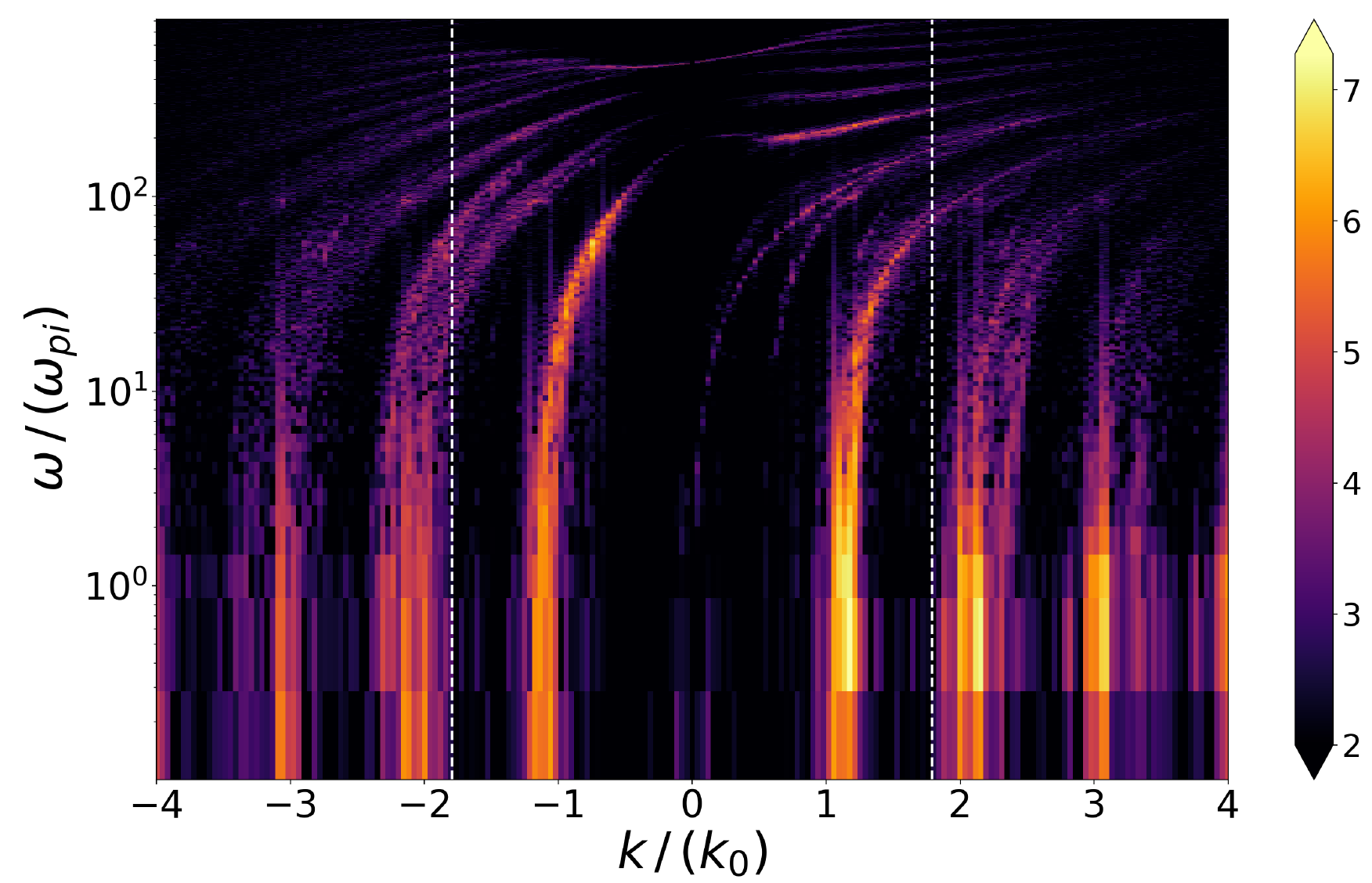}
\label{fig:8b}
\end{subfigure}
\begin{subfigure}[b]{0.44\textwidth}
\centering
\caption{ Phase 3}
\includegraphics[width=\textwidth]{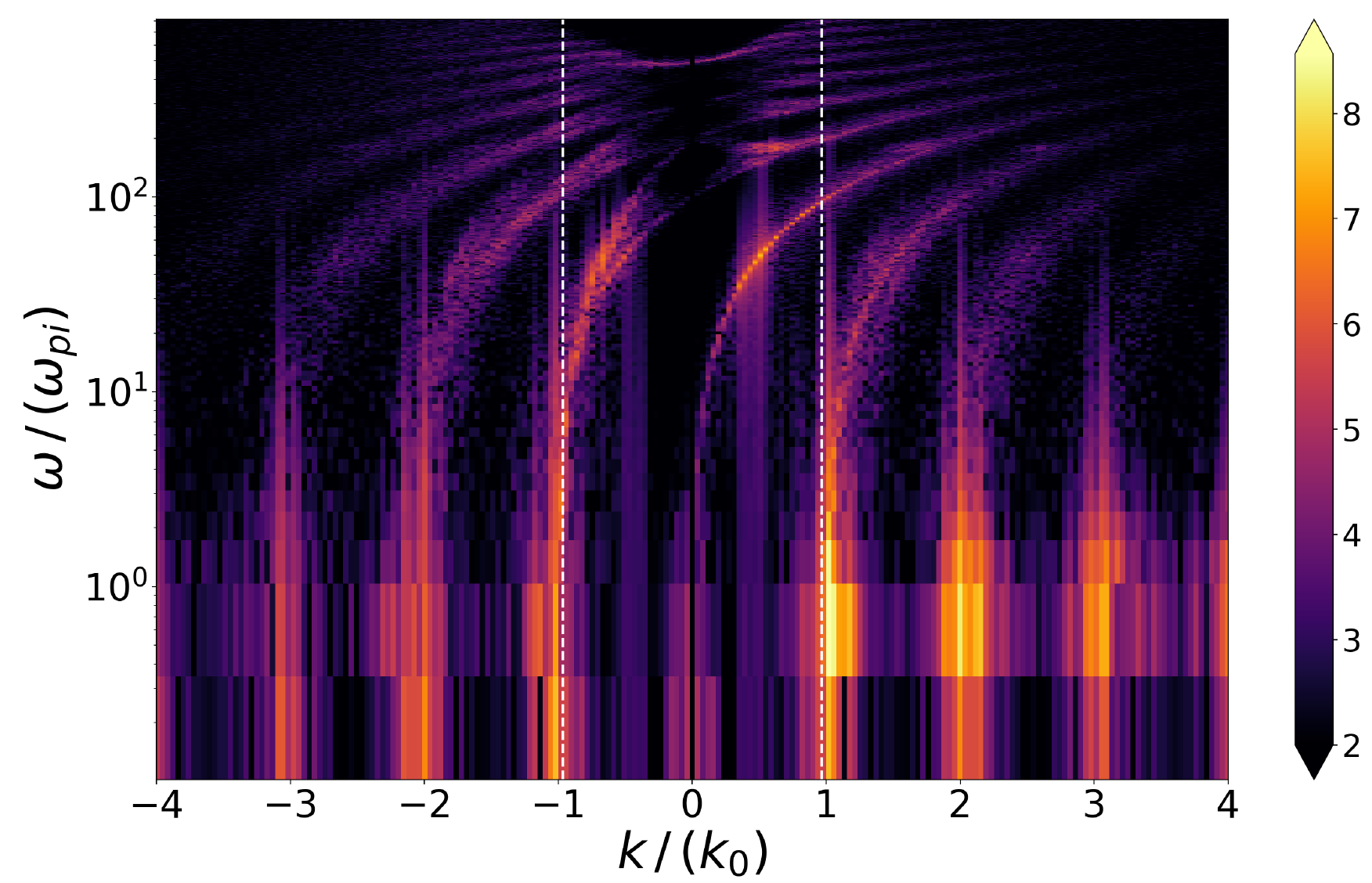}
\label{fig:8c}
\end{subfigure}
\begin{subfigure}[b]{0.44\textwidth}
\centering
\caption{ Phase 4}
\includegraphics[width=\textwidth]{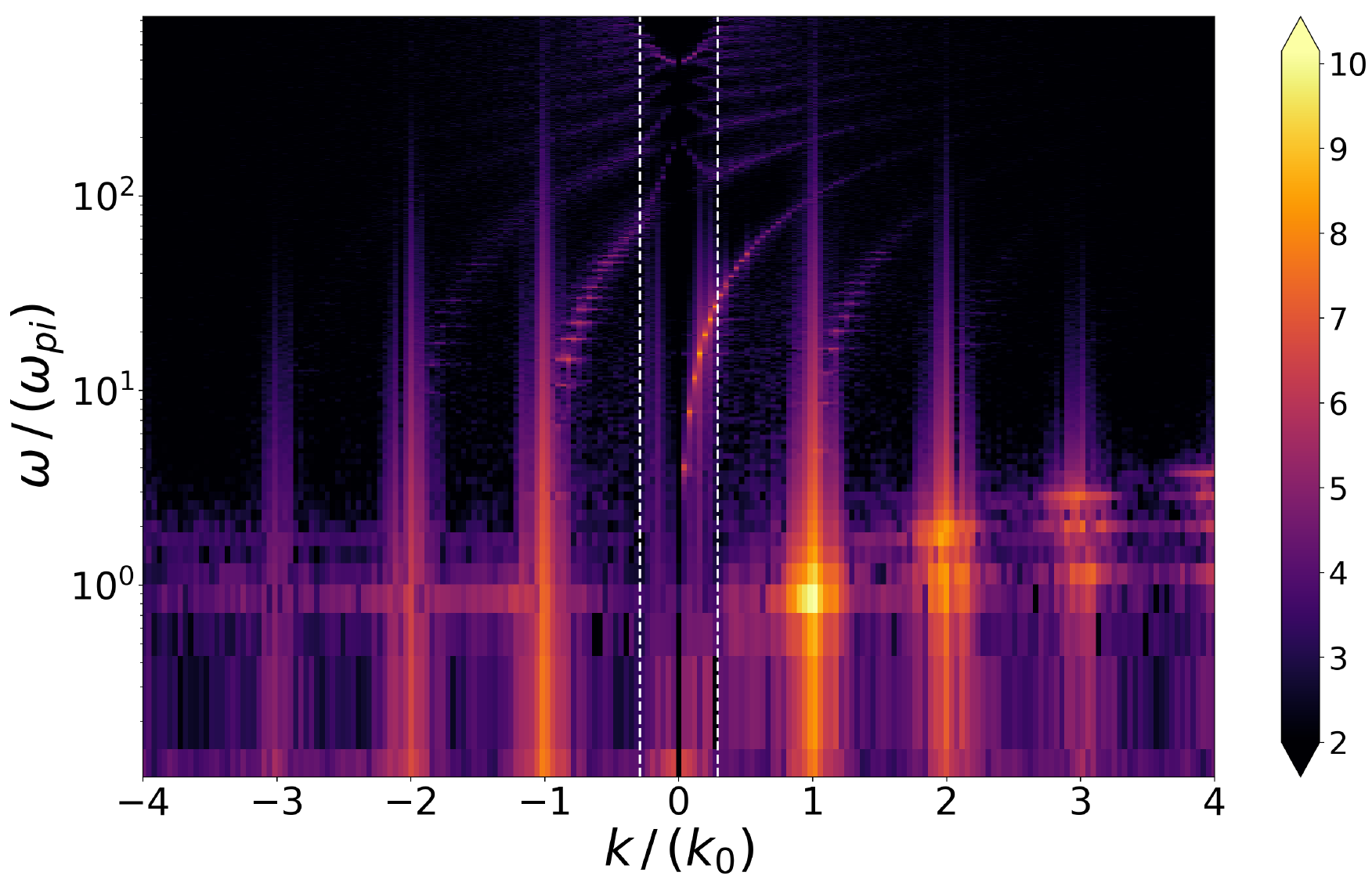}
\label{fig:8d}
\end{subfigure}
\end{figure*}
\begin{figure*}[htbp]\ContinuedFloat
\centering
\begin{subfigure}[b]{0.44\textwidth}
\centering
\caption{ Phase 5}
\includegraphics[width=\textwidth]{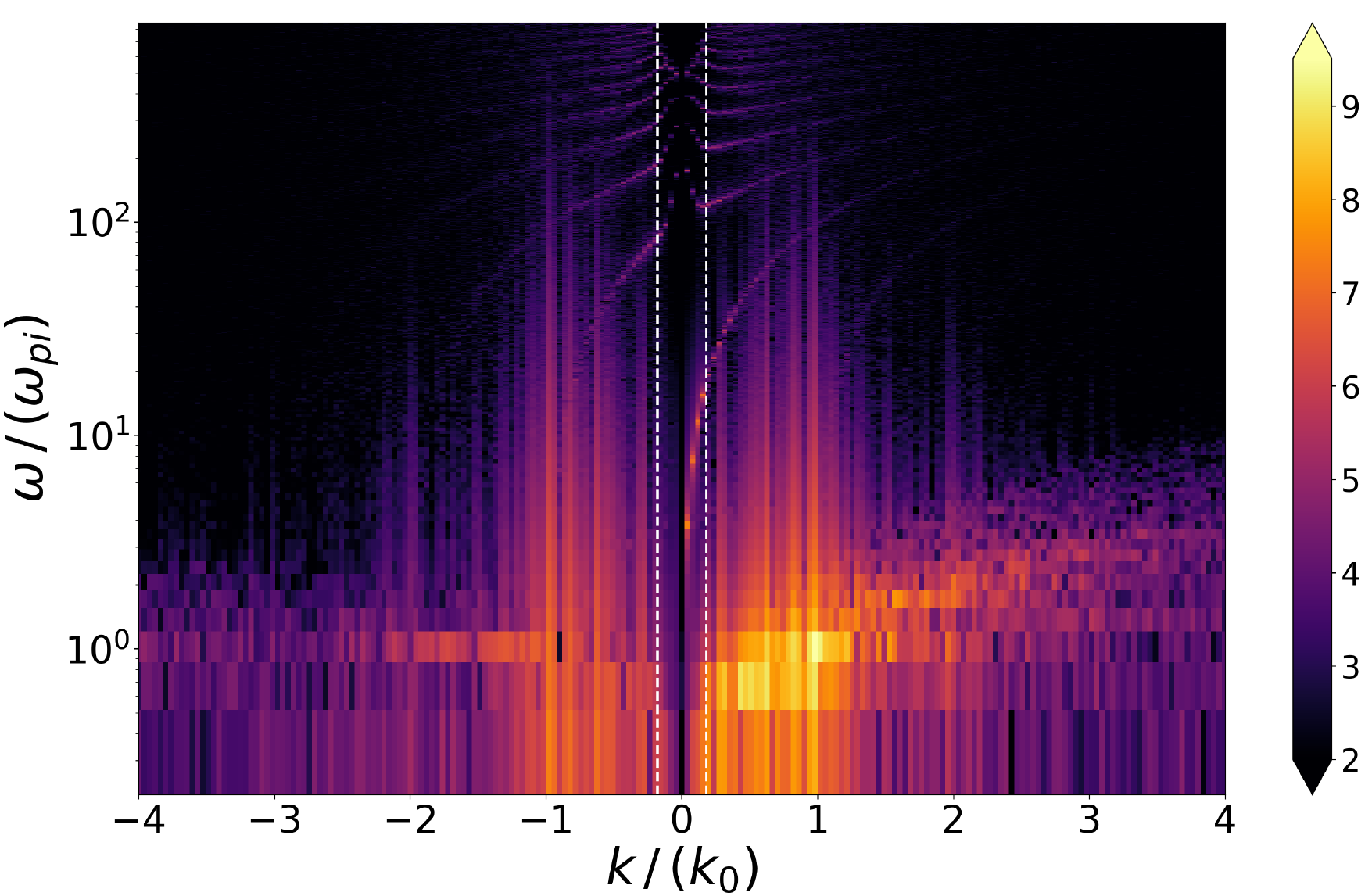}
\label{fig:8e}
\end{subfigure}
\caption{Power spectrum $\tilde{P} (\omega, k)$ given by Eq.~\eqref{eq:2dfft_power} for five different time ranges. (a) Phase 1 ($0 < t < \mathit{T_1}$), (b) Phase 2 ($\mathit{T_1} < t < \mathit{T_2}$), (c) Phase 3 ($\mathit{T_2}$ < t < $\mathit{T_3}$), (d) Phase 4 ($\mathit{T_3}$ < t < $\mathit{T_5}$), and (e) Phase 5 ($t \gg \mathit{T_5}$). The vertical axis is in logscale to show the distinction between the power in the lower- and upper-frequency spectrum and to cover all frequencies from $\omega_{pi}$ up to $750\,\omega_{pi} (\sim 8\omega_{ce})$. The two vertical white dotted lines in Figs.~\ref{fig:8a}--~\ref{fig:8e} correspond to $k = \pm k_s$ values obtained at the end of each phase.}
\label{fig:8}
\end{figure*}


Before proceeding further, we give a brief overview of Figs.~\ref{fig:3}--\ref{fig:8}, characterizing different quantities derived from the PIC simulation data for Case 1A. 
Fig.~\ref{fig:3} shows the time evolution of the spectral coefficients for the 1D Fourier transform of the electric field $E_z$ in Fig.~\ref{fig:3a}, the ion density $n_i$ in Fig.~\ref{fig:3b}, and the electron density $n_e$ in Fig.~\ref{fig:3c}; the  $T_1,T_2,\dots,T_5$ times are depicted by white horizontal dotted lines. The white curve starting at $t=0$ and $k/k_0=2.7$ shows the evolution of the value $k_s=1/\sqrt{2} \lambda_{De}$ representing the wavelength of the maximum growth rate of the ion-acoustic mode calculated with the electron temperature at a given time.
Fig.~\ref{fig:4} shows snapshots of the evolution of the ion phase space in column 1, the ion distribution function in column 2, and the electrons phase space in column 3. The snapshots are provided for six different times from Phase 1 to Phase 5.
Figs.~\ref{fig:5a}--\ref{fig:5d} shows the flattening of the electron distribution function in comparison with the Maxwellian distribution at several times up to the end of Phase 3, and Fig.~\ref{fig:5e} shows the further evolution of the electron distribution function for later times up to the end of Phase 4.

The normalized electron $\tilde{f}_e$ and ion $\tilde{f}_i$ distribution functions (in Figs.~\ref{fig:4}--\ref{fig:5}) are defined as
\begin{equation}
\label{eq:fe}
\tilde{f}_{e,i} (v_z, t) = \frac{1}{n_0 L_z} \int_{-L_z/2}^{L_z/2} f_{e,i}(z,v_z,t) dz.
\end{equation}

Figs.~\ref{fig:6}--\ref{fig:8} show the linear and non-linear evolution of the $\omega$--$k$ spectra, in both the low ($\sim \omega_{pi}$) and high ($\sim \omega_{ce}$) frequency range. Fig.~\ref{fig:6} shows the $\omega$--$k$ spectra obtained from the logarithm of the spectral coefficients from the 2D Fourier transform of the electric field $E_z$. These spectra show the evolution of the dispersion relation from the initial linear stage (Phase 1) to the deeply non-linear and saturated state (Phase 5). Figs.~\ref{fig:7a} and \ref{fig:7b} provide a magnified view of the $\omega$--$k$ spectra near $\omega_{pi}$ frequencies for Phases 4 and 5, corresponding to the wider-range spectra shown in Figs.~\ref{fig:6d} and \ref{fig:6e}. 

It is important to note that, unlike in Fig.~\ref{fig:6}, the spectral coefficients in Fig.~\ref{fig:7} are not plotted on a logarithmic scale, allowing for the enhancement of the spectral features at $\omega \sim \omega_{pi}$ in these later phases. The two vertical white lines in Figs.~\ref{fig:6} through \ref{fig:8} correspond to $k = \pm k_s$ values obtained at the end of each time range used in the 2D Fourier transforms. Figs.~\ref{fig:8a}--\ref{fig:8e} display the power spectra across the five different phases, correlating directly with the $\omega$--$k$ spectra from Figs.~\ref{fig:6a}--\ref{fig:6e}. In Fig.~\ref{fig:8}, we adopt a logarithmic scale for the vertical $\omega$ axis to effectively capture the full range of frequencies observed in the simulations, ranging from $\omega_{pi}$ to approximately $8\,\omega_{ce}$. This scaling choice highlights the concentration of energy  at the cyclotron harmonics $k/k_0 = 1, 2, 3, \ldots$ throughout the simulation.

The 1D discrete FFT, $\tilde{E}_z (k, t)$ in Fig.~\ref{fig:3a}, 2D discrete FFT, $\log(\tilde{E}_z (\omega, k))$ in Fig.~\ref{fig:6}, and $\tilde{E}_z (\omega, k)$ in Fig.~\ref{fig:7}, and the power spectrum $\tilde{P} (\omega, k)$ in Fig.~\ref{fig:8} are computed as follows:
\begin{equation}
\label{eq:1dfft}
\tilde{E}_{z} (k, t) = \frac{2}{N_z} \sum_{ \ell = 0}^{N_z - 1} E_{z}(z, t) e^{-i 2\pi k \ell/N_z},
\end{equation}
\begin{equation}
\label{eq:2dfft}
\tilde{E}_z (\omega, k) = \frac{2}{N_t} \frac{2}{N_z} \sum_{\kappa = 0}^{N_t - 1} \sum_{\ell = 0}^{N_z - 1} E_z(z, t) e^{-i2 \pi \omega \kappa/N_t} e^{-i 2\pi k\ell/N_z},
\end{equation}
\begin{equation}
\label{eq:2dfft_power}
\tilde{P} (\omega, k) = log(\tilde{E}_z^{2} (\omega, k)),
\end{equation}
where $N_z$ represents the total number of discrete points in the spatial domain and $N_t$ represents the total number of discrete points in the temporal domain considered for the 2D FFT. Whereas $\tilde{E}_z (k, t)$ is calculated for the entire evolution together from Phase 1 to Phase 5, $\tilde{E}_z (\omega, k)$ is calculated for a time range of the signal $E_z$, for Phases 1--5 separately. The 1D discrete FFTs for ion ( $\tilde{n}_i (k, t)$ in Fig.~\ref{fig:3b}) and electron ( $\tilde{n}_e(k ,t)$ in Fig.~\ref{fig:3c}) density perturbations are calculated from Eq.~\eqref{eq:1dfft} by changing the signal $E_z(z, t)$ to $n_i(z ,t)$ and $n_e(z ,t)$, respectively.


\subsection{\label{sec3:level2} Phases 1--3: Linear stage and resonance between the ECDI $m=1$ mode and the ion-acoustic modes}

To discuss the linear stage of ECDI observed in Phase 1, it is useful to recall the basic features of the linear dispersion relation for ECDI driven by the electron $\mathbf{E}\times \mathbf{B}$ drift in the $z$-direction. The linear dispersion relation of the ECDI for this case is given by

\begin{equation}
1+\chi _{i}+\chi _{e}=0,  \label{deq}
\end{equation}
where $\chi _{i}$ and $\chi _{e}$ are the ion and electron susceptibilities, respectively. The unmagnetized ion susceptibility is given by

\begin{equation}
\label{eq:chii}
\chi _{i}=-\frac{\omega _{pi}^{2}}{2k^{2}v_{ti}^{2}}Z^{\prime }\left( \frac{%
\omega }{\sqrt{2}kv_{ti}}\right) \simeq -\frac{{\omega _{pi}^{2}}}{\omega
^{2}},
\end{equation}
where $\omega _{pi}=\sqrt{n_{0}e^{2}/m_{i}\epsilon _{0}}$, $Z$ is the usual plasma dispersion function, $n_{0}$ is the plasma density, $e$ is the electron charge, and $m_{i}$ is the ion mass. We also assume cold ions in the approximation in Eq.~\eqref{eq:chii}. We write the electron susceptibility for warm magnetized electrons for 1D fluctuations propagating in the direction of the current as

\begin{eqnarray}
&&\chi _{e}=\frac{1}{k^{2}\lambda _{De}^{2}}\left[ 1-\exp \left( -k^{2}\rho
_{e}^{2}\right) I_{0}\left( k^{2}\rho _{e}^{2}\right) \right.   \notag \\
&& \hspace{25mm}\left. -2\left( \omega -kV_{de}\right) ^{2}\sum\limits_{m=1}^{\infty }%
\frac{\exp \left( -k^{2}\rho _{e}^{2}\right) I_{m}\left( k^{2}\rho
_{e}^{2}\right) }{\left( \omega -kV_{de}\right) ^{2}-m^{2}\Omega _{ce}^{2}}%
\right],   \label{chie}
\end{eqnarray}%
where $k=k_{z}$ is the wave vector in the direction of the electron current, $I_{m}$ is the modified Bessel function of the second kind of order $m$, $ b=k^{2}\rho _{e}^{2}$, $\rho_{e} ^{2}=T_{e}/m_{e}\omega _{ce}^{2}$ is the electron Larmor radius, and $V_{de}$ is the electron drift velocity.

For high frequencies, the ion contribution can be neglected, and then from $1+\chi_{e}=0$, one has electron Bernstein modes centered around the cyclotron frequencies $\omega \simeq n\omega _{ce}$. When the electron drift is included, these frequencies are shifted by the Doppler shift $kV_{de}$.

Neglecting all effects of the magnetic field in the electron response so that   $\chi_{e}\simeq \left( k^{2}\lambda _{De}^{2}\right) ^{-1}$,  one can obtain  from (\ref{deq}) the ion-acoustic mode,

\begin{equation}
\omega ^{2}=\frac{\omega _{pi}^{2}}{1+\left( k^{2}\lambda _{De}^{2}\right) ^{-1}}.
\end{equation}

For low frequencies $\omega \ll \omega _{ce}$, the resonant terms in the electron response (the last term in Eq. (\ref{chie})) can be neglected, and one can obtain the ion-acoustic branch modified by the magnetic field~\cite {ArefevTechPhys1970} as
\begin{equation}
\omega ^{2}=\frac{\omega _{pi}^{2}}{1+\left[ 1-\exp \left( -b\right)
I_{0}\left( b\right) \right] /k^{2}\lambda _{De}^{2}}.  \label{mias}
\end{equation}
For moderate values of the electron Larmor radius, $b<1,$ $1-\exp \left( -b\right) I_{0}\left( b\right) \rightarrow b$, and the modified ion-acoustic mode (\ref{mias}) reduces to the lower-hybrid mode
\begin{equation}
\omega ^{2}=\frac{\omega _{pi}^{2}}{1+\ \omega _{pe}^{2}/\omega _{ce}^{2}}%
\simeq \omega _{ce}\omega _{ci}.
\end{equation}%
We note that for the chosen parameters, $\omega _{pe}\gg \omega_{ce}$.

For $B\rightarrow 0$, $b\rightarrow \infty $,  $\exp \left( -b\right) I_{0}\left( b\right) \sim 1 \sqrt{2\pi b}\rightarrow 0,$ equation (\ref{mias}) becomes the  standard ion-acoustic mode for an unmagnetized plasma, which reduces to $\omega_{pi}$ for short-wavelength fluctuations
\begin{equation}
\omega ^{2}=\frac{\omega _{pi}^{2}}{1+1/k^{2}\lambda _{De}^{2}}\rightarrow
\omega _{pi}^{2}\text{ for }k^{2}\lambda _{De}^{2}\gg 1.
\end{equation}%

Normally, the Bernstein and ion-acoustic mode are well-separated in frequencies and do not overlap. However, in plasma with drifting electrons, the Doppler-shifted electron Bernstein waves and the low frequency ion-acoustic mode may overlap, resulting in reactive instability near the crossing points defined by the condition

\begin{equation*}
\omega -m\omega _{ce}-k_{z}V_{de}\simeq 0.
\end{equation*}

This represents the ECDI forming a discrete set of narrow-band unstable modes centered around $k/k_{0}=1,2,3,\ldots $, $k_{0}\equiv \omega _{ce}/V_{de}$. The growth rates $\gamma $ and frequencies $ \omega $ typically fall within a range of those associated with $\omega _{pi}$. We note that $\omega _{pi}\ll \omega _{ce}$.

For the chosen parameters, the linear solution of the dispersion equation (\ref{deq})  predicts that the growth rate of the $m=2$ mode is slightly larger than $m=1,3$, and other modes.\cite{tavassoli2023electron} In the linear Phase 1 stage, our simulations show the growing modes at cyclotron resonances $k/k_{0}=1,2,3,\ldots $ and $k/k_{0}=-1,-2,-3,\ldots $ as seen in the power spectrum of Phase 1 (Fig.~\ref{fig:8a}), with $m=2$ being the most active mode, followed by the $m=1,3$ modes. The dispersion relation obtained from the highly resolved simulations for Phase 1 (Fig.~\ref{fig:6a}) also shows the higher-frequency branches of electron Bernstein waves modified by the Doppler shift $k_{z}V_{de}$ due to the electron drift, $\omega =m\omega_{ce}-k_{z}V_{de}$. Additionally, we identify a distinct upper-hybrid resonance $\omega ^{2}=\omega _{pe}^{2}+\omega _{ce}^{2},$  near $\omega $ $\approx 5\omega _{ce}$ in  Fig.~\ref{fig:6a}. As is typical for such simulations,~\cite{TavassoliPoP2021} in the linear stage, we observe a rather coherent periodic profile of the density perturbations (Fig.~\ref{fig:4a}). The sharp peaks can be attributed to some signature of higher $m$ harmonics. The electron density perturbations are rather smooth and look linear. In this phase, the electron and ion distribution functions largely remain unchanged and Maxwellian as seen from Figs.~\ref{fig:4a} and \ref{fig:5a}.

In Phase {2}, the cyclotron resonances continue to grow (Fig.~\ref{fig:8b}), with the most activity in $m = 1,2$ followed by $m=3$ modes. As in Phase 1, fluctuations are concentrated in the long wavelength part of the spectrum near  $k/k_0 = \pm 1, \pm 2, \pm 3$. The electron phase space shows signs of clump formation and resonance broadening as seen in Fig.~\ref{fig:3c} (below the horizontal white line $t = T_2$) and in Fig.~\ref{fig:4b}. The electron distribution function shows signs of quasi-linear deformation modified by the magnetic field as it starts to flatten as indicated in Fig.~\ref{fig:5b}. A prominent backward wave at $k=-k_{0}$ also emerges in this phase, as observed in the power spectrum for Phase 2 (Fig.~\ref{fig:6b}). The ion dynamics shows signs of a modulational instability, observed in the ion phase space in Fig.~\ref{fig:4b}. The growing electron temperature $T_e$ and increase of $\lambda_{De}$ result in decrease of the values $k_{s} \equiv 1/\sqrt{2}\lambda_{De}$ from $k_s/k_0 \sim 2.7$ at the end of Phase 1 to $k_s/k_0 \sim 1.8$ at the end of Phase 2, as shown by the vertical white lines in Figs.~\ref{fig:6a}--\ref{fig:6b}.

\begin{figure*}[htbp]
\centering
\begin{subfigure}[b]{0.48\textwidth}
\centering
\caption{$E_z = 20000$ V/m$ $, $B_y = 200$ G}
\includegraphics[width=\textwidth]{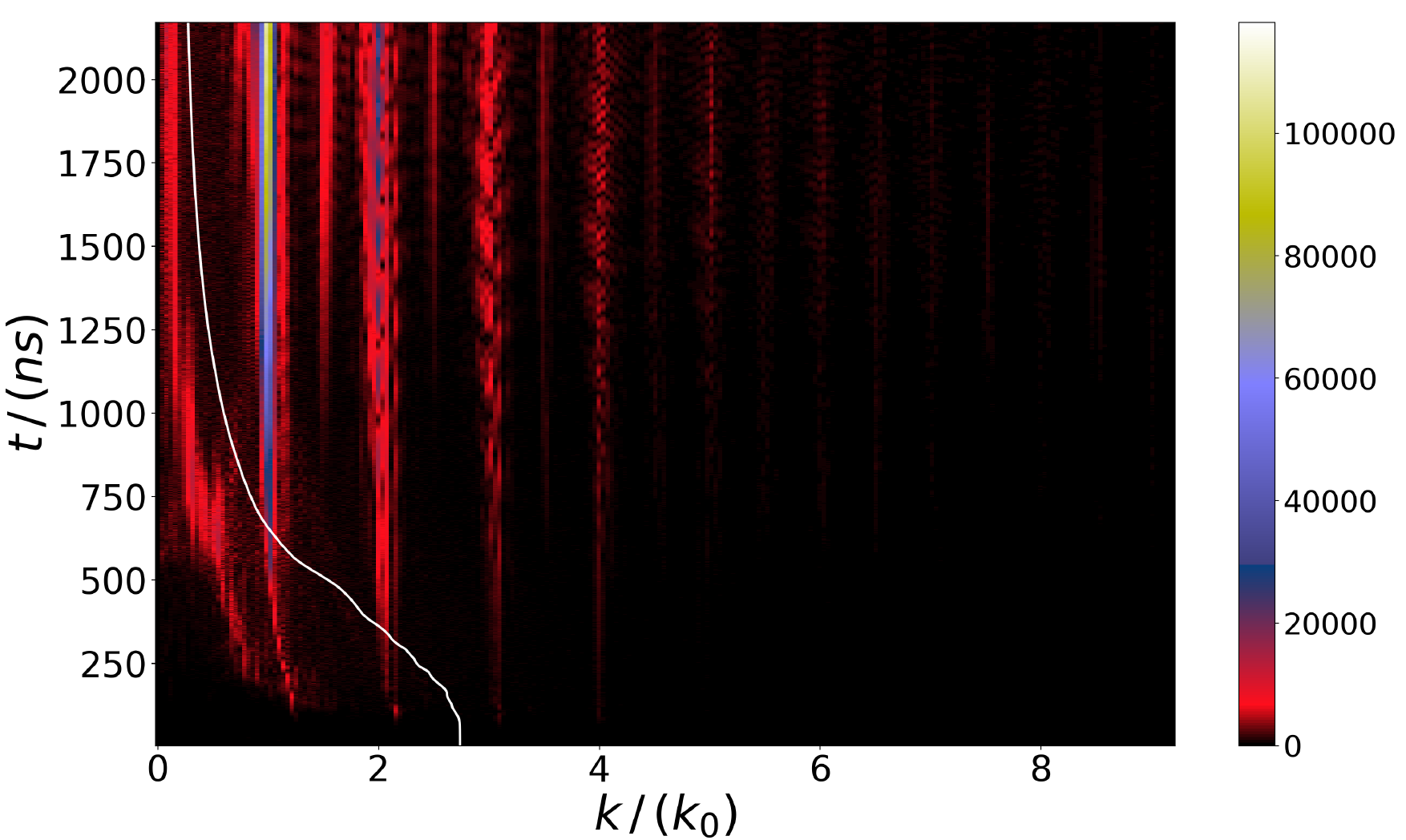}
\label{fig:9a}
\end{subfigure}
\begin{subfigure}[b]{0.48\textwidth}
\centering
\caption{$E_z = 20000$ V/m$ $, $B_y = 250$ G}
\includegraphics[width=\textwidth]{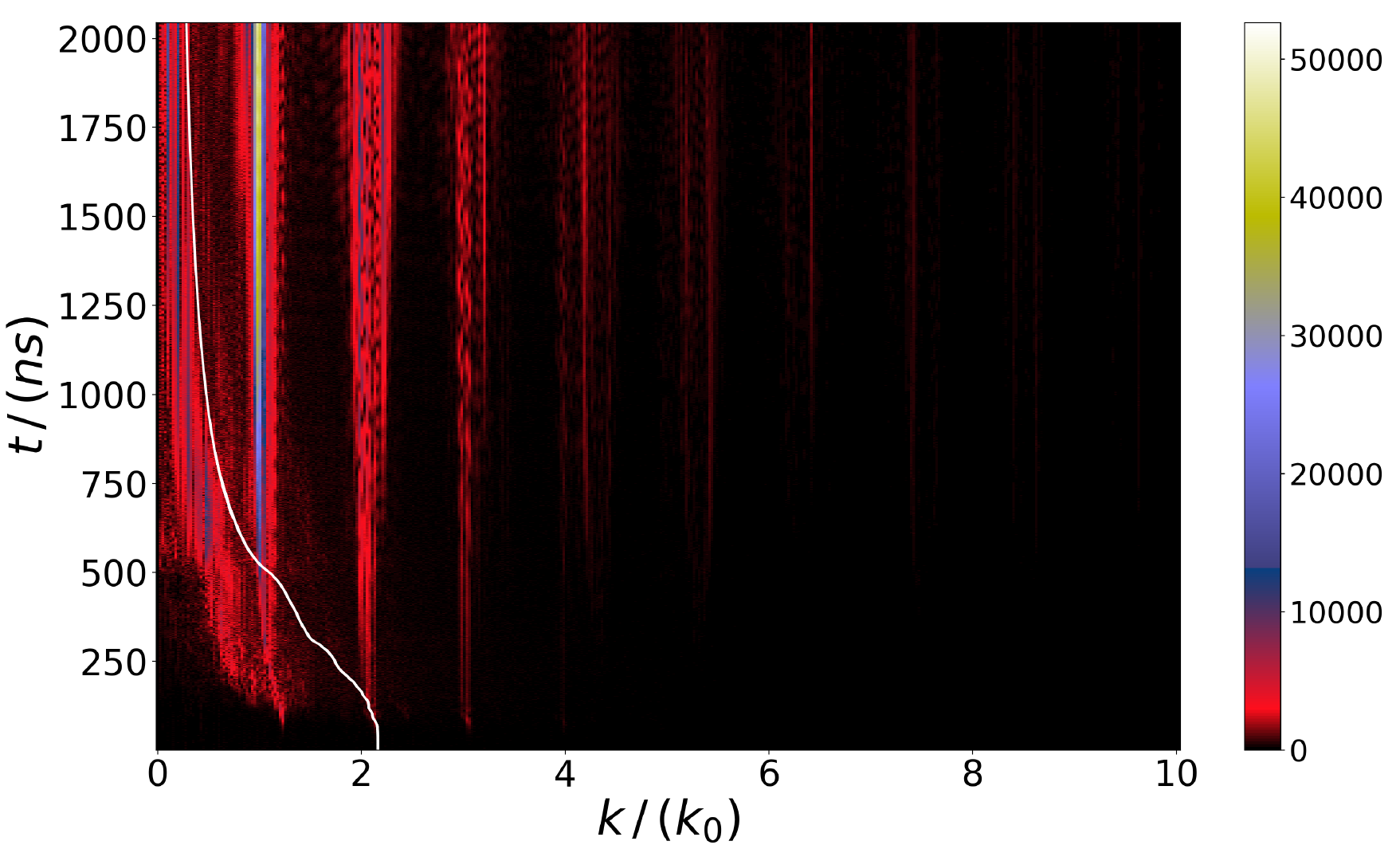}
\label{fig:9b}
\end{subfigure}
\begin{subfigure}[b]{0.48\textwidth}
\centering
\caption{$E_z = 16000$ V/m$ $, $B_y = 200$ G}
\includegraphics[width=\textwidth]{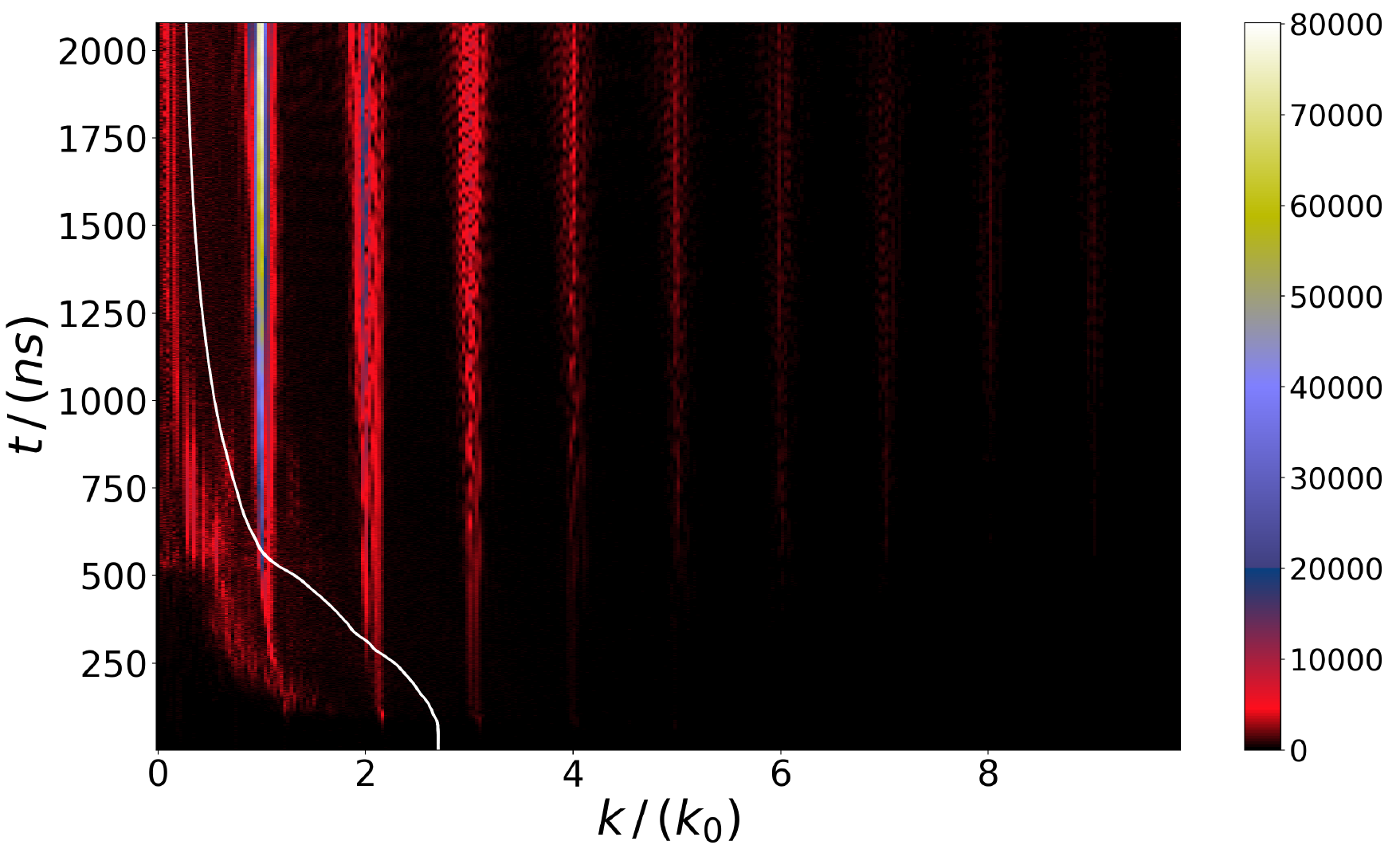}
\label{fig:9c}
\end{subfigure}
\caption{The depiction of the non-linear resonance between the most dominant ion-acoustic branch with $k_s$ (depicted by the white curve) and $m=1$ ECDI mode for three different scenarios : (a) $E_z = 20000$ V/m$ $, $B_y = 200$ G, (b) $E_z = 20000$ V/m , $B_y = 250$ G, (c) $E_z = 16000$ V/m , $B_y = 200$ G.}
\label{fig:9}
\end{figure*}

Phase {3} characterizes a newly discovered resonant interaction between the $m=1$ ECDI mode and the ion-acoustic mode at wavelength $k_{s}=1/\sqrt{2}\lambda_{De}$ corresponding to the maximum growth rate of the unmagnetized ion-acoustic mode driven by the electron beam at $V_{de}$. During Phase 3, we see that the wavelength of the most unstable ion-acoustic modes (shown by the white curve in Fig.~\ref{fig:3a}) approaches the wavelength of the ECDI $m=1$ cyclotron resonance at $k/k_0 = 1$ so that, at the end of Phase 3 at $t = T_3$,  $k_{s}=k_{0}$. It is clearly visible from the power spectrum of Phase 3 in Fig.~\ref{fig:8c} (where the vertical lines indicate the value of $k_s/k_0 = \pm 1$) coincide with the cyclotron resonances at $k/k_0 = 1$. From this moment and up to saturation, the $k=k_{0}$ mode is the most dominant mode in the spectra, as seen in Figs.~\ref{fig:3a}--\ref{fig:3c}.

To confirm the validity of the resonant interaction between the ECDI $m=1$ mode and the ion-acoustic branch, we ran three different test cases with different values of $\mathbf{E}$ and $\mathbf{B}$. Figs.~\ref{fig:9a}--\ref{fig:9c} show the evolution of the spectral coefficients from the 1D FFT of the electric field $E_z$, obtained for three test cases. In all setups, after the dominant ion-acoustic mode with $k_s$ crosses with the ECDI $m=1$ mode, $k=k_0$, we observe an intense growth of the $m=1$ mode, which eventually becomes dominant. In each case, the ion-acoustic curve starts with a slightly different initial temperature because we tried to keep the ratio $v_{th}/V_{d}$ constant for all the cases.

\subsection{\label{sec3:level3} Phase 4: Ion trapping, growth of the ion-acoustic modes, and intense electron heating}

\begin{figure*}[htbp]
\centering
\begin{subfigure}[b]{0.48\textwidth}
\centering
\caption{ Phase 4}
\includegraphics[width=\textwidth]{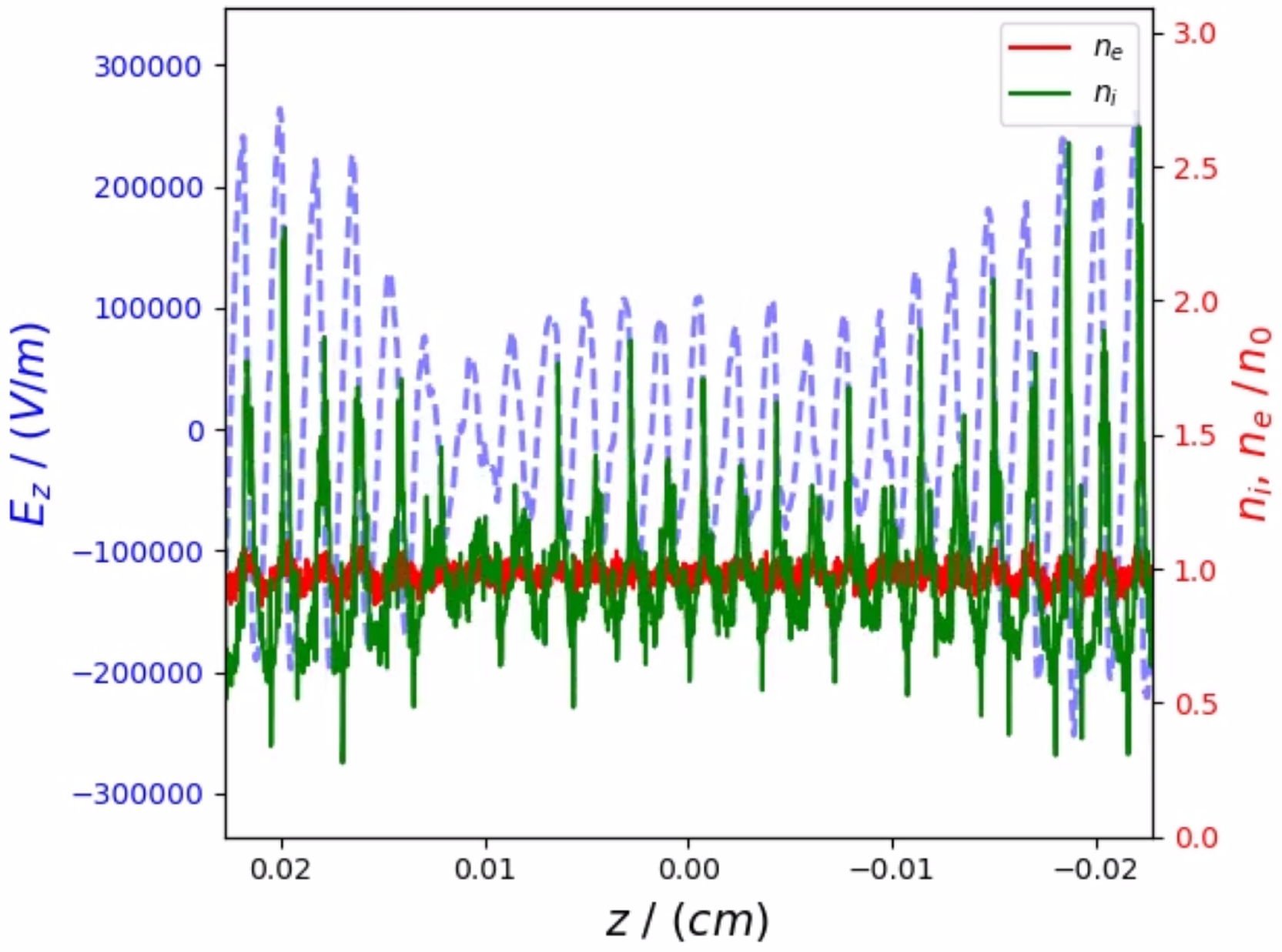}
\label{fig:10a}
\end{subfigure}
\begin{subfigure}[b]{0.48\textwidth}
\centering
\caption{ Phase 5}
\includegraphics[width=\textwidth]{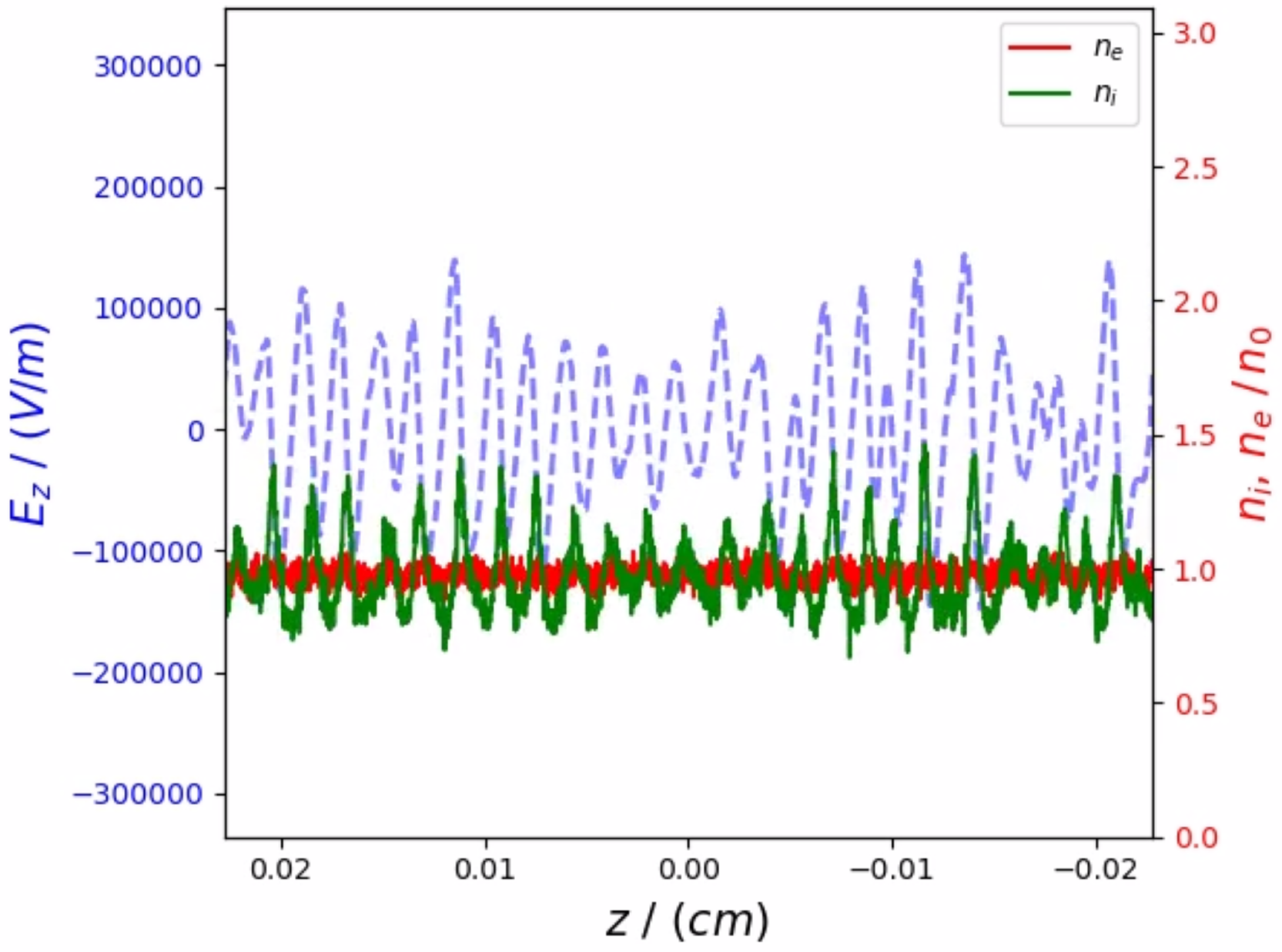}
\label{fig:10b}
\end{subfigure}
\caption{(a) Phase 4: $n_i$, $n_e$, and $E_z$, depicting periodic cnoidal wave structure (sharp crests and broad trough), correlating to multiple harmonics present in $\tilde{n}_i(k, t)$ Fig.~\ref{fig:3b}) in the ion density perturbations, and $E_z$, depicting the increase in $\delta E/ E_0$ to $\sim 15$, and (b) Phase 5.}
\label{fig:10}
\end{figure*}

Phase {4} ($T_3 < t < T_5$) is characterized by intense interactions of the ECDI $m=1$ mode and ion-acoustic fluctuations that occur mostly at $k/k_0 = 1$ and $\omega/\omega_{pi}=1$ and achieve a maximum amplitude during this stage, as seen in Figs.~\ref{fig:3a}--\ref{fig:3c} (between times $T_3 < t < T_5$) and Fig.~\ref{fig:7a}, with weaker fluctuations at higher resonances $k/k_0 = 2,3$ and $\omega/\omega_{pi}=2,3$, as seen in Figs.~\ref{fig:7a} and \ref{fig:8d}. The electron distribution function flattens and saturates by the end of Phase 4, Fig.~\ref{fig:5e}. The higher $m$ modes in the electron response are broadened further and partially demagnetized, as seen in Fig.~\ref{fig:3c}, but the $m=1$ resonance is still intense, as seen in Figs.~\ref{fig:3c} and \ref{fig:8d}. Long wavelength ($k \rho_e < 1$) electron dynamics shows the signatures of inverse cascade, which can be seen in the electric field and electron density fluctuations, Figs.~\ref{fig:3a} and \ref{fig:3c}, as well as the emergence of long wave modes in the electron phase space in Figs.~\ref{fig:4d}--\ref{fig:4e}. Ion dynamics during this phase, shows the existence of spectral energy at several cyclotron harmonics $k/k_0 = 1,2,3,\dots$ (Fig.~\ref{fig:3b}), driven by ion-acoustic type fluctuations at $\omega \sim n \omega_{pi},\ n=1,2,\dots$ (Fig.~\ref{fig:7a}), with the most energy in ECDI $m=1$ mode at $k = k_0, \omega = \omega_{pi}$. Simultaneously, the ion density shows the development of wave focusing and sharp peaks, resulting in a periodic cnoidal wave structure, which is a signature of multiple harmonics existing together, as seen in Fig.~\ref{fig:10a}. There are obvious signatures of ion trapping, as seen in the ion phase space and the ion distribution function in Figs.~\ref{fig:4d} and \ref{fig:4e}.

In Fig.~\ref{fig:10a}, during Phase 4, we also see the growth of electrostatic fluctuations in $E_z$, increasing by an order of magnitude to the quasi-stationary applied electric field $\delta E/E_0 \sim 15$, with maximum amplitudes of $E_z$ reaching $\sim 300,000$ eV, with an applied electric field of $E_0 = 20,000$ eV. It is in this stage, where the magnitude of the anomalous current $J_{xe}$, after some initial growth, reaches its maximum at $t \, = \, T_4$ and starts to decrease and eventually gets suppressed, as seen in Fig.~\ref{fig:11a} ($T_3 < t < T_5$).

\subsection{\label{sec3:level4} Phase 5: Turbulence saturation and suppression of anomalous transport}  

\begin{figure*}[htbp]
\centering
\begin{subfigure}[b]{0.5\textwidth}
\centering
\caption{ }
\includegraphics[width=\textwidth]{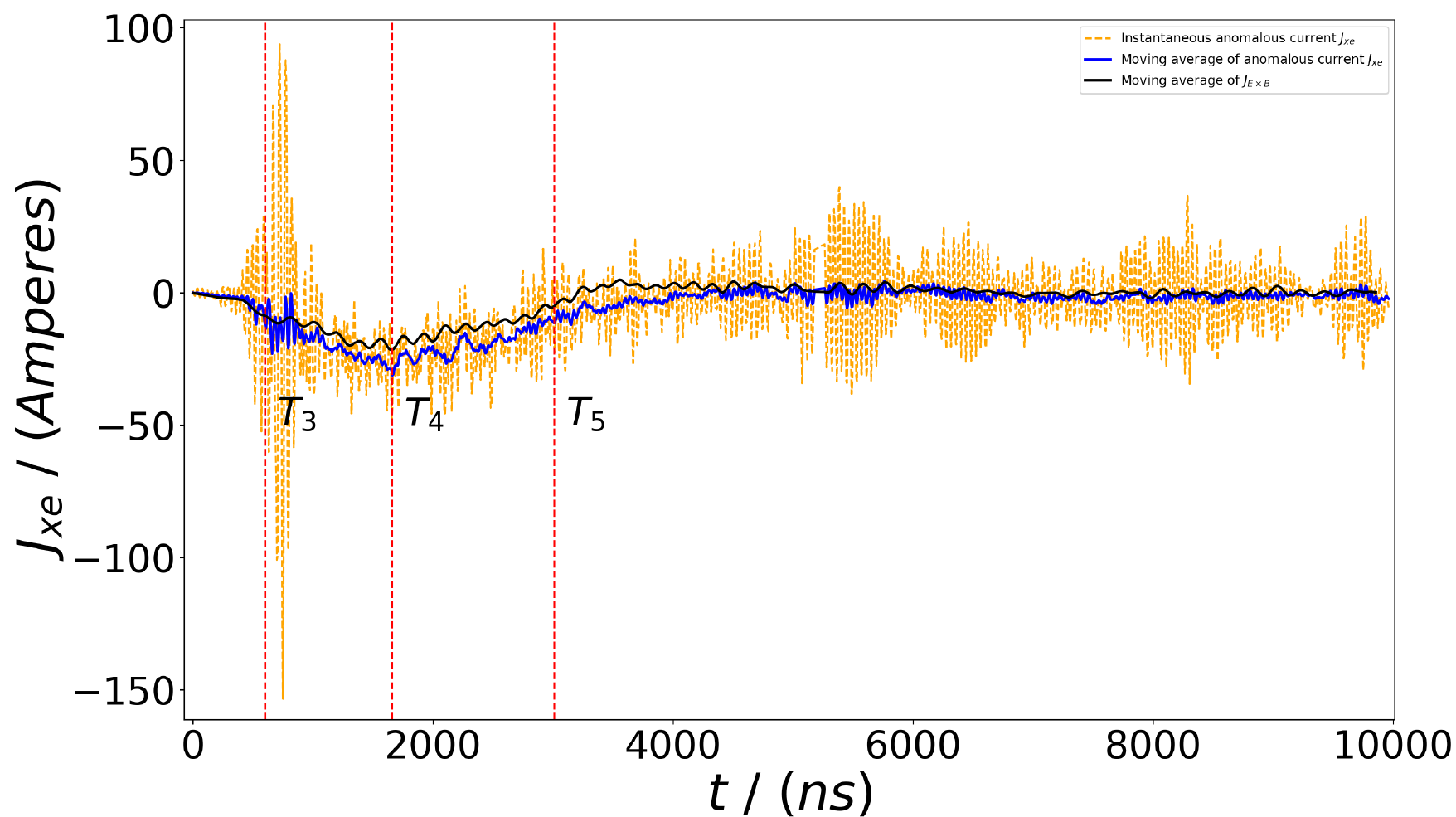}
\label{fig:11a}
\end{subfigure}
\begin{subfigure}[b]{0.48\textwidth}
\centering
\caption{ }
\includegraphics[width=\textwidth]{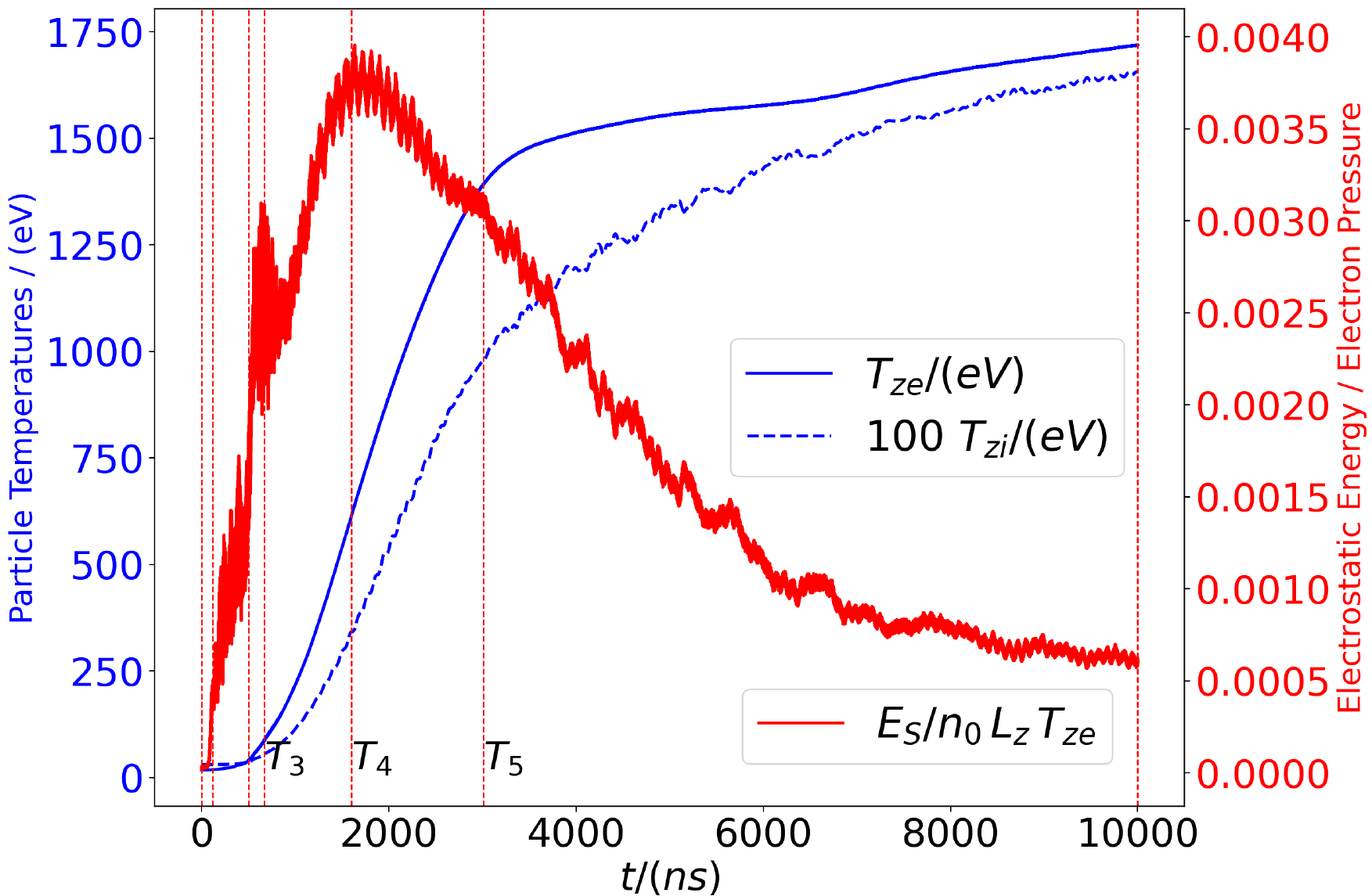}
\label{fig:11b}
\end{subfigure}
\begin{subfigure}[b]{0.4\textwidth}
\centering
\caption{ }
\includegraphics[width=\textwidth]{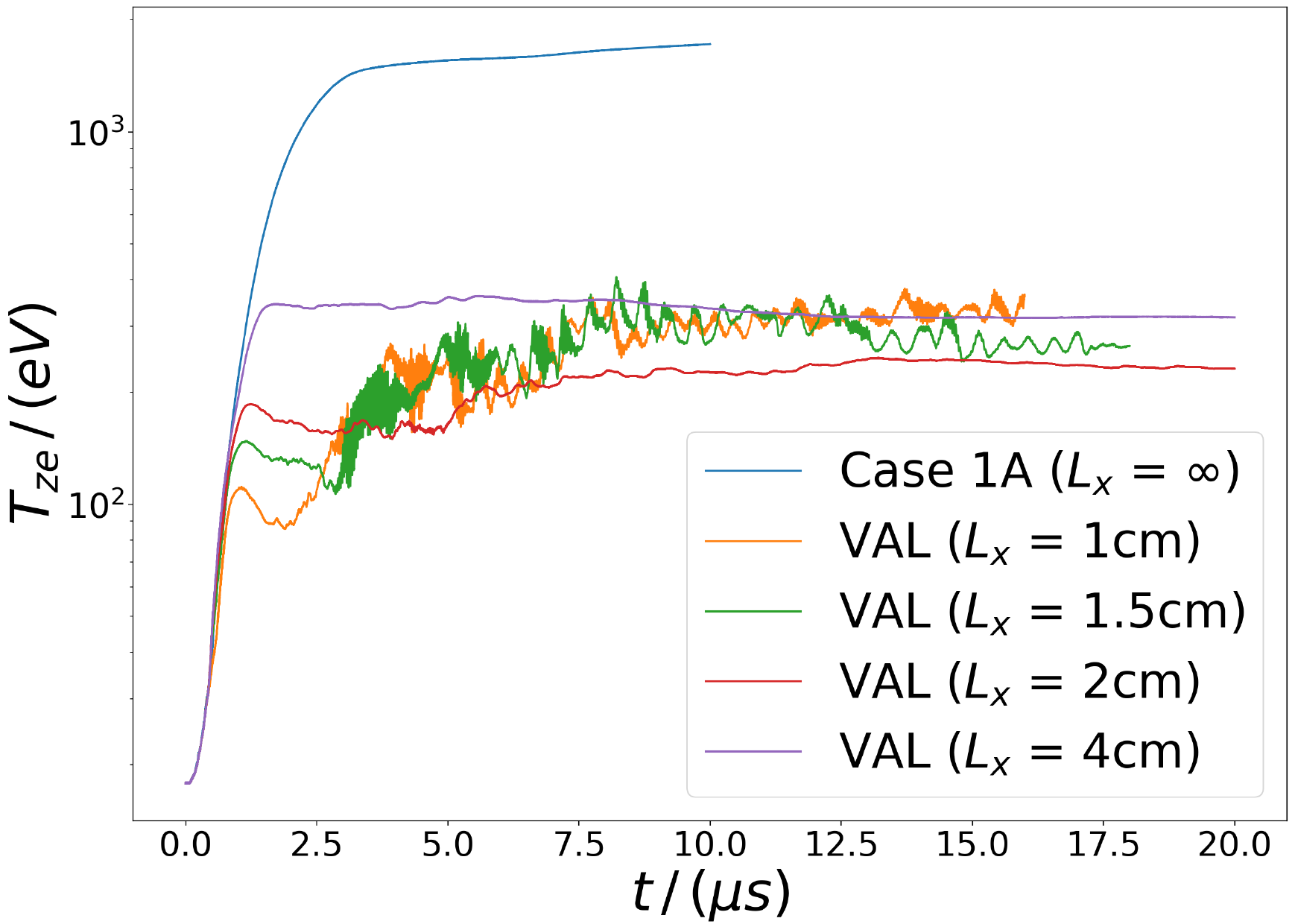}
\label{fig:11c}
\end{subfigure}
\begin{subfigure}[b]{0.4\textwidth}
\centering
\caption{ }
\includegraphics[width=\textwidth]{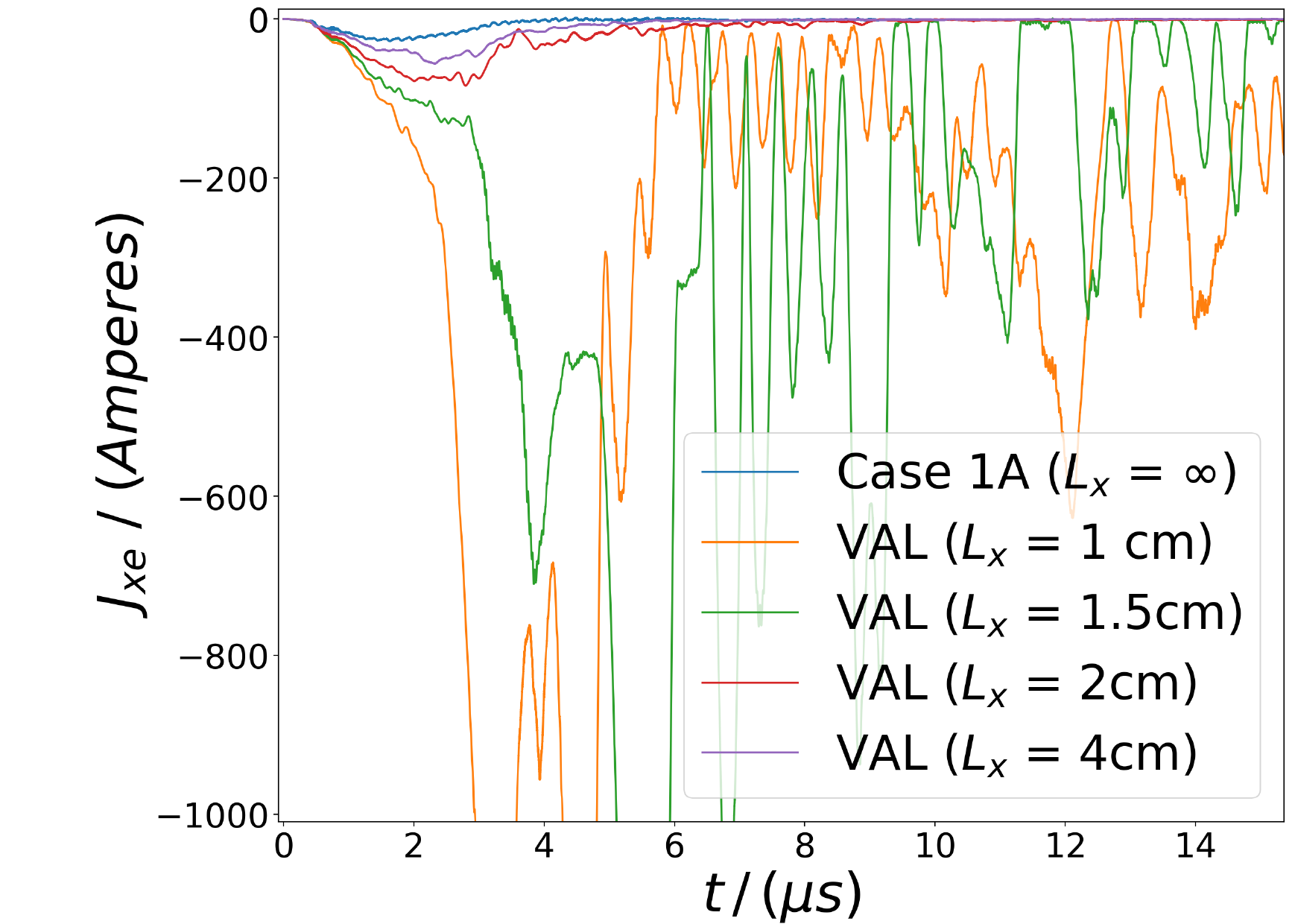}
\label{fig:11d}
\end{subfigure}
\begin{subfigure}[b]{0.4\textwidth}
\centering
\caption{ }
\includegraphics[width=\textwidth]{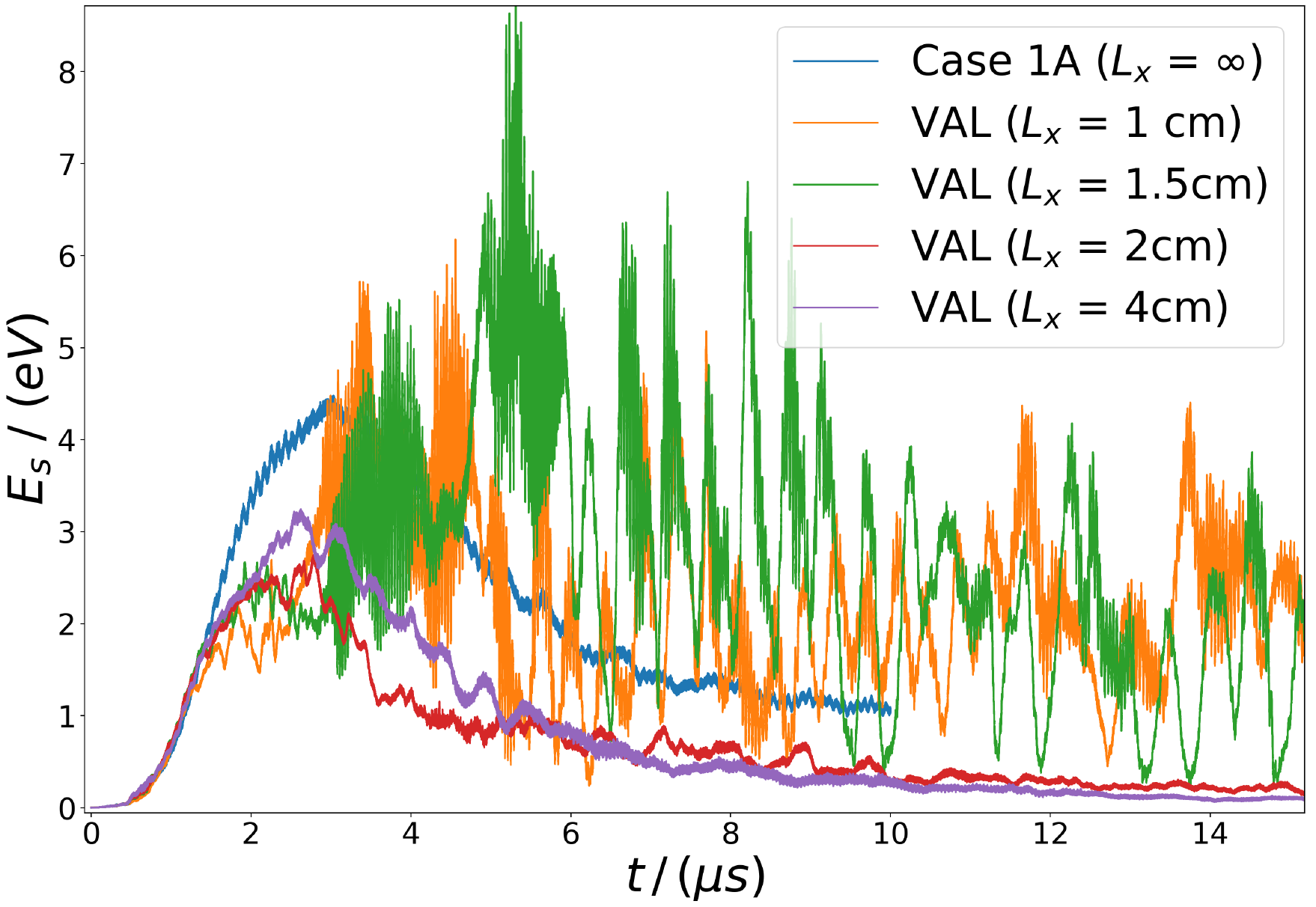}
\label{fig:11e}
\end{subfigure}
   \caption{(a) Anomalous electron current $J_{xe}$ for Case 1A; (b) Comparison of electron and ion temperatures and the ratio of the  electrostatic energy to the  electron pressure. The  time evolution of (c) the electron energy $\mathit{T_{ze}}$, (d) the anomalous current $J_{xe}$, and (e) electrostatic energy $\epsilon_s$ for different values of virtual axial length, $\mathit{L_x} = 1 \, \text{cm},\ 1.5 \, \text{cm},\ 2 \, \text{cm},\ \text{and} \  4 \, \text{cm}$.}
\label{fig:11}
\end{figure*}

By the end of Phase 4 at $t=T_5$, the fluctuation energy $\epsilon_s$ reaches its maximum and starts to decrease towards the saturation value, Fig.~\ref{fig:2b}. Comparing Fig.~\ref{fig:10a}, and \ref{fig:10b}, one can observe a decrease in the amplitude of electrostatic fluctuations going from Phase 4 to Phase 5. By this time, the anomalous electron current $J_{xe}$, calculated as
\begin{equation}
J_{xe} =- e\int_{-\infty}^{\infty} v_x f_{e}(z, \mathbf{v}, t) d^3\mathbf{v},
\label{eq:J}
\end{equation}
is substantially decreased and gets fully suppressed in Phase 5 (Fig.~\ref{fig:11a} after $t = T_5$). In this stage, the fluctuation spectra start to most closely resemble the ion-acoustic mode. One can see the frequency of fluctuations near $\omega_{pi} $ spread along the ion-acoustic mode dispersion curve, Figs.~\ref{fig:7b} and \ref{fig:8e}. One can also observe the energy accumulation at two additional acoustic-like dispersion curves at multiples of $\omega_{pi}$ frequency, Fig.~\ref{fig:7b}. There are some remnant fluctuations around $k_0$ that are notably broadened by non-linear effects. In this stage, the phase space dynamics of ions has changed from intermittent trapping-detrapping evident in Figs.~\ref{fig:4d}--\ref{fig:4e} to a fully mixed state as in Fig.~\ref{fig:4f}, consistent with a typical quasi-linear distribution function for ions in Fig.~\ref{fig:4f} exhibiting a long tail of high energy ions.  

Essentially, we observe a quasi-linearly saturated turbulent state of the ion-acoustic fluctuations with a strongly flattened electron distribution function and an ion distribution function with a long quasi-linear tail. We note that at this stage, the wavelength of the ion-acoustic fluctuations is much smaller than $2\pi/k_s$, as seen in Fig.~\ref{fig:3a}. It is noteworthy that the backward wave generated earlier in previous stages at $k=k_0$ during Phase 5 also spreads along the mirrored ion-acoustic like dispersion curve around $\omega_{pi}$ and $k = -3k_0/2$ (Fig.~\ref{fig:7b}).

\subsection{\label{sec3:level5} General features of the non-linear evolution, inverse cascade, and the effect of virtual axial length (VAL)}

An important general aspect of electron dynamics is the tendency of the inverse cascade toward the long-wavelength modes, $k \rho_e < 1$ (Fig.~\ref{fig:3c}), whereas no such tendency is seen in the ion density (Fig.~\ref{fig:3b}). The inverse cascade toward the long-wavelength modes is increased when we increase the intensity of the magnetic field, as seen in Fig.~\ref{fig:9b}, which is consistent with stronger magnetization  of electrons in a larger magnetic field. 

As can be seen from the evolution of the power spectrum (Figs.~\ref{fig:8a}--\ref{fig:8e}), the spectral energy is primarily concentrated in the lower frequency range $\sim \omega_{pi}$ driven by ECDI cyclotron resonance at $k=k_0$. In addition, during Phases 1--3, we also observe the high-frequency Doppler-modified electron Bernstein waves $\omega \sim \omega_{ce}$, as shown in Figs.~\ref{fig:6a}--\ref{fig:6c}. We observe  resonance broadening of these Bernstein modes  ($\omega_{ce}$--$8 \omega_{ce}$) and an increase in the intensity of fluctuations concentrated in the lower ($\omega / \omega_{ce} =1,2$) Bernstein branches.

By the end of Phase 3, and further in Phases 4 and 5, we observe substantially reduced spectral energy in the higher frequency part of the spectrum, Figs.~\ref{fig:8c}--\ref{fig:8e}, but  the signatures of skewed electron Bernstein waves at $\sim \omega_{ce}$ are still observed in the final quasi-equilibrium stage as well (Phase 5, Fig.~\ref{fig:6e}). During Phases 3--5 (Figs.~\ref{fig:8c}--\ref{fig:8e}), one can observe strong broadening of electron cyclotron modes as well as some kind of inverse cascade toward the longer wavelength and the saturation in the region of $k/k_0 <1$ and lower frequencies $ \omega/\omega_{pi} \sim 20$ (Fig.~\ref{fig:8c}).

During Phase 5 (after $t = T_5$),  the anomalous current existing in the intermediate stages is quenched to zero as discussed previously (Fig.~\ref{fig:11a}). The previous studies~\cite{LafleurJAP2021,LafleurPoP2016b,SmolyakovPPR2020,TaccognaPSST2019} have used VAL models to limit the rise in the electron temperature. In such models,  electrons that travel a finite distance $L_x$ in the axial $x$-direction are re-sampled again from the initial Maxwellian distribution (with $T_e = 10$ eV and $V_{de} = 10^6$ m/s), thus  mimicking a finite residence time in  the 2D simulations and suppressing electron heating at some level which depends on the VAL value $L_x$.  The results of using the VAL in our simulations is shown in  Figs.~\ref{fig:11c} and \ref{fig:11d}. As seen in these figures, finite residence times associated with finite  $L_x$ result in lower values of the electron energy and finite values of the anomalous current.

\section{\label{sec4:level1}Results: ECDI driven by Xe ion beam (Case 1B)}

\subsection{Comparison of long term dynamics of Case 1B with Case 1A}

\begin{figure}[htbp]
\centering
\begin{subfigure}[b]{\linewidth}
\centering
\caption{ }
\includegraphics[width=\linewidth]{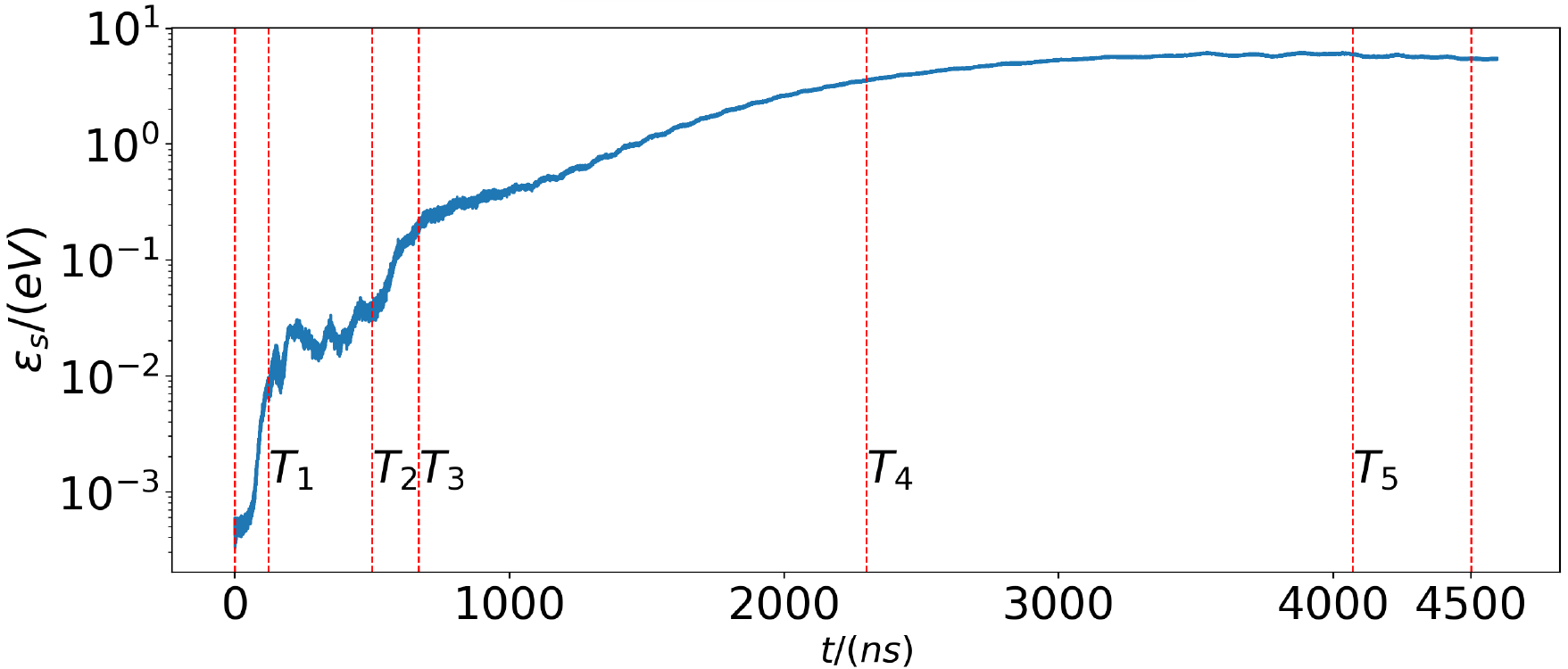}
\label{fig:12a}
\end{subfigure}
\hfill
\begin{subfigure}[b]{\linewidth}
\centering
\caption{ }
\includegraphics[width=\textwidth]{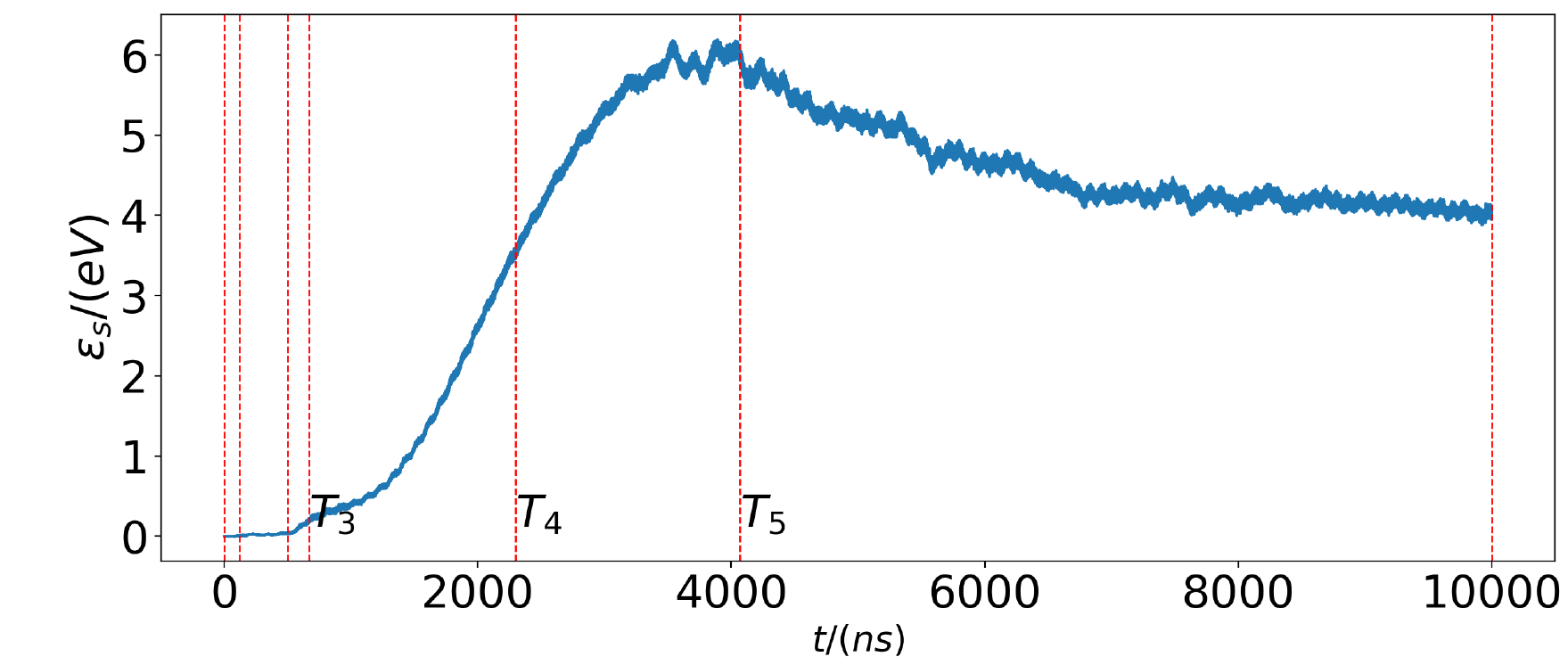}
\label{fig:12b}
\end{subfigure}
\caption{Time evolution of the electrostatic field energy ${\epsilon_s}$ (Eq.~\eqref{eq:es}) for Case 1B. Vertical dotted lines shows time moments $\mathit{T_1}, T_2,\dots,\mathit{T_5}$ for the transitions between different phases as described in the text. (a) $\epsilon_s$ in logscale from $0 < t < 4500$ ns; (b) Evolution of ${\epsilon_s}$ for the entire simulation from $0 < t < 10000$ ns (Phases 1--5).}
\label{fig:12}
\end{figure}

\begin{figure}[htbp]
  \centering
   \includegraphics[width=0.75\linewidth]{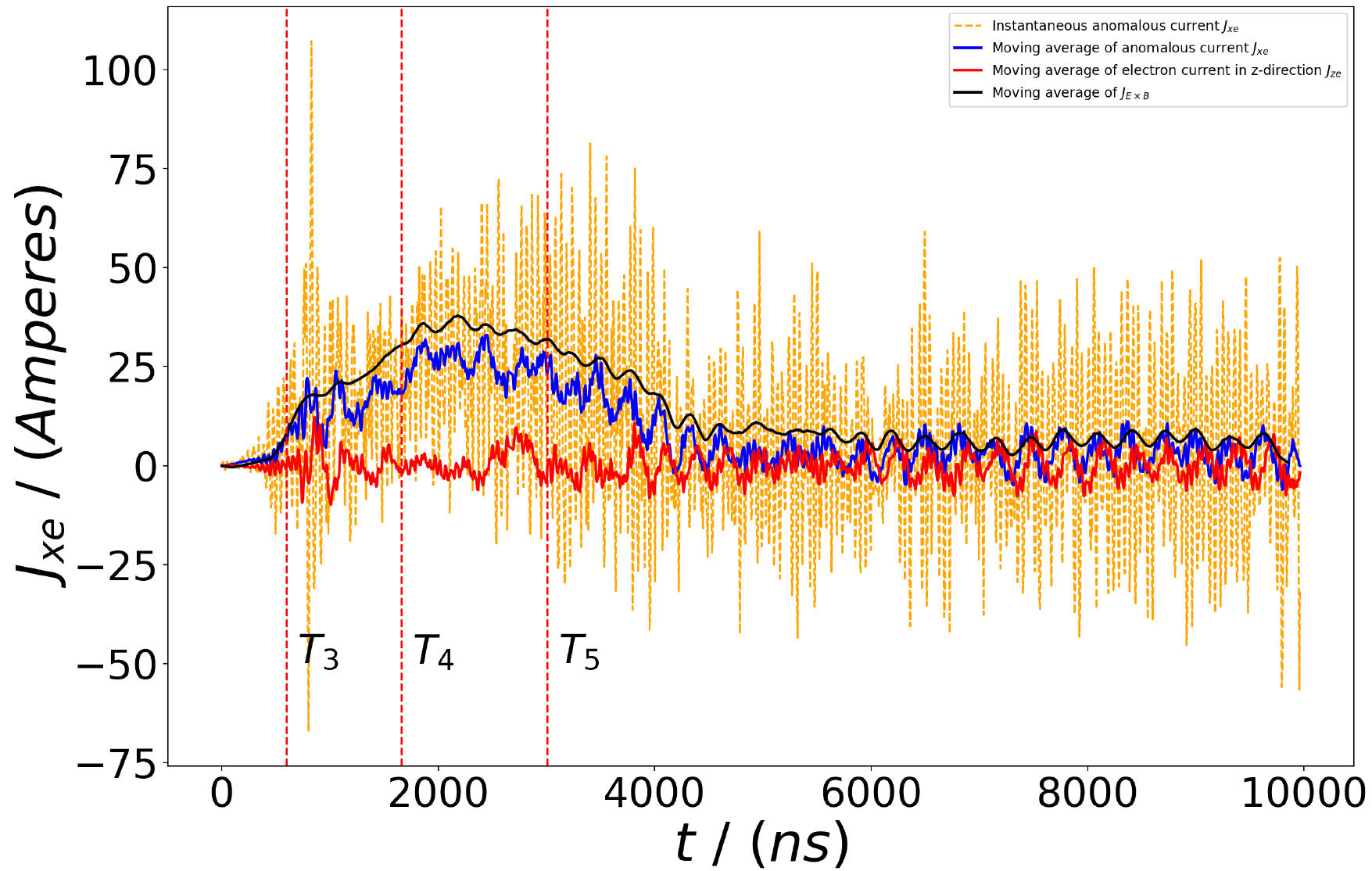}
   \caption{\label{fig:13} Time evolution of the anomalous electron current $J_{xe}$, given by Eq.~\eqref{eq:J} for Case 1B.}
\end{figure}

\begin{figure*}[htbp]
\centering
\begin{subfigure}[b]{0.75\textwidth}
\centering
\caption{ }
\includegraphics[width=\textwidth]{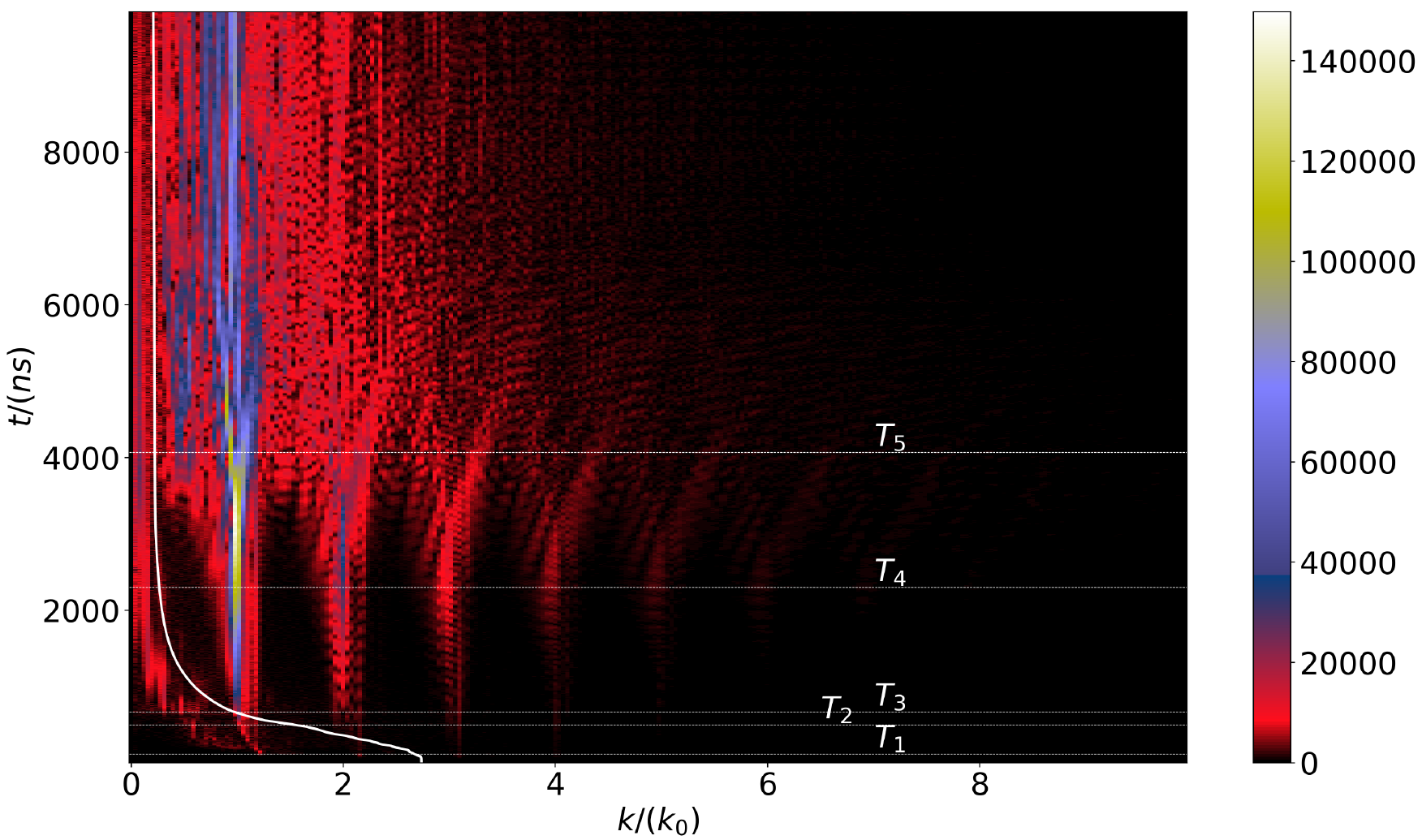}
\label{fig:14a}
\end{subfigure}
\begin{subfigure}[b]{0.48\textwidth}
\centering
\caption{ }
\includegraphics[width=\textwidth]{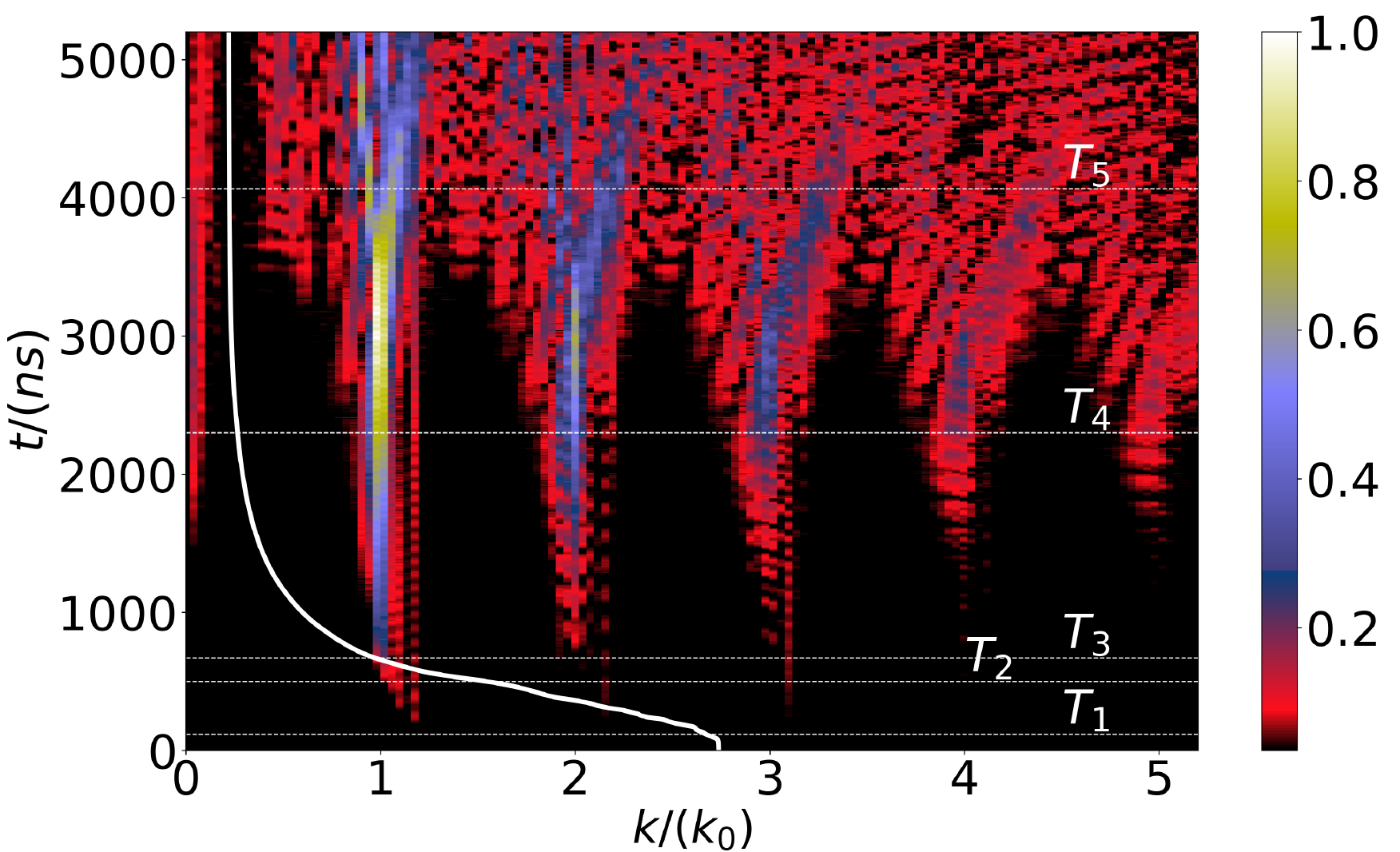}
\label{fig:14b}
\end{subfigure}
\hfill
\begin{subfigure}[b]{0.48\textwidth}
\centering
\caption{ }
\includegraphics[width=\textwidth]{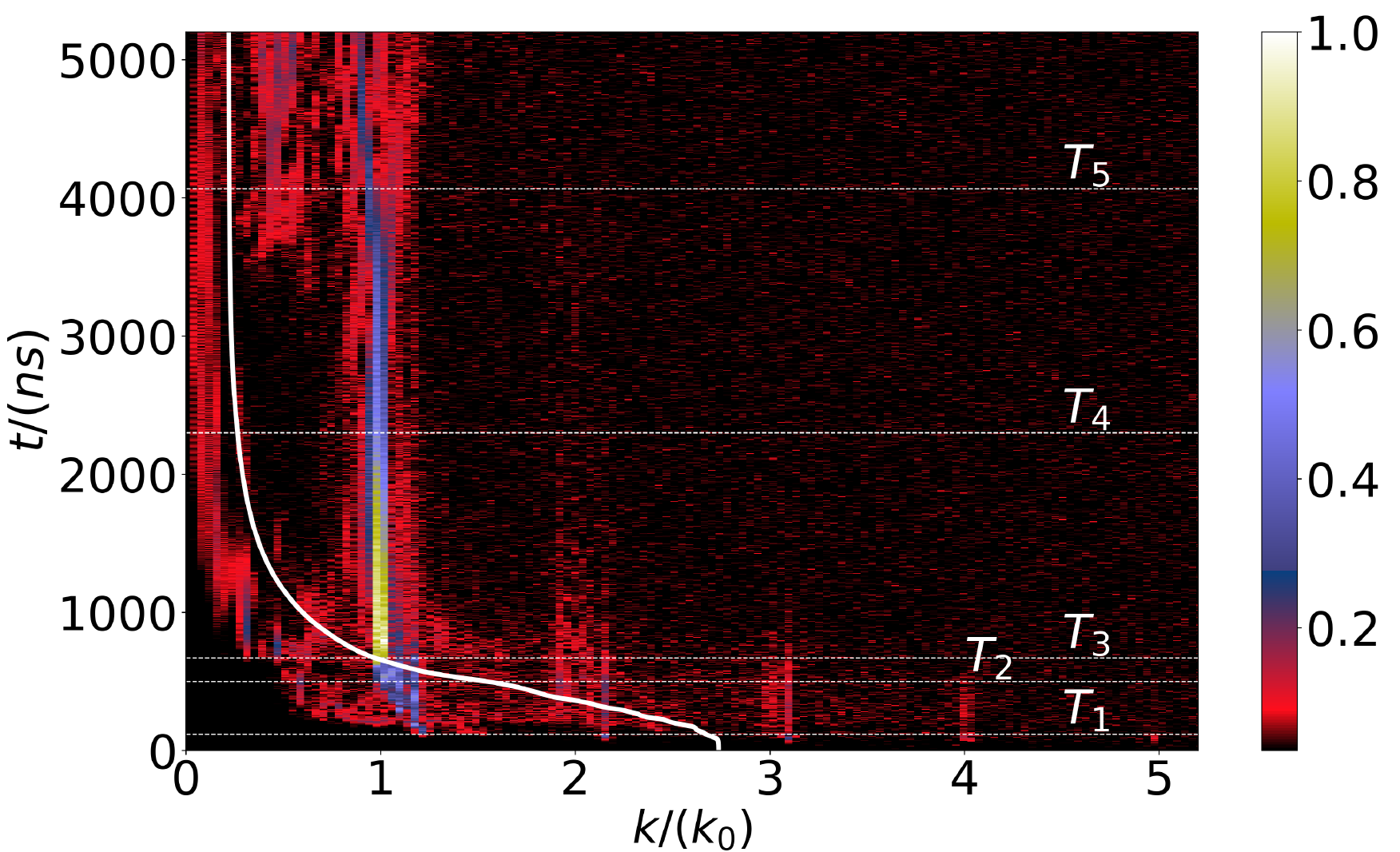}
\label{fig:14c}
\end{subfigure}
\caption{One-dimensional Fourier transform of (a) Electric field in the $z$-direction $\tilde{E}_z (k ,t)$, as given by Eq.~\eqref{eq:1dfft}, (b) Ion density in the $z$-direction $\tilde{n}_i (k ,t)$, and (c) Electron density in the $z$-direction $\tilde{n}_e (k ,t)$ for Case 1B. Horizontal dotted lines show the time instants $\mathit{T_1}, T_2, \dots, \mathit{T_5}$ for the transitions between different phases as described in the text.}
\label{fig:14}
\end{figure*}

\begin{figure*}[htbp]
\centering
\begin{subfigure}[b]{0.48\textwidth}
\centering
\caption{Phase 1 }
\includegraphics[width=\textwidth]{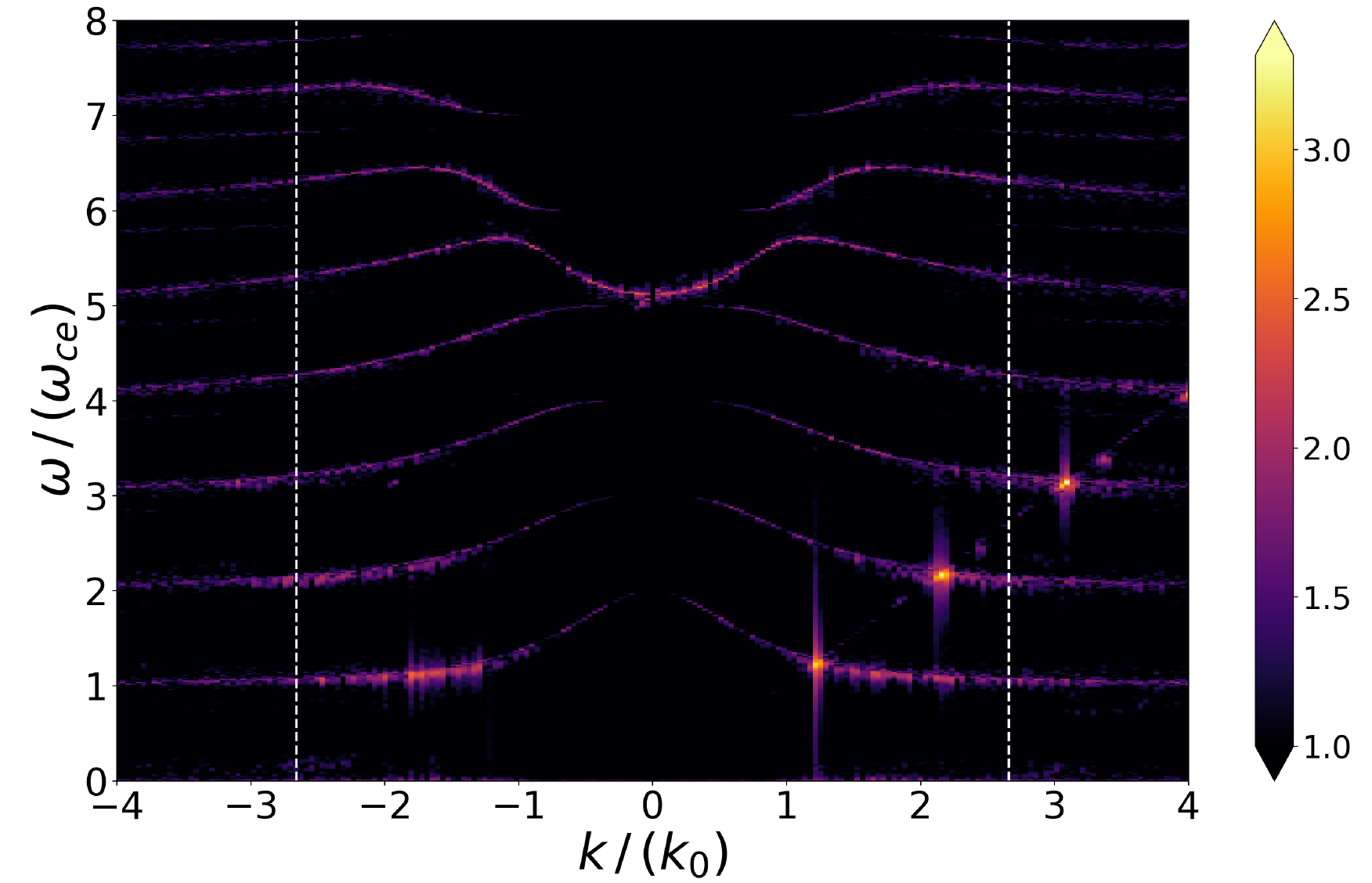}
\label{fig:16a}
\end{subfigure}
\begin{subfigure}[b]{0.48\textwidth}
\centering
\caption{ Phase 2}
\includegraphics[width=\textwidth]{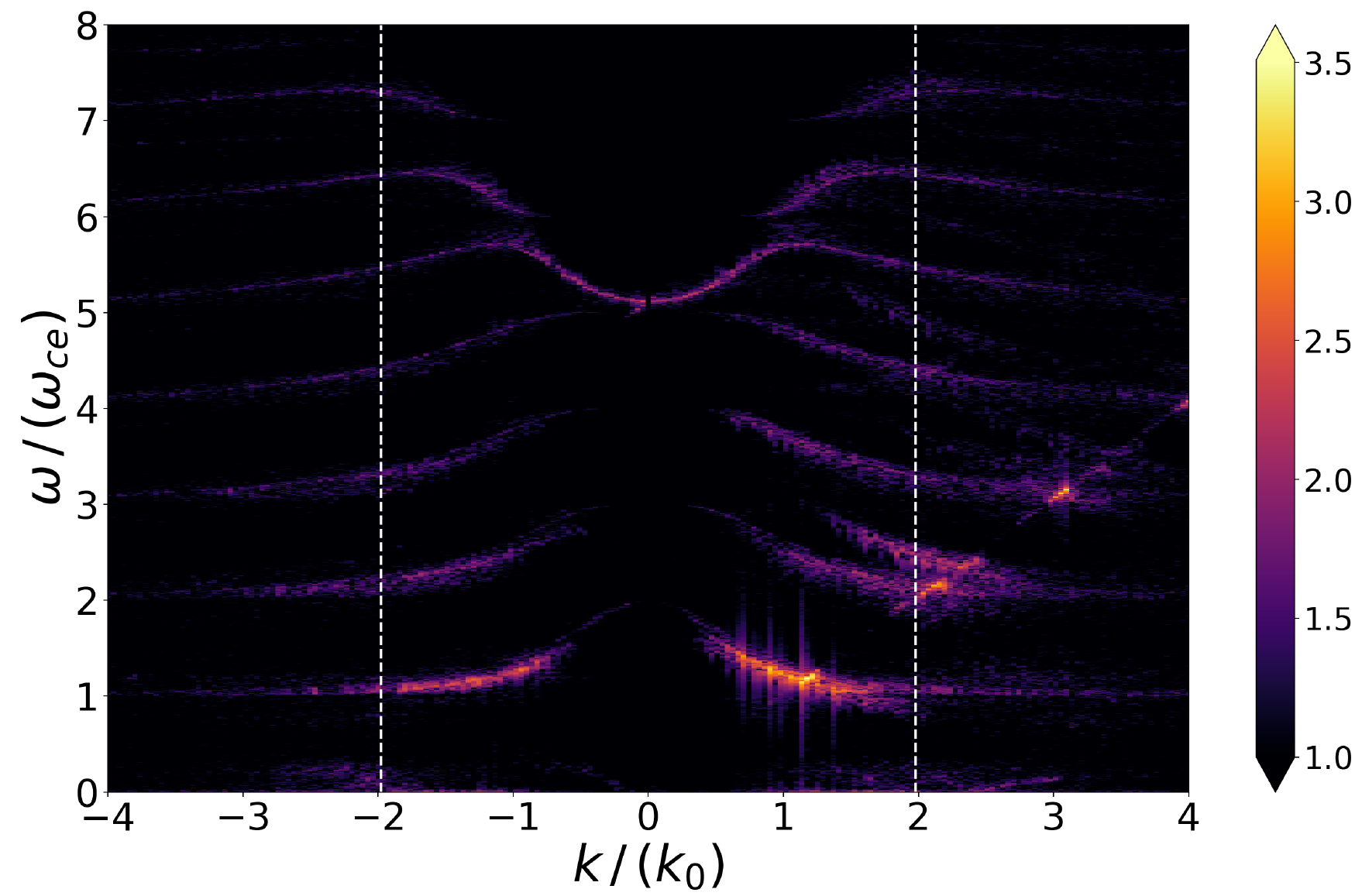}
\label{fig:16b}
\end{subfigure}

\begin{subfigure}[b]{0.48\textwidth}
\centering
\caption{ Phase 3}
\includegraphics[width=\textwidth]{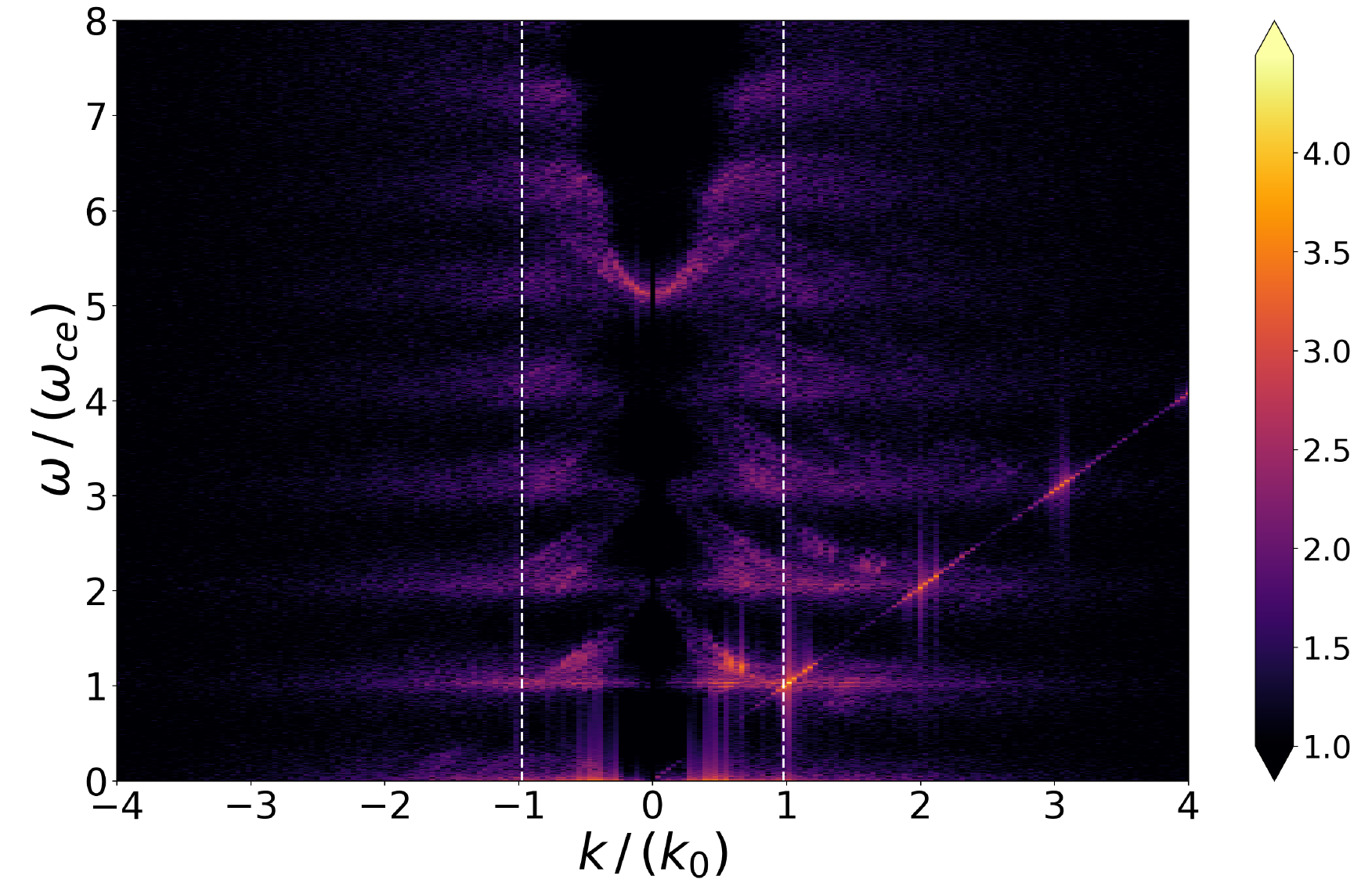}
\label{fig:16c}
\end{subfigure}
\begin{subfigure}[b]{0.48\textwidth}
\centering
\caption{ Phase 4}
\includegraphics[width=\textwidth]{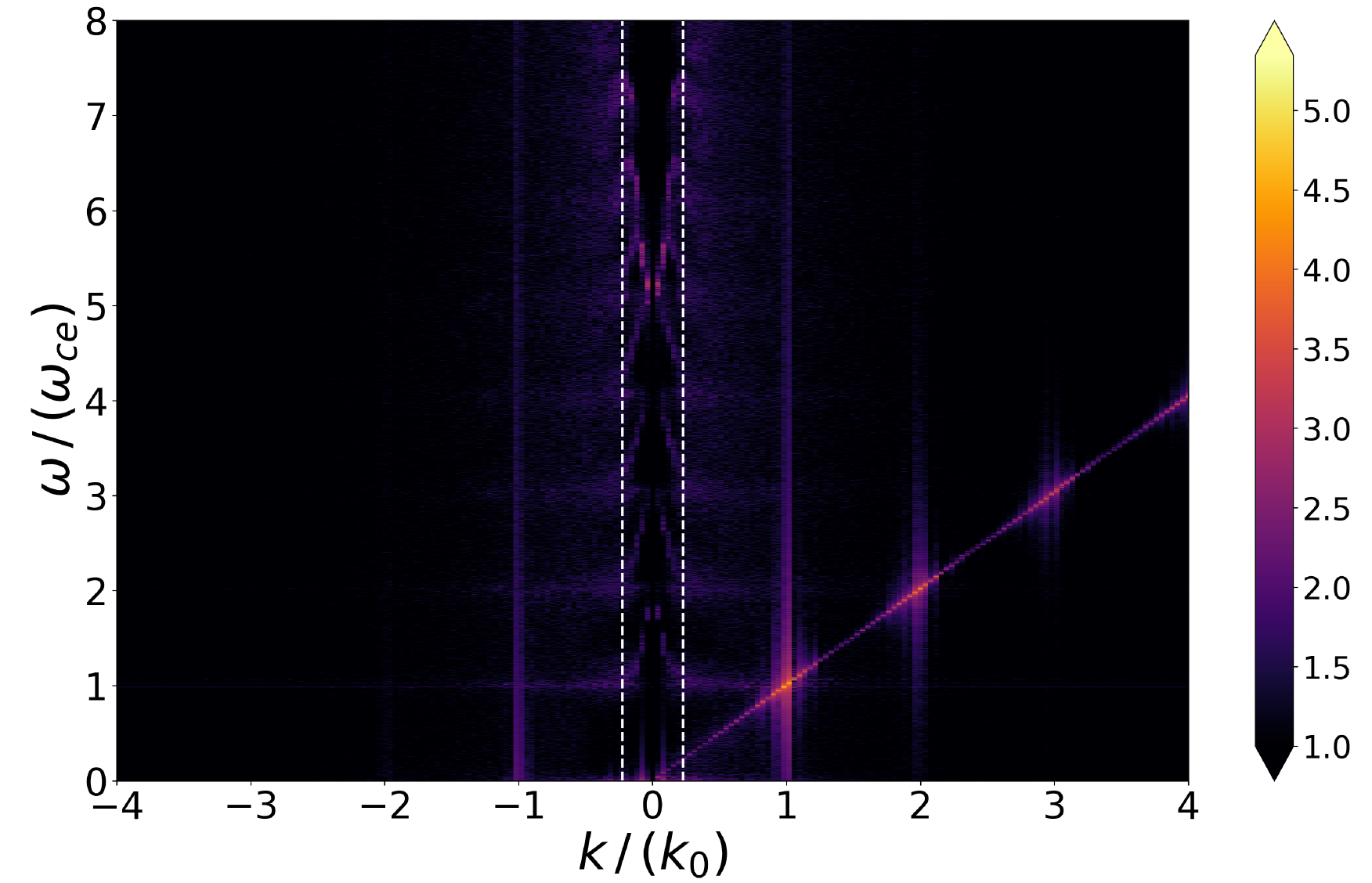}
\label{fig:16d}
\end{subfigure}
\begin{subfigure}[b]{0.48\textwidth}
\centering
\caption{ Phase 5}
\includegraphics[width=\textwidth]{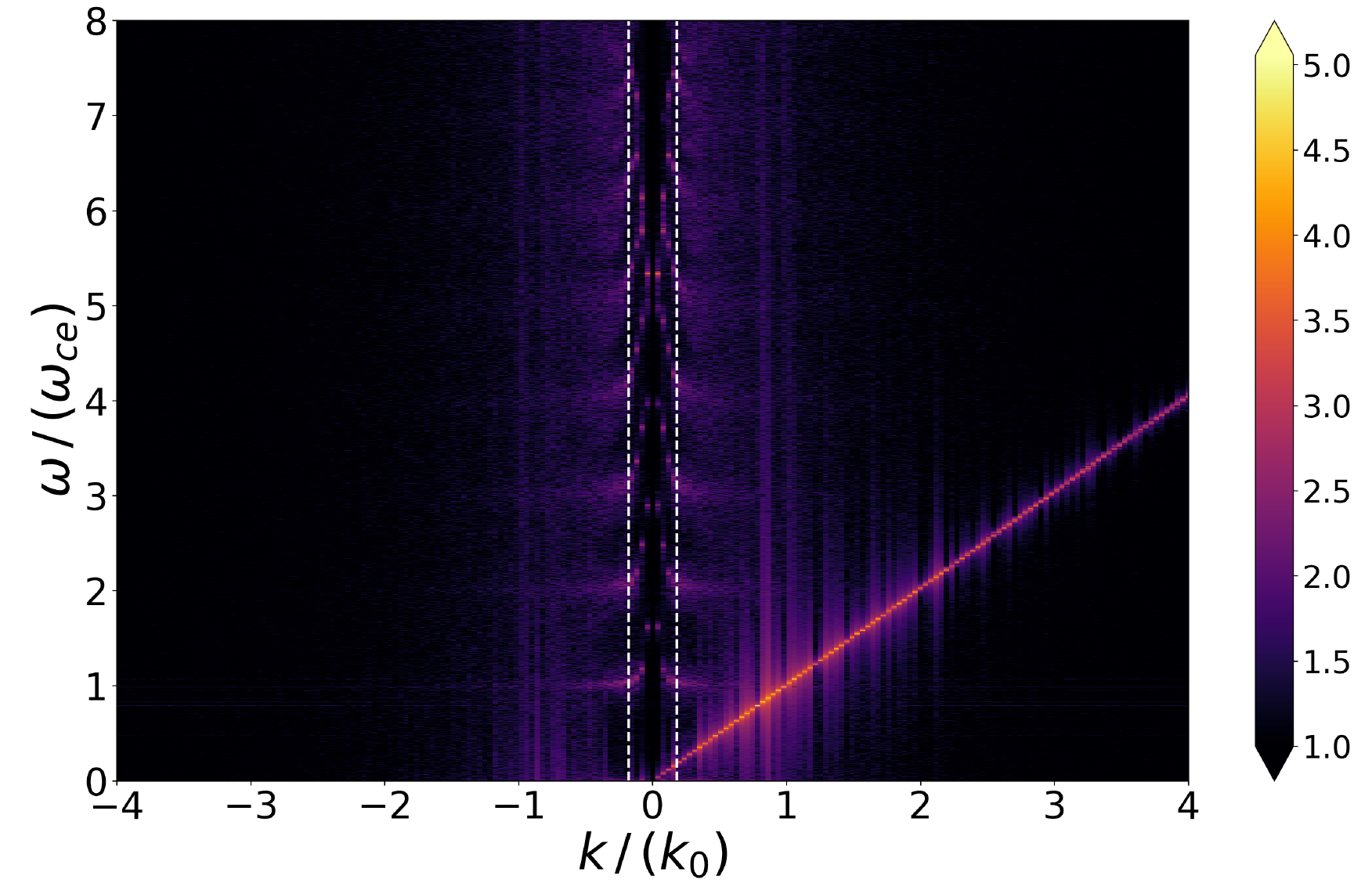}
\label{fig:16e}
\end{subfigure}
\caption{The logarithm of the 2D Fourier transform of electric field $\log(\tilde{E}_z (w, k))$, given by Eq.~\eqref{eq:2dfft}, for five different time ranges for Case 1B. (a) Phase 1 (0 < t < $\mathit{T_1}$), (b) Phase 2 ($\mathit{T_1} < t < \mathit{T_2}$), (c) Phase 3 ($\mathit{T_2}$ < t < $\mathit{T_3}$), (d) Phase 4 ($\mathit{T_3}$ < t < $\mathit{T_5}$), and (e) Phase 5 ($t \gg \mathit{T_5}$). The two vertical dotted white lines in Figs.~\ref{fig:16a}--~\ref{fig:16e} correspond to $k = \pm k_s$ values obtained at the end of each phase.}
\label{fig:16}
\end{figure*}

The characteristic transitions from Phase 1 to Phase 5, as observed in the evolution of electrostatic field energy $\epsilon_s$, the 1D Fourier transform of the electric field $\tilde{E}_z (k, t)$, the ion $\tilde{n}_i (k, t)$ and electron $\tilde{n}_i (k, t)$ densities, and the electron $\tilde{f}_e$ and ion $\tilde{f}_i$ distribution functions, exhibit similar behaviour for both Case 1A and Case 1B. The complete evolution of Case 1B can be fully described from Fig.~\ref{fig:12}--\ref{fig:16}.

The evolution of the electrostatic field energy $\epsilon_s$, given by Eq.~\eqref{eq:es}, for Case 1B is shown in Fig.~\ref{fig:12}. Fig.~\ref{fig:12a} shows the evolution of $\epsilon_s$ on logscale and up to $t = 4500$ ns and can be divided at five distinct times ($\mathit{T_1}-\mathit{T_5}$), similar to Case 1A. Case 1B is run for $t=10000$ ns, and the complete evolution is shown in Fig.~\ref{fig:12b}. We quantify different times $\mathit{T_1}-\mathit{T_5}$ for Case 1B as $T_1$ = 110 ns, $T_2$ = 480 ns, $T_3$ = 680 ns, $T_4$ = 2300 ns, and $T_5$ = 4040 ns. From Fig.~\ref{fig:2a} and \ref{fig:12a}, we note that times $T_1$ to $T_3$ for both Case 1A and Case 1B are similar, but times $T_4$ and $T_5$ for Case 1B are prolonged to 2300 ns (for Case 1A: $T_4=1800$ ns) and 4040 ns (for Case 1A: $T_5=3100$ ns), respectively. All the phases (Phase 1 to Phase 5) are again characterized by the same time ranges ($\mathit{T_1}-\mathit{T_5}$) as used for Case 1A and as described in Sec.~\ref{sec3:level1}. From Fig.~\ref{fig:12}, one also observes that in the final quasi-equilibrium state after $T_5$ (in Phase 5), the energy in electrostatic fluctuations saturates at a much higher value of about $4$ eV, as compared to about $1$ eV for Case 1A.

Figure \ref{fig:13} shows the anomalous current $J_{xe}$ (calculated from Eq.~\eqref{eq:J}) in the $x$-direction for Case 1B. It is important to note that, in Case 1B, the ion beam is driven in the same direction as the $\mathbf{E} \times \mathbf{B}$ electron drift observed in Case 1A. Consequently, the wave vector $k$ is oriented in opposite directions for Case 1A and Case 1B. This results in the electrons experiencing anomalous drift in the positive $x$-direction for Case 1B, in contrast to in the negative $x$-direction observed in Case 1A (Figure \ref{fig:11a}). Additionally, we observe a higher level of fluctuation about the moving average of the anomalous current for Case 1B, in comparison to Case 1A.

\subcaptionsetup{font=large}

Fig.~\ref{fig:14} shows the evolution of the spectral coefficients for the 1D Fourier transform of the electric field $E_z$ (Fig.~\ref{fig:14a} $\tilde{E}_z (k, t)$), the ion density $n_i$ (Fig.~\ref{fig:14b} $\tilde{n}_i (k, t)$), and the electron density $n_e$ (Fig.~\ref{fig:14c} $\tilde{n}_e (k, t)$) for Case 1B. Similarly to Fig.~\ref{fig:3} for Case 1A, we derive Fig.~\ref{fig:14} for Case 1B, from Eq.~\eqref{eq:1dfft} as described in Sec.~\ref{sec3:level1}. Comparing Fig.~\ref{fig:3} and Fig.~\ref{fig:14}, we observe that the complete evolution is similar for both Case 1A and Case 1B. One distinctive feature is the growth of half-integer cyclotron harmonics at $k/k_0 = 1/2, 3/2, 5/2,\dots$ for Case 1A, as seen in the electric field and ion density fluctuations (Figs.~\ref{fig:3a} and \ref{fig:3b}), which are clearly absent for Case 1B (Figs.~\ref{fig:14a}, and \ref{fig:14b}). The spectra for electron density perturbations for Case 1A (Fig.~\ref{fig:3c}) are similar to Case 1B (Fig.~\ref{fig:14c}).

Fig.~\ref{fig:15} shows the evolution of the ion phase space in column 1, the ion distribution function in column 2, and the electrons phase space in column 3 for Case 1B. The snapshots are provided for 6 different times from Phase 1 to Phase 5. The ion distribution function $\tilde{f}_i$ for Case 1B is calculated similarly as for Case 1A from Eq.~\eqref{eq:fe}. The evolution of all the three quantities in Fig.~\ref{fig:15} for Phases 1--3 (Fig.~\ref{fig:15a}--\ref{fig:15c}) is comparable to their counterparts from Case 1A (Figs.~\ref{fig:4a}--\ref{fig:4c}). For Phases 4 and 5 (Figs.~\ref{fig:15d}--\ref{fig:15f}), we see that the quasi-linear distortion to the ion distribution function is caused by the low-energy ions comprising the long tail in the opposite direction to the long tail observed due to non-linear ion trapping for Case 1A, where the high-energy ions were responsible for the long tail formation.

\subcaptionsetup{font=small}
\begin{figure*}[htbp]
\centering
\begin{subfigure}[b]{\textwidth}
\centering
\caption{ $\sim$ End of Phase 1}
\includegraphics[width=\textwidth]{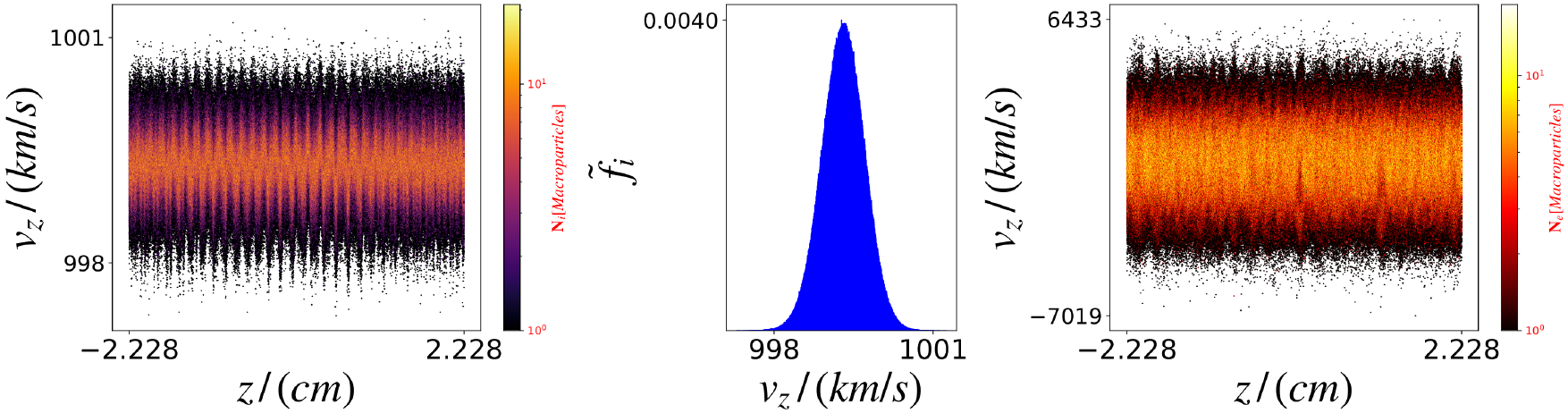}%
\label{fig:15a}
\end{subfigure}
\begin{subfigure}[b]{\textwidth}
\centering
\caption{ $\sim$ End of Phase 2}
\includegraphics[width=\textwidth]{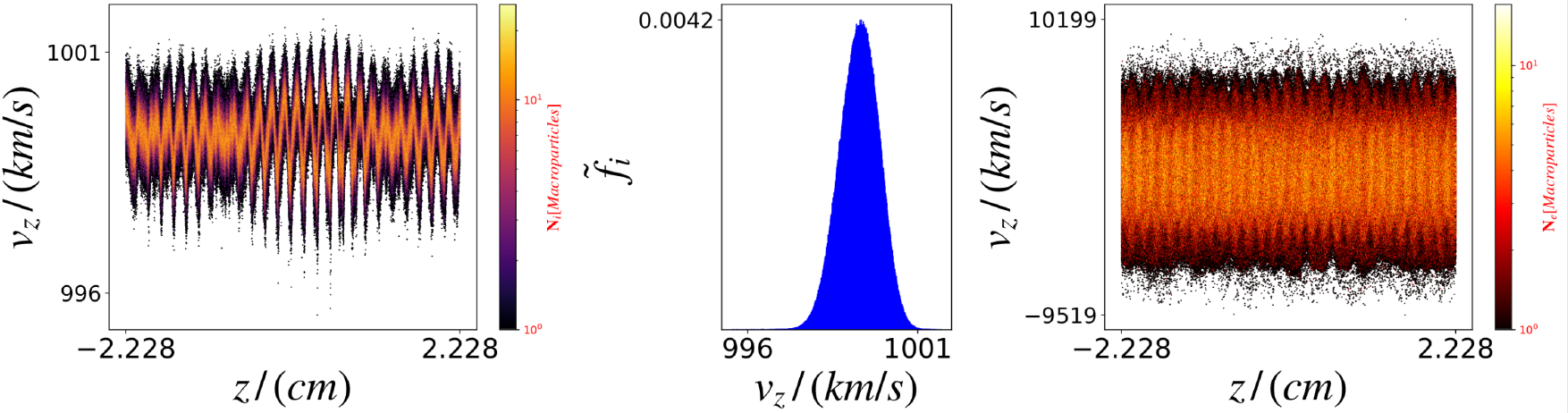}
\label{fig:15b}
\end{subfigure}
\begin{subfigure}[b]{\textwidth}
\centering
\caption{ $\sim$ End of Phase 3}
\includegraphics[width=\textwidth]{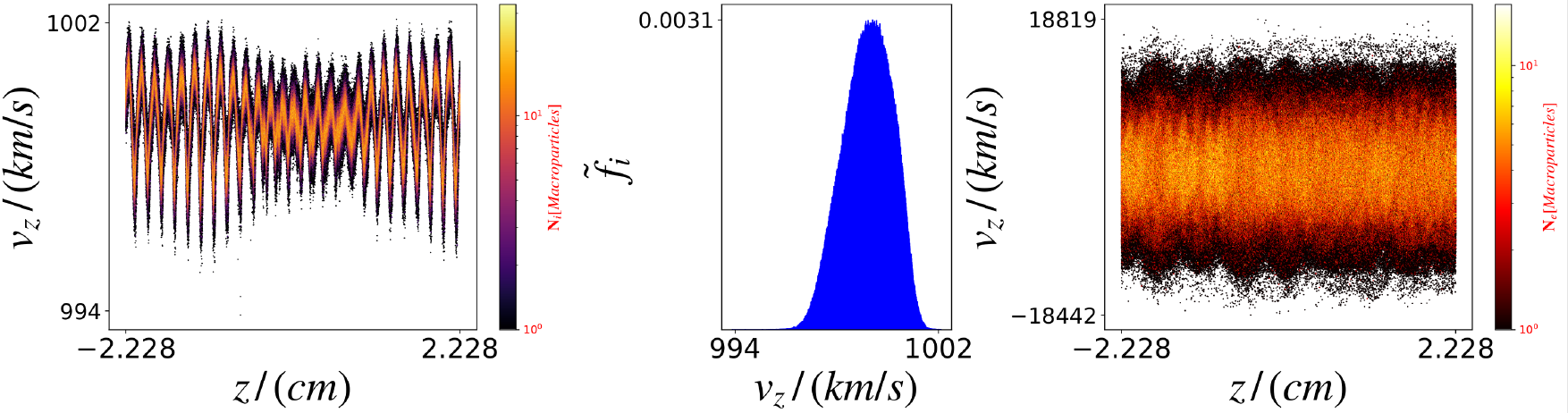}
\label{fig:15c}
\end{subfigure}
\begin{subfigure}[b]{\textwidth}
\centering
\caption{ $\sim$ Middle of Phase 4}
\includegraphics[width=\textwidth]{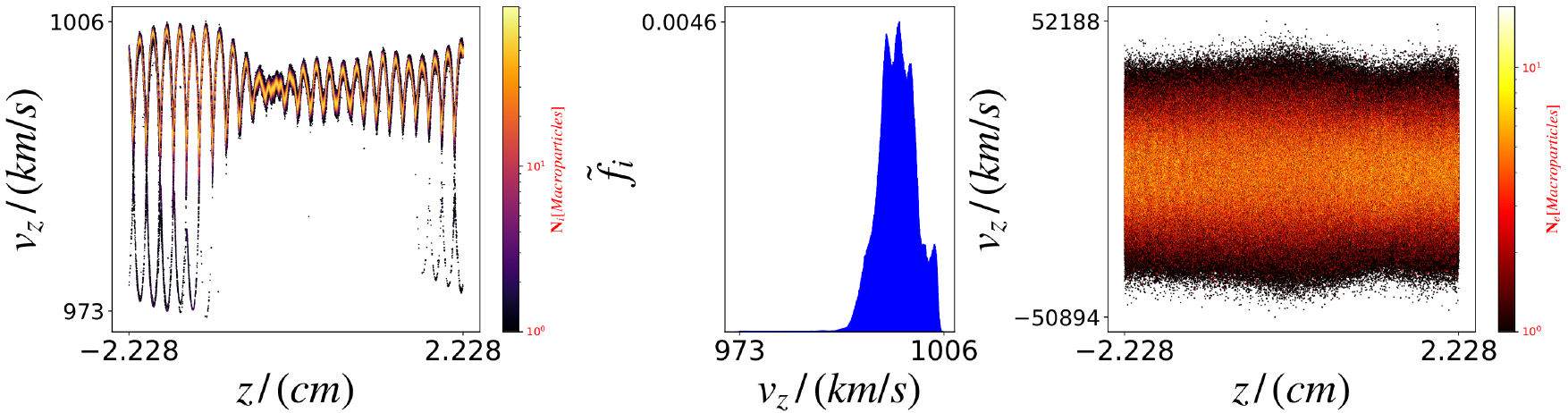}
\label{fig:15d}
\end{subfigure}
\end{figure*}

\begin{figure*}[htbp]\ContinuedFloat
\begin{subfigure}[b]{\textwidth}
\centering
\caption{ $\sim$ End of Phase 4}
\includegraphics[width=\textwidth]{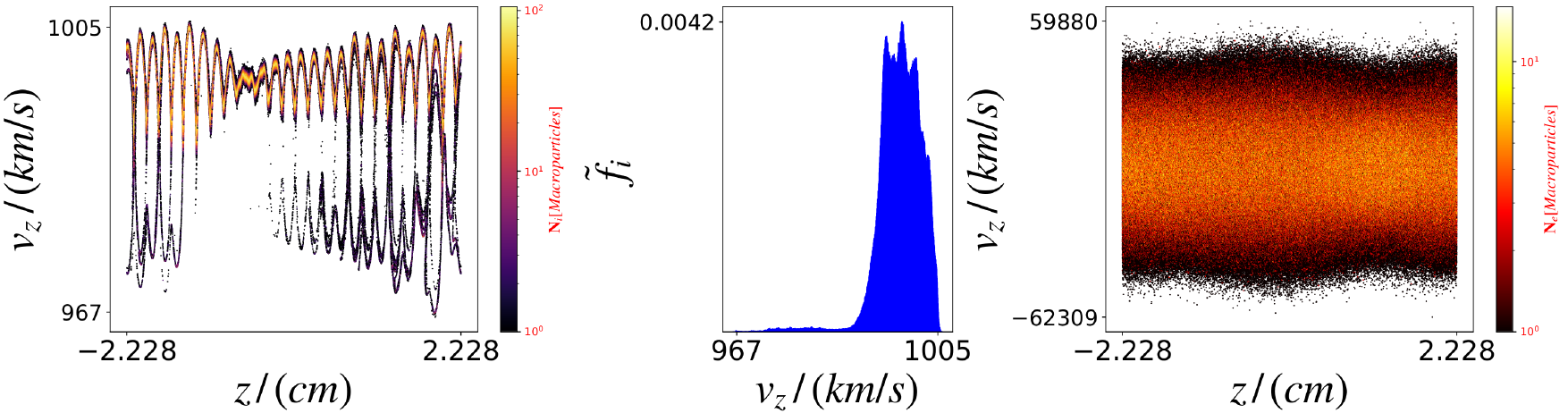}
\label{fig:15e}
\end{subfigure}
\begin{subfigure}[b]{\textwidth}
\centering
\caption{ $\sim$ End of Phase 5}
\includegraphics[width=\textwidth]{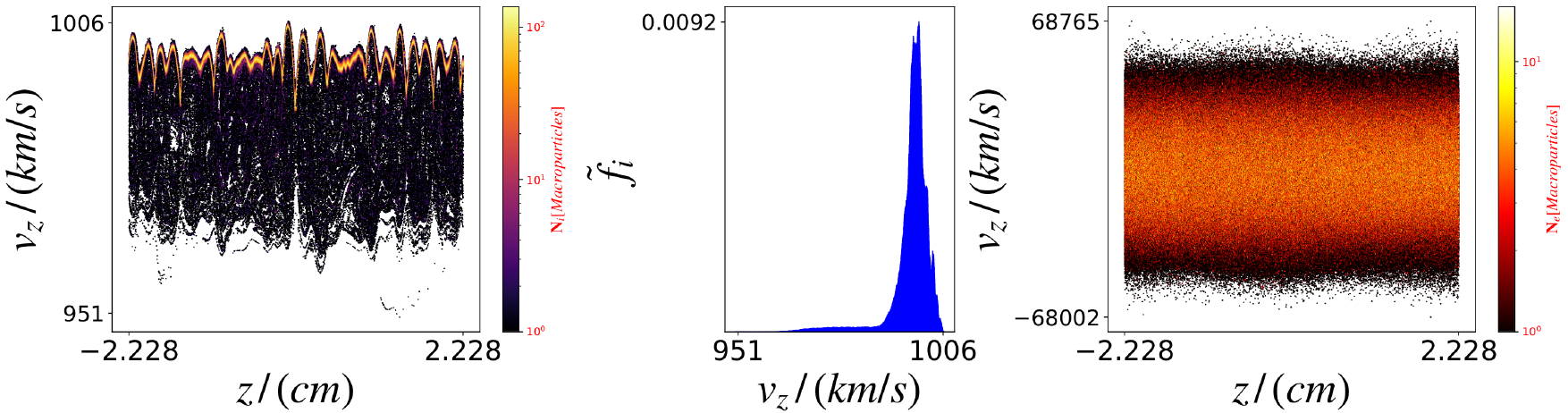}
\label{fig:15f}
\end{subfigure}
\caption{The snapshots of the evolution of three quantities shown for six different timeframes from Phase 1 to Phase 5 for Case 1B. Column 1: Ion phase space in the $z$-direction, Column 2: Ion distribution function $\mathit{\tilde{f_i}}$ as given by Eq.~\eqref{eq:fe}, and Column 3: Electron phase space in the $z$-direction.}
\label{fig:15}
\end{figure*}

\subsection{Comparison of the ECDI driven by the electron $\mathbf{E} \times \mathbf{B}$ current (Case 1A) with the ECDI driven by the ion beam (Case 1B).}

The situation of the ECDI driven by an electron $\mathbf{E}\times \mathbf{B}$ drift current as considered in Case 1A is typical for $\mathbf{E}\times \mathbf{B}$ plasma devices such as Hall thrusters. Other laboratory experiments~\cite{StenzelPRL1973,StenzelRSI1973w} as well as space applications consider the ECDI driven by an ion beam perpendicular to the magnetic field. In the linear approximation, these two situations are identical and can be transformed to each other by a change of  reference frame. The energy of the ions moving in Case 1B with velocity equivalent to the electron $\mathbf{E}\times \mathbf{B}$ drift in Case 1A is much larger than that of the electron flow. It is of interest to compare the non-linear behavior of these two cases. The linear dispersion relation for ECDI driven by an ion beam in the z-direction (with velocity $V_{di}$) can be obtained from Eq.~\eqref{deq} by replacing the wave frequency in the expression for the ion susceptibility $\chi_{i}$, Eq.~\eqref{eq:chii}, with a Doppler shifted frequency $\omega \rightarrow -k V_{di}$, and removing the $\mathbf{E}\times \mathbf{B}$ drift in the electron susceptibility, $\chi_{e}$, Eq.~\eqref{chie}. The dispersion relation thus obtained is slightly different from Case 1A (Eq.~\eqref{deq}) such that the Doppler shift is introduced in ion susceptibility rather than the electron susceptibility.






When the ECDI is driven by an ion beam current ($V_{di}$), the ion beam modes are Doppler-shifted by $k_z V_{di}$ into higher frequencies of electron Bernstein modes ($\sim \omega_{ce}$), resulting in a reactive instability due to the interaction between the electron Bernstein modes and the Doppler shifted ion beam mode ($\omega/k = V_{di}$). Therefore, the resonance condition is similar to Eq.~\eqref{eq:2dfft_dop}, given by
\begin{equation*}
\omega -k_{z}V_{di}\simeq m\omega _{ce}.
\end{equation*}

This is the ECDI driven by an ion beam, forming a discrete set of narrow-band unstable modes centered around $k/k_{0}=1,2,3,\ldots $, $k_{0}\equiv \omega _{ce}/V_{di}$. The associated frequencies $ \omega $ typically fall within a range of $\omega _{ce}$. We note that $\omega _{ce} \gg \omega _{pi}$. 

Fig.~\ref{fig:16} shows the evolution of the spectral coefficients for the 2D Fourier transform of the electric field $E_z$ derived using  Eq.~\eqref{eq:2dfft} and represents the evolution of the dispersion relation ($\omega$--$k$ spectra) starting from the linear stage (Phase 1) up to the deeply non-linear stage (Phase 5) (Figs.~\ref{fig:16a}--\ref{fig:16e}). In Phase 1 (Fig.~\ref{fig:16a}), one sees the generation of typical high-frequency electron Bernstein modes (from $\omega_{ce}$ to $8 \omega_{ce}$) and the Doppler-shifted ion beam mode $\omega / k = V_{di}$ crossing these Bernstein branches and giving rise to unstable modes centered around cyclotron harmonics ($k/k_0 = 1,2,3,\dots$) at frequencies ($\omega/\omega_{ce} = 1,2,3,\dots$). Similar to Case 1A, during Phase 2 (Fig.~\ref{fig:16b}) and Phase 3 (Fig.~\ref{fig:16c}), we observe the resonance broadening of the Bernstein branches, and after the end of Phase 3, when the ion-acoustic mode $k_s$ resonates with the ECDI $m=1$ mode (at $k/k_0 = 1$, shown by the vertical white line at $k/k_0 = 1$ in Fig.~\ref{fig:16c}), we observe that the fluctuation energy is primarily concentrated at cyclotron harmonics ($k/k_0 = 1,2,3,\dots$) (Fig.~\ref{fig:14} after $t=T_3$ and Fig.~\ref{fig:16d}).

Similar to Case 1A, in Phase 4 (Fig.~\ref{fig:16d}), we see the enhancement of fluctuation energy concentrated at cyclotron harmonics ($k/k_0 = 1,2,3,\dots$). In Phase 5 (Fig.~\ref{fig:16e}), we see that the energy is concentrated at cyclotron harmonics ($k/k_0 = 1,2,3,\dots$) and disperse along the curve $\omega / k = V_{di}$. Trivially, this curve does not in any way resemble the ion-acoustic like dispersion curves obtained for Phase 5 (Fig.~\ref{fig:7b}) for Case 1A.

\begin{figure*}[htbp]
\centering
\begin{subfigure}[b]{0.48\textwidth}
\centering
\caption{Phase 1 }
\includegraphics[width=\textwidth]{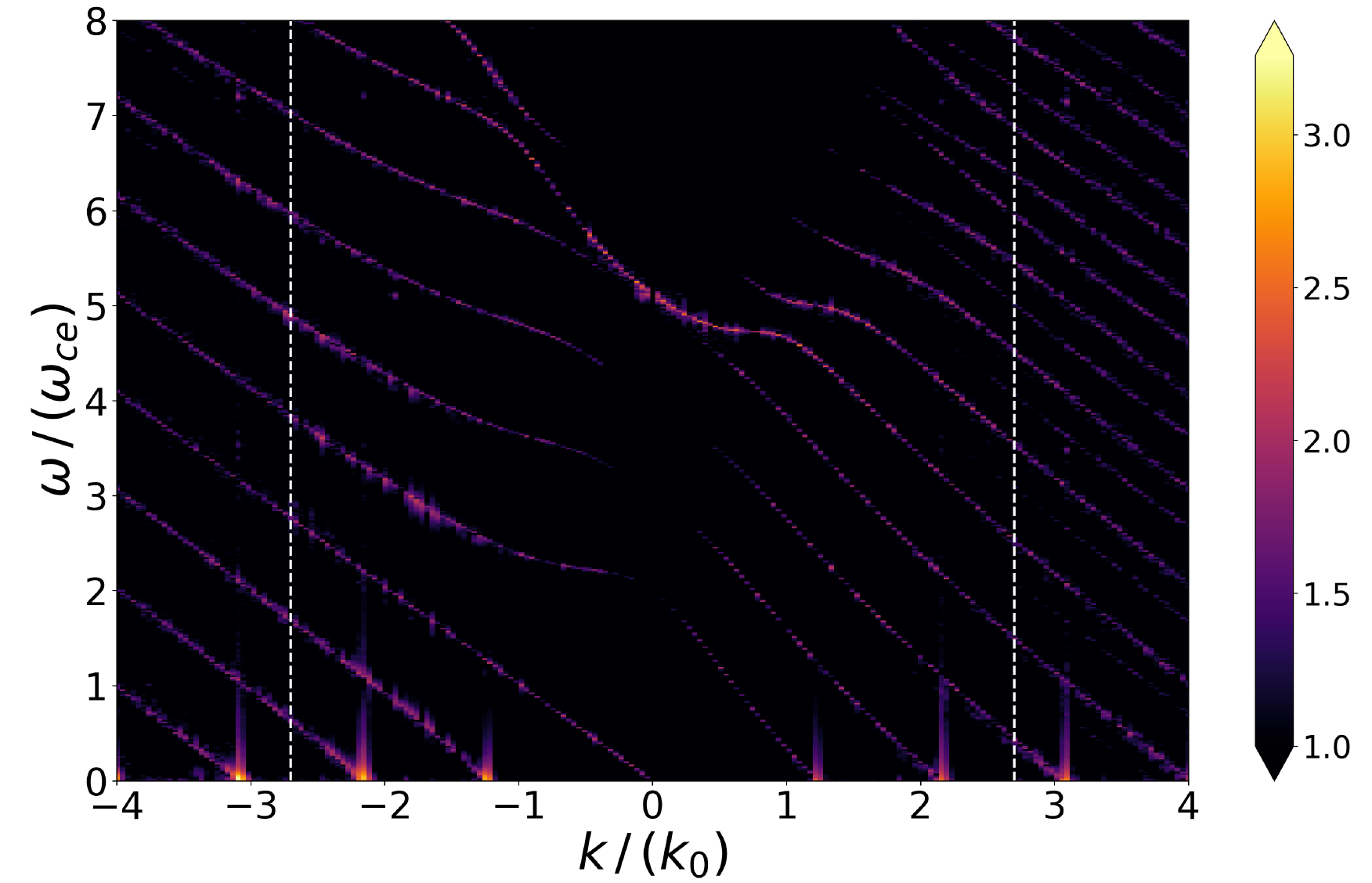}
\label{fig:17a}
\end{subfigure}
\begin{subfigure}[b]{0.48\textwidth}
\centering
\caption{ Phase 5}
\includegraphics[width=\textwidth]{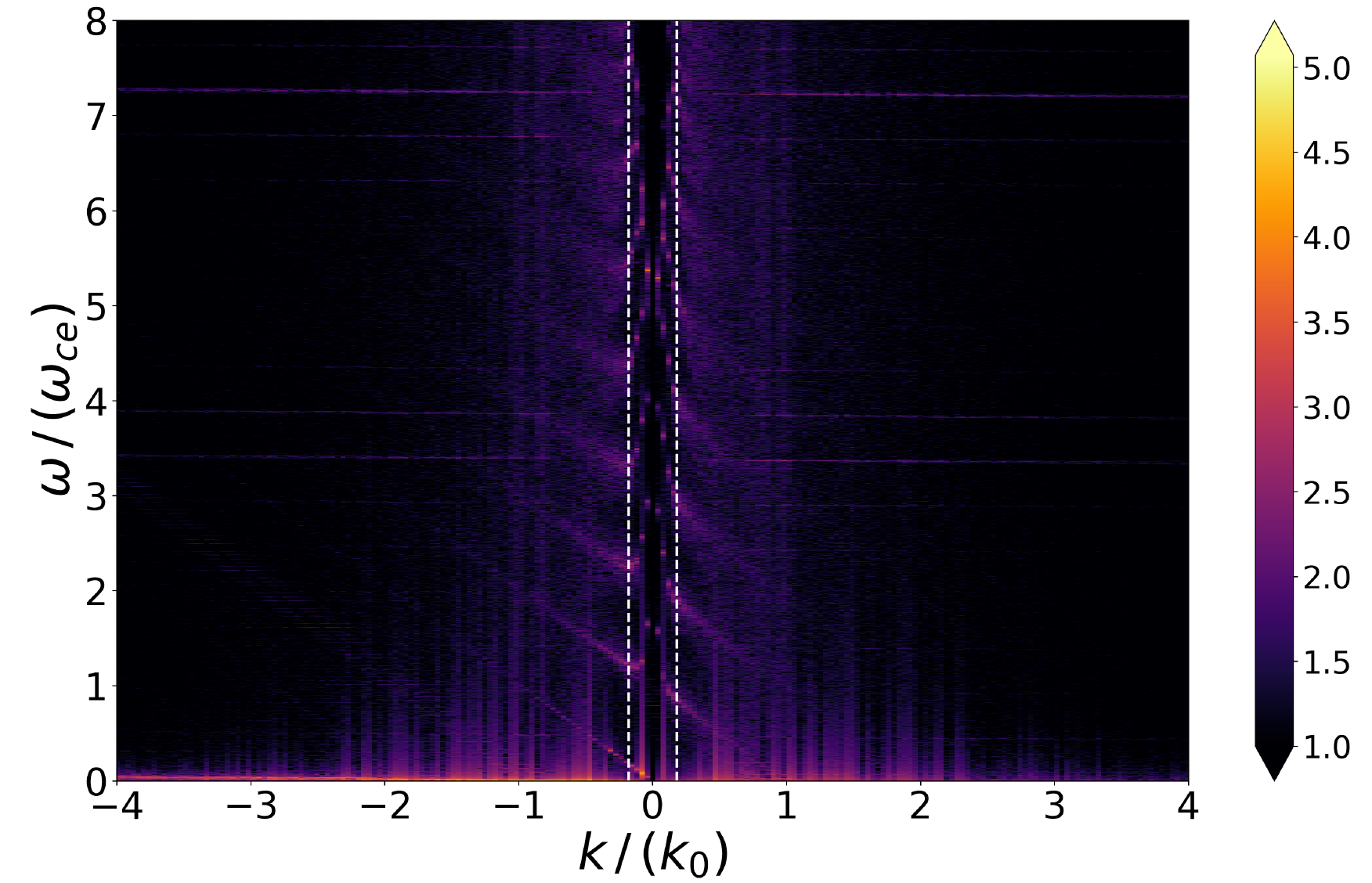}
\label{fig:17b}
\end{subfigure}
\begin{subfigure}[b]{0.48\textwidth}
\centering
\caption{ Phase 5}
\includegraphics[width=\textwidth]{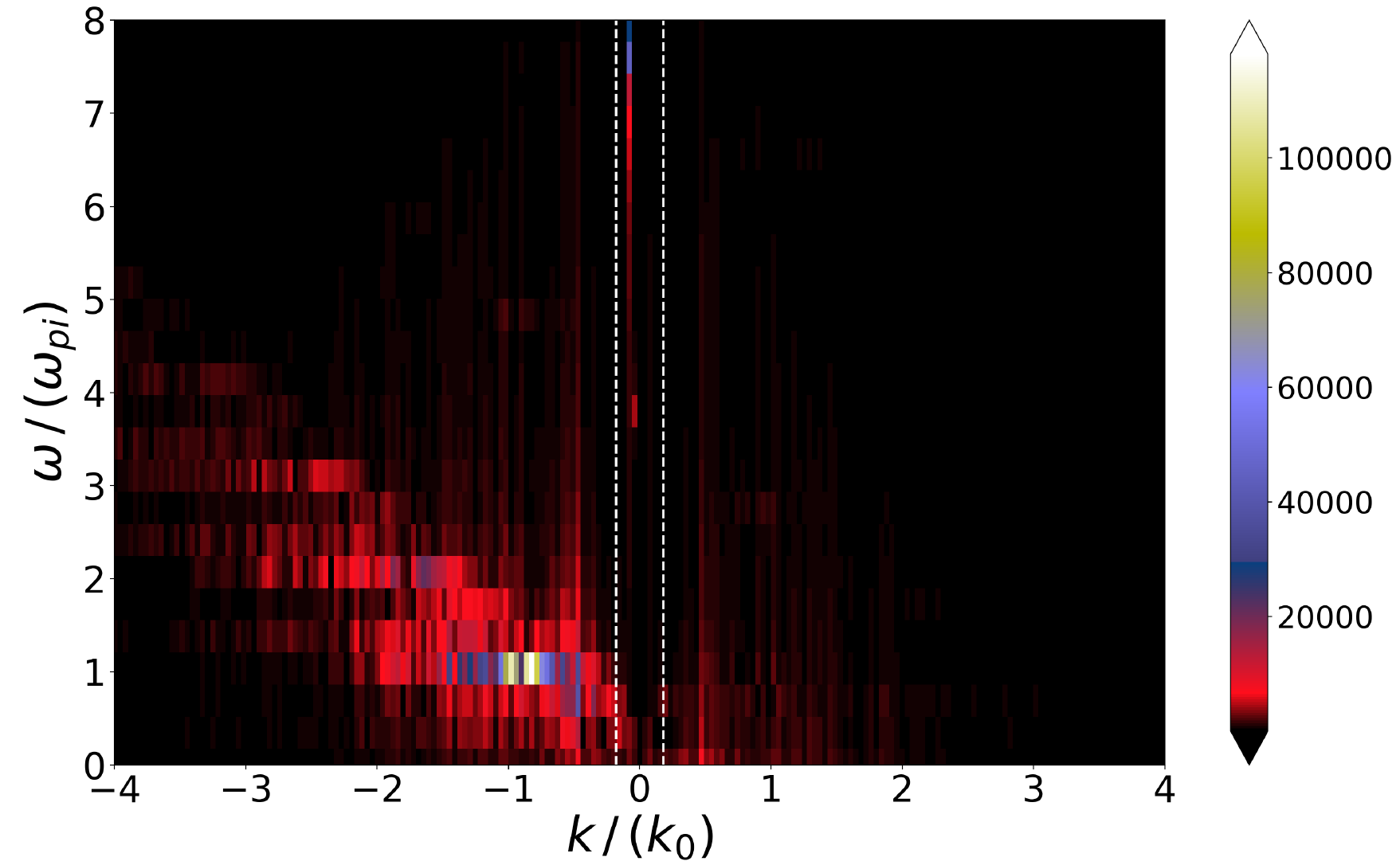}
\label{fig:17c}
\end{subfigure}
\caption{The 2D Fourier transform for the Doppler-shifted electric field signal $\tilde{E}_z^{\text{DOP}} (\omega, k)$, given by Eq.~\eqref{eq:2dfft_dop}; (a) $\log (\tilde{E}_z^{\text{DOP}} (\omega, k))$ for linear stage, Phase 1 (0 < t < $\mathit{T_1}$) and (b) $\log (\tilde{E}_z^{\text{DOP}} (\omega, k))$ for deeply non-linear stage, Phase 5 ($t \gg \mathit{T_5}$). (c) $\tilde{E}_z^{\text{DOP}} (\omega, k)$ for Phase 5, plotted on normal scale to highlight the dynamics around $\omega_{pi}$.}
\label{fig:17}
\end{figure*}

When comparing the non-linear evolution of derived quantities from Figs.~\ref{fig:3}--\ref{fig:8} for Case 1A with those of Case 1B (Figs.~\ref{fig:12}--\ref{fig:16}), one observes the distinct resemblance in the evolution of various integral quantities throughout the course of the simulations. Although the non-linear evolution of the dispersion relation for Case 1A and Case 1B cannot be trivially related,  the observed resemblance between the two cases motivated us to Doppler shift the electric field $E_z$ signal by $V_{di}$, obtained for Case 1B, and then derive the $\omega$--$k$ spectra for the linear stage (Phase 1, Fig.~\ref{fig:17a}) and the deeply non-linear stage (Phase 5, Fig.~\ref{fig:17b}) from this Doppler-shifted signal $E_{z}^{\text{DOP}}$.

The Doppler shift is applied to the electric field $E_z$ time-series signal by transforming it into a moving reference frame corresponding to the drift velocity $V_{di}$. This process is performed iteratively for each timestep of the signal. At each timestep, the current time is used to calculate a new set of spatial positions $z'$, adjusted by the velocity $V_{di}$ through the relation $z' = z - V_{di} \,\, t$. A key aspect of the transformation is the circular shift applied to ensure that the new positions remain within the spatial domain, which is handled by the operation $z'' = \mod(z' + L_z/2, L - L_z/2)$ and which also imposes periodic boundary conditions and ensures the spatial coordinates wrap around the domain size ($-L_z/2 \, , \, L_z/2$). For each transformed spatial coordinate, the nearest original position within the domain ($-L_z/2 \, , \, L_z/2$) is identified, and the corresponding value of the electric field from the original signal is mapped onto the Doppler-shifted signal array. This procedure is repeated for each timestep, yielding a time series of Doppler-shifted electric field data, $\tilde{E}_z^{\text{DOP}} (\omega, k)$, for Case 1B.



The 2D discrete FFT on the Doppler-shifted electric field signal $\tilde{E}_z^{\text{DOP}}$ can be expressed as

\begin{equation}
\label{eq:2dfft_dop}
\tilde{E}_z^{\text{DOP}} (\omega, k) = \frac{2}{N_t} \frac{2}{N_z} \sum_{\kappa = 0}^{N_t - 1} \sum_{\ell = 0}^{N_z - 1} E_z^{\text{DOP}}(z'', t) e^{-i2 \pi \omega \kappa/N_t} e^{-i 2\pi k\ell/N_z}.
\end{equation}

Fig.~\ref{fig:17} shows the results obtained from transformation Eq.~\eqref{eq:2dfft_dop}. As mentioned previously, the ion beam is driven in the same positive $z$-direction (for Case 1B) as the $\mathbf{E} \times \mathbf{B}$ electron drift is applied for Case 1A. Consequently, the wave vector $k$ is oriented in opposite directions for Case 1A and Case 1B. With this fact in mind, one can see that the $\omega$--$k$ spectra obtained for Phase 1 (Fig.~\ref{fig:17a}) are similar to the $\omega$--$k$ spectra obtained for Phase 1 of Case 1A (Fig.~\ref{fig:6a}), with the primary difference being the reversal of the wave vector $k$ direction.

The deeply non-linear stage (Phase 5, Figs.~\ref{fig:17b} and \ref{fig:17c}) can be related similarly to Phase 5 obtained for Case 1A, Fig.~\ref{fig:6e} and \ref{fig:7b}, respectively. In Fig.~\ref{fig:17c}, the $\omega , k$ spectra obtained for Phase 5 are zoomed at lower frequencies ($\sim \omega_{pi}$) and not plotted on a logarithm scale to enhance features near $\omega_{pi}$. Comparing the negative wave vector part ($k/k_0 < 0$) of Fig.~\ref{fig:17c} and the positive wave vector part ($k/k_0 > 0$) of Fig.~\ref{fig:7b}, one can observe the resemblance in the ion-acoustic curves in Fig.~\ref{fig:7b} and the dispersion relation obtained at lower frequencies ($\sim \omega_{pi}$, Fig.~\ref{fig:17c}, although less resolved). With this observation and taking into consideration the distinct similarities between the evolution of various integral quantities in Case 1A and Case 1B, we conclude that the ECDI driven by the $\mathbf{E} \times \mathbf{B}$ electron drift and the ECDI driven by an ion beam are non-trivially related as the same instability seen from different Doppler-shifted frames. Fig.~\ref{fig:16e} thus refers to the dispersion relation that should be observed in the laboratory frame for ECDI driven by a xenon ion beam with the ion-acoustic like dispersion curves Doppler-shifted to lie along the Doppler-shifted ion beam mode ($\omega/ k = V_{di}$) alongside a subtle trace of skewed electron Bernstein wave modes.

\section{\label{sec5:level1}Results, Case 2: ECDI driven by magnetized H ion beam}

In applications to electric propulsion where heavy ions are used, such as Hall thrusters, the effects of the magnetic field on ion motion can be neglected as in our simulations  of  the ECDI driven by an unmagnetized ion beam (Case 1B) and direct comparison  with the ECDI driven by an $\mathbf{E} \times \mathbf{B}$ electron drift. The focus of these long-term, highly resolved simulations was to capture detailed transitions from the linear to the deeply non-linear stages of turbulence.


In space conditions, where much lighter ion species such as hydrogen prevail, the effects of the magnetic field on the ion trajectories could be important. The effects of the magnetic field on particle trajectories have to be included in many other laboratory applications of energetic particles beams   used for plasma heating and diagnostics.  \cite{KumarLPB2012,FujisawaNF1996,WelchPoP2006,DavidsonNI2009} It is therefore  of interest to 
consider the ECDI excitation with a fully magnetized ion beam.
Here, we select a hydrogen (H) ion beam that allows us to track turbulence development over  several ion cyclotron periods.

\begin{figure*}[htbp]
\centering
\begin{subfigure}[b]{0.49\textwidth}
\centering
\caption{ }
\includegraphics[width=\textwidth]{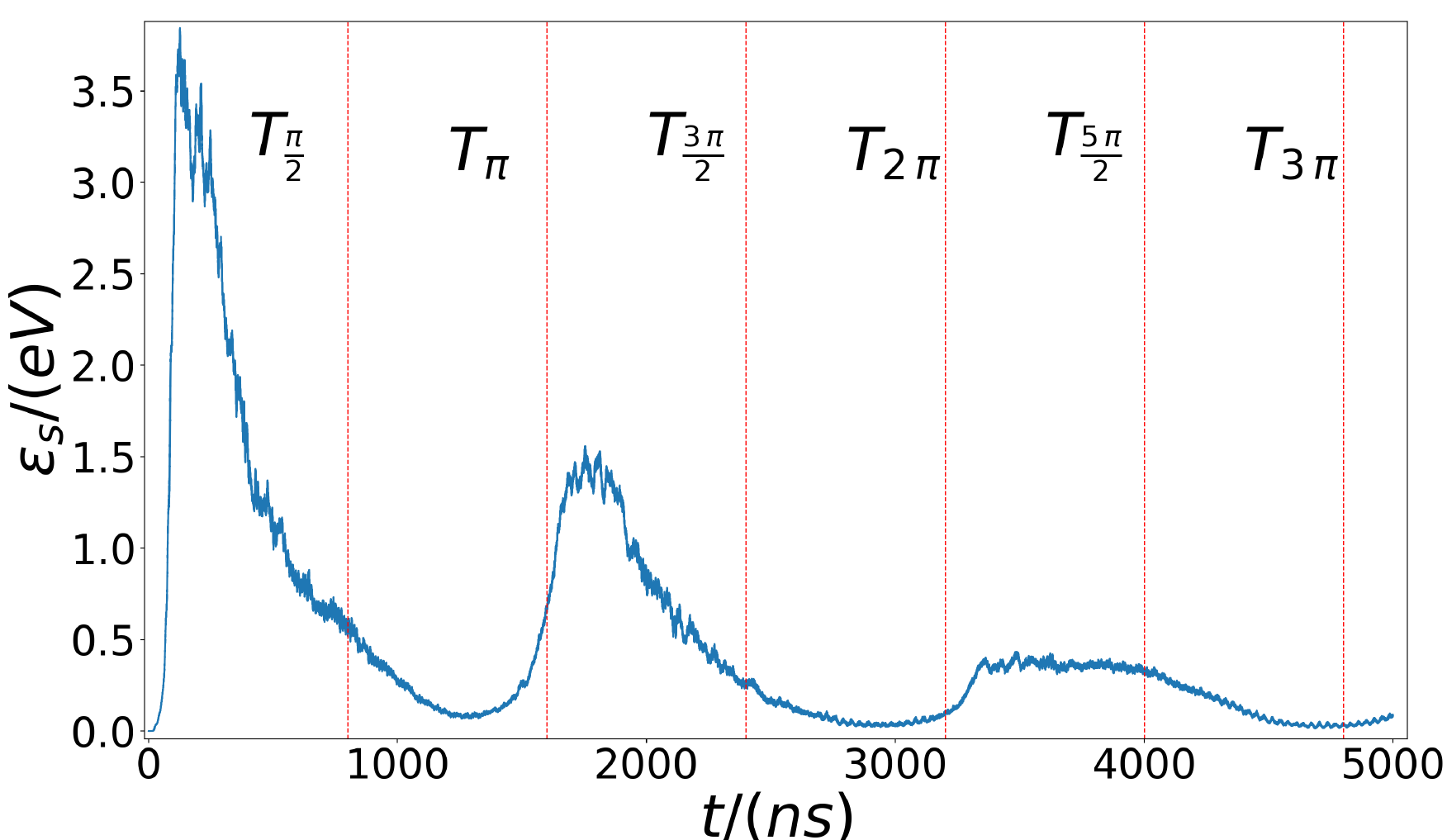}
\label{fig:18a}
\end{subfigure}
\begin{subfigure}[b]{0.49\textwidth}
\centering
\caption{ }
\includegraphics[width=\textwidth]{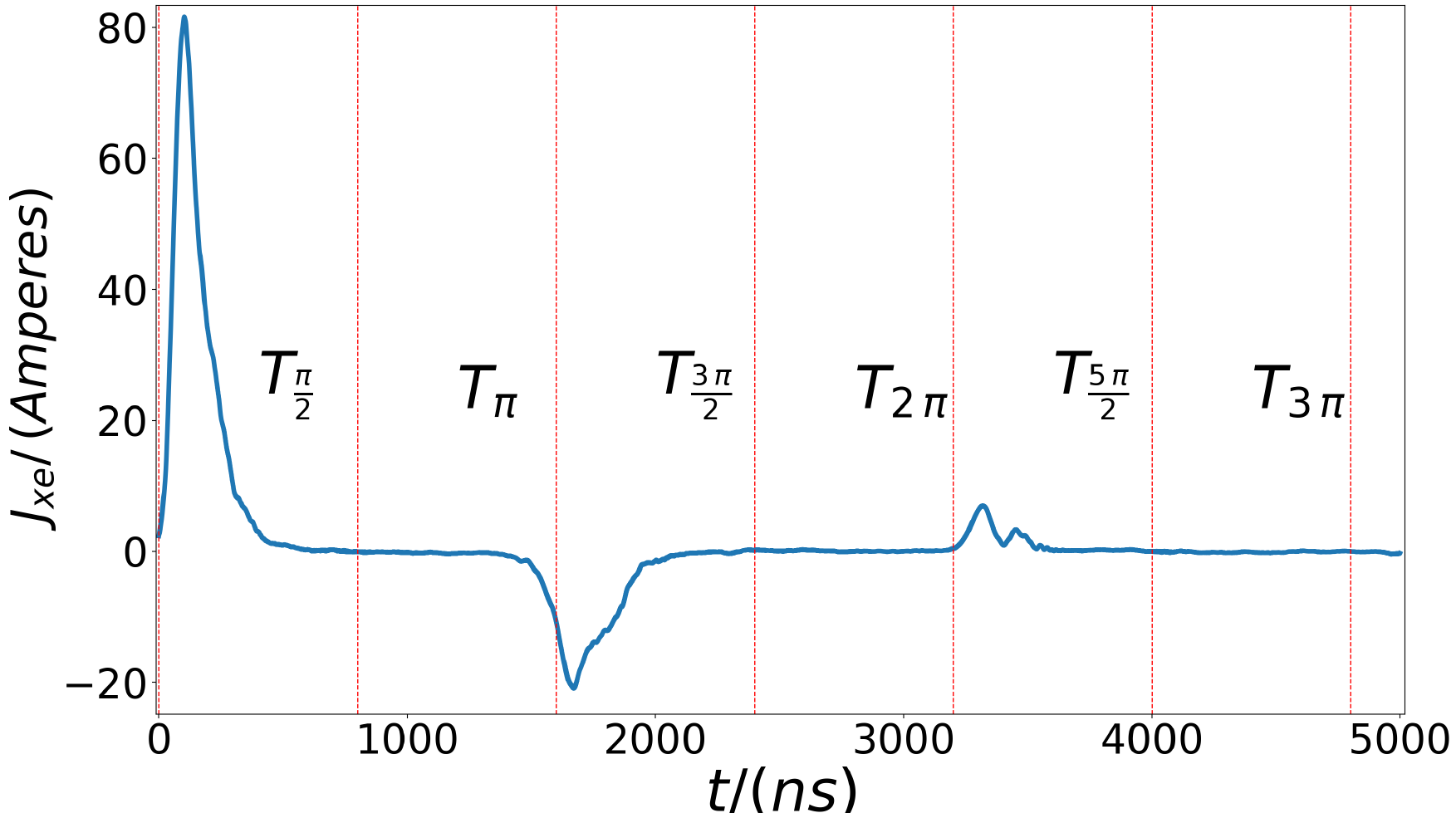}
\label{fig:18b}
\end{subfigure}
\begin{subfigure}[b]{0.6\textwidth}
\centering
\caption{ }
\includegraphics[width=\textwidth]{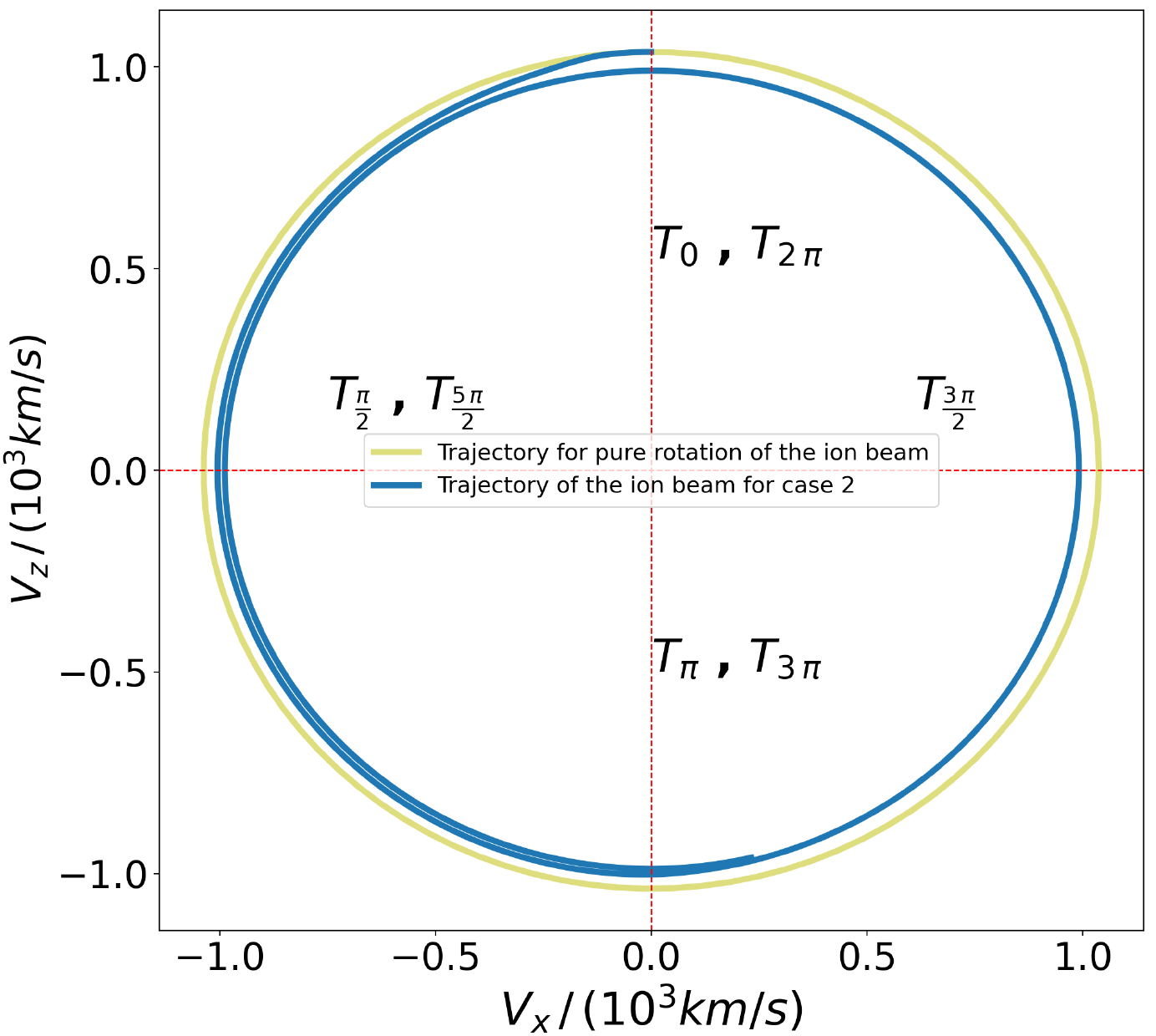}
\label{fig:18c}
\end{subfigure}
\caption{(a) Time evolution of the electrostatic field energy $\epsilon_s$ for Case 2, given by Eq.~\eqref{eq:es}, (b) Time evolution of the anomalous current $J_{xe}$ for Case 2, given by Eq.~\eqref{eq:J}, and (c) Trajectory of the ion beam in $\mathit{V_x}$-$\mathit{V_z}$ plane.}
\label{fig:18}
\end{figure*}

\begin{figure*}[htbp]
\centering
\begin{subfigure}[b]{0.8\textwidth}
\centering
\caption{ }
\includegraphics[width=\textwidth]{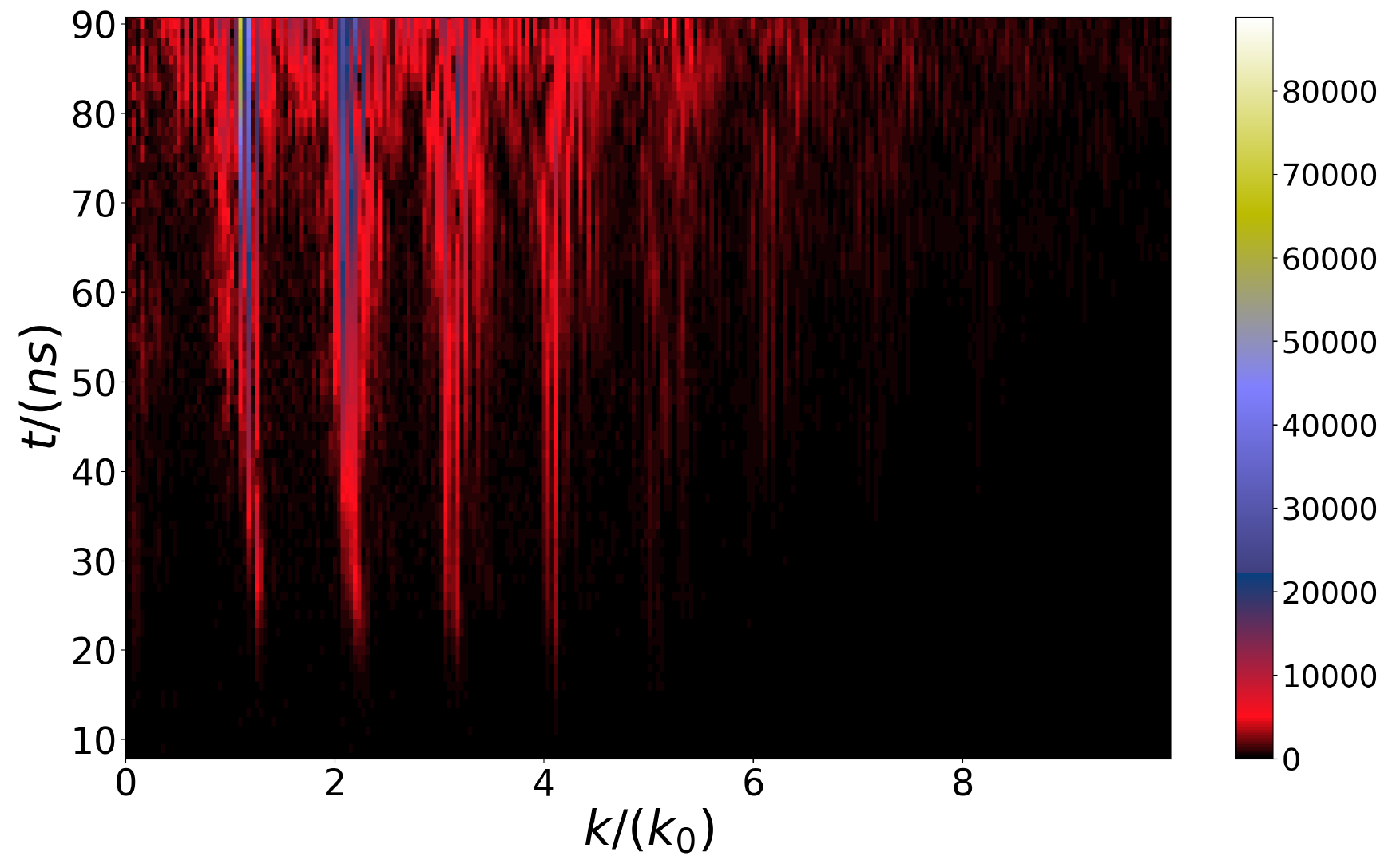}
\label{fig:19a}
\end{subfigure}
\begin{subfigure}[b]{0.8\textwidth}
\centering
\caption{ }
\includegraphics[width=\textwidth]{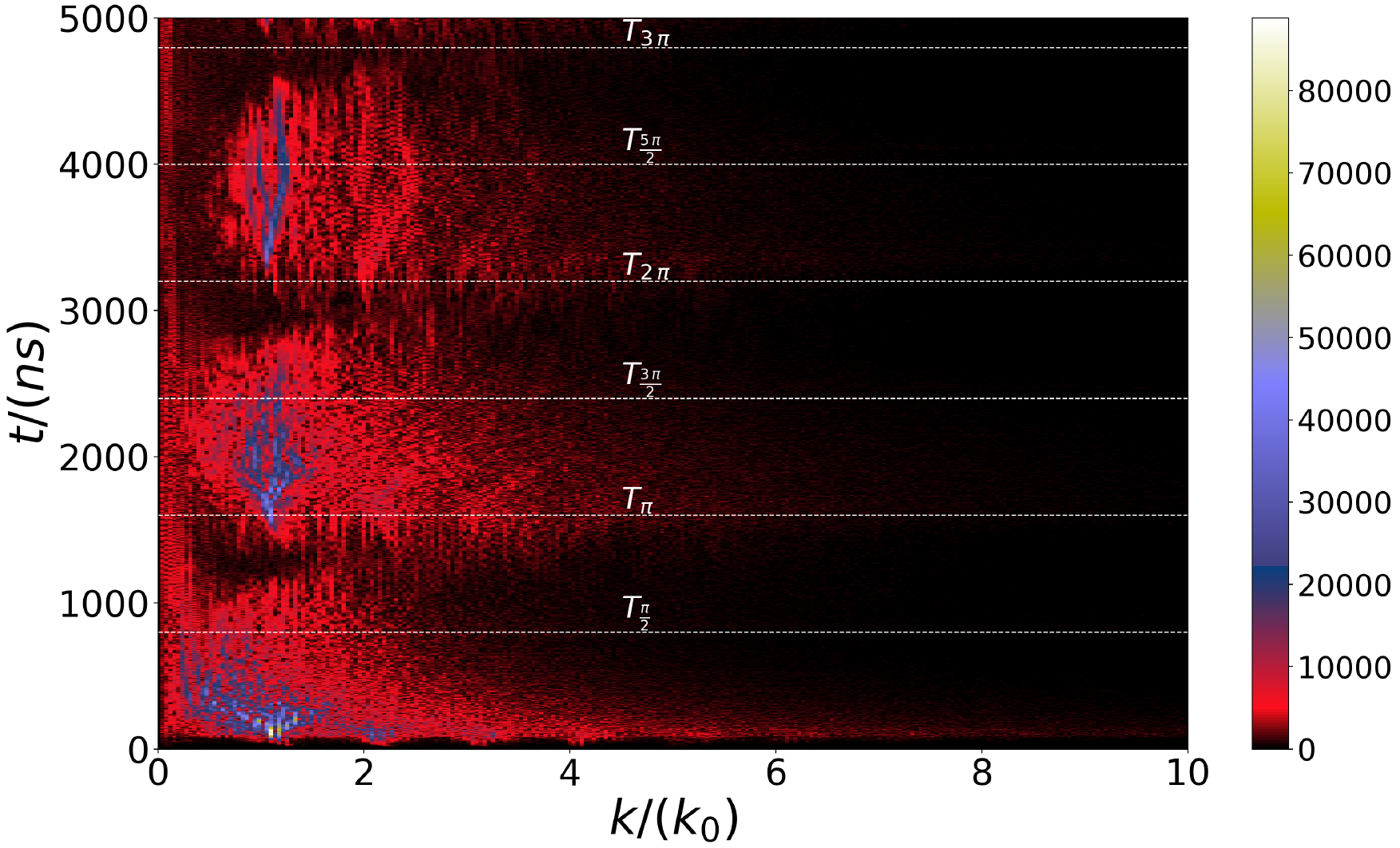}
\label{fig:19b}
\end{subfigure}
\caption{One-dimensional Fourier transform of electric field $E_z$ for (a) $0 < t < 100$ ns depicting initial growth from the first cycle and (b) $0 < t < 5000$ ns depicting three periodic burst cycles of growth and saturation of ECDI during 1.5 full beam inversion $(0 < t < T_{3 \pi})$.}
\label{fig:19}
\end{figure*}

The complete evolution of the ECDI driven by a magnetized ion beam (Case 2) is shown to develop a full beam inversion in the plane perpendicular to the applied magnetic field, accompanied by periodic burst cycles of growth and saturation of ECDI, along the direction of the beam current. Case 2 is run for $t=5000$ ns ($\sim$ 1.5 cyclotron periods for H ions), and the complete evolution is shown in Figs.~\ref{fig:18}--\ref{fig:19}.

Fig.~\ref{fig:18a} shows the evolution of the electrostatic field energy $\epsilon_s$ for Case 2 derived from Eq.~\eqref{eq:es}. Fig.~\ref{fig:18b} shows the anomalous current $J_{xe}$ (from Eq.~\eqref{eq:J}) generated from $t = 0$ to $t = 5000$ ns. Fig.~\ref{fig:18c} shows the trajectory of the ion beam in the $V_x$-$V_z$ plane for the complete simulation. Vertical dotted lines show different times from $0 < t < 5000$ ns, where $T_0$, $T_{\frac{\pi}{2}}$, $T_{\pi}$, $T_{\frac{3 \pi}{2}}$, $T_{2 \pi}$, $T_{\frac{5 \pi}{2}}$, and $T_{3 \pi}$ represent the times for $0^{\circ}, 90^{\circ}, 180^{\circ}, 270^{\circ}, 360^{\circ}, 450^{\circ}$, and $540^{\circ}$  rotation of the ion beam in the $V_x$-$V_z$ plane, respectively. From Figs.~\ref{fig:18a} and \ref{fig:18b}, one recognizes the first cycle of growth and saturation of ECDI from $0 < t < T_{\pi}$, the second cycle from $T_{\pi} < t < T_{2 \pi}$, and the third cycle from $T_{2 \pi} < t < T_{3 \pi}$. In the first cycle, the ion beam is directed in the positive $z$-direction, and we observe the generation of anomalous current $J_{xe}$ in the positive $x$-direction (similar to $J_{xe}$ for Case 1B, Fig.~\ref{fig:13}). After $t = T_{\pi}$, the ion beam is inverted by $180^{\circ}$, and now the wave vector $k$ points in the negative $z$-direction, and the anomalous current $J_{xe}$ is generated in the negative $x$-direction (Fig.~\ref{fig:18b}). After $t = T_{2 \pi}$, the ion beam experiences a full beam inversion, and the wave vector $k$ is again directed in the positive $z$-direction, with the anomalous current generated in this cycle in the positive $x$-direction. Each subsequent cycle of ECDI exhibits decreasing magnitude of electrostatic fluctuation energy, with the generated anomalous current having little energy in the last cycle observed during our simulations ($T_{2 \pi} < t < T_{3 \pi}$). From Fig.~\ref{fig:18c}, one recognizes that after one complete rotation of the ion beam ($0 < t < T_{2 \pi}$), the reduction in the $z$-momentum of the ion beam due to the energy transfer to two subsequent ECDI bursts (from $0 < t < T_{\pi}$ and $T_{\pi} < t < T_{2 \pi}$).

Fig.~\ref{fig:19} shows the 1D Fourier transform of the electric field $E_z$. In Fig.~\ref{fig:19a}, we observe the initial growth of ECDI in the first cycle, with the growth of spectral energy at the typical cyclotron resonances (at $k/k_0 = 1,2,3,\dots$), similar to the ones generated in Case 1B (Fig.~\ref{fig:14a}, below $t=T_5$). Fig.~\ref{fig:19b} shows the periodic generation of ECDI for all the three cycles observed between $t=0$ and $t=T_{3 \pi}$, noting the reduction of spectral energy for each subsequent burst of ECDI. Each individual burst of the growth-saturation cycle of ECDI (from Fig.~\ref{fig:19b}) can be attributed to the generation of ECDI-driven modes (growth phase) and their subsequent transition to ion-acoustic modes (the saturation phase).

\section{\label{sec6:level1}Energy Conservation}
In this section, we analyze energy conservation in our simulations. The analysis is somewhat different between the  ECDI driven by an $\mathbf{E} \times \mathbf{B}$ electron drift in Xe plasma (Case 1A) and ECDI driven by an unmagnetized Xe ion beam (Case 1B).

\begin{figure}[htbp]
  \centering
   \includegraphics[width=0.75\linewidth]{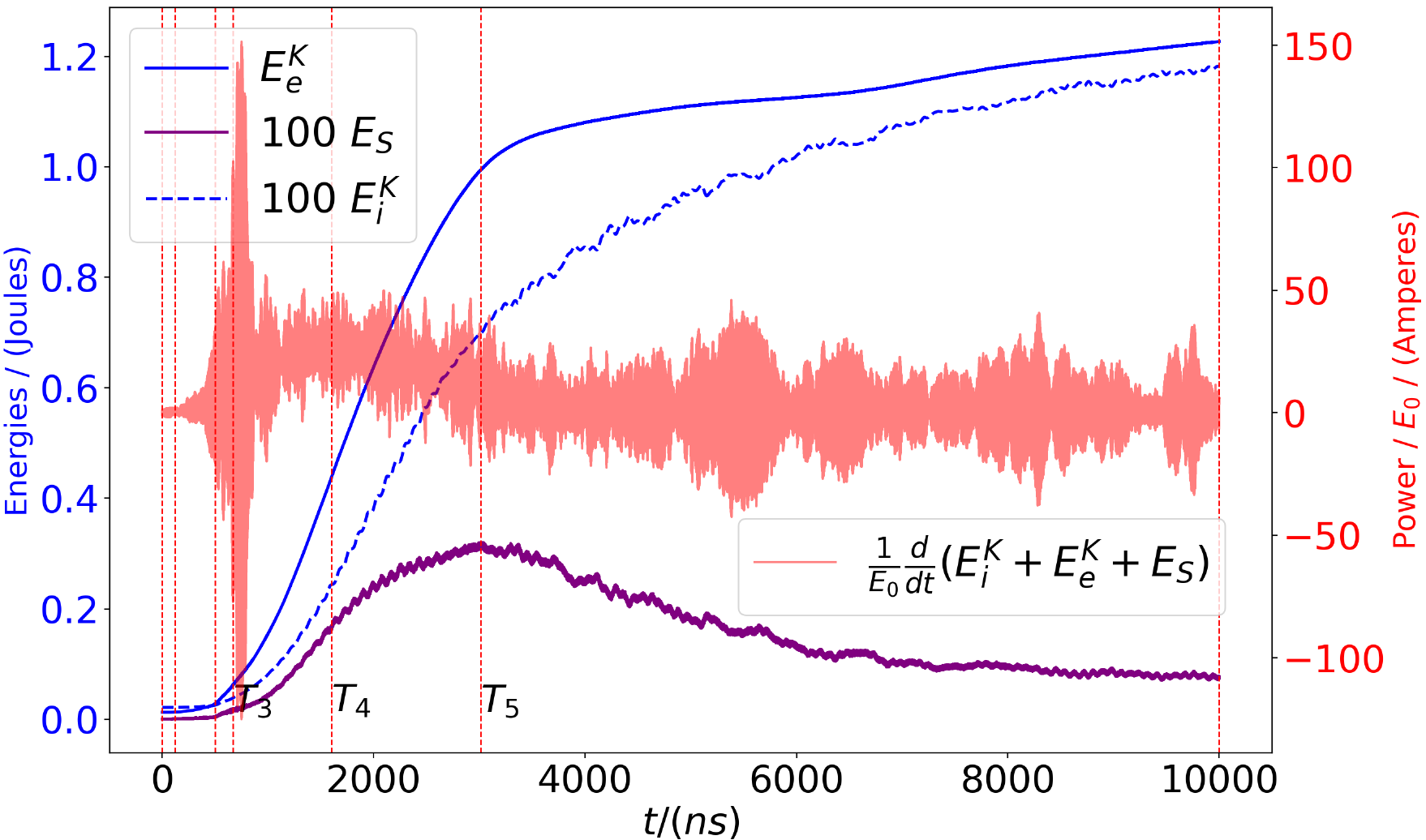}
   \caption{\label{fig:20} Energy conservation for Case 1A:  $100$ $\times$ electrostatic energy $E_s$, total ion kinetic energy $100$ $\times$ $E^K_i$, and total electron kinetic energy $E^K_e$. The derivative of the total energy, shown with a red transparent curve, matches exactly with the anomalous current ${J}_{xe}$ profile in Fig.~\ref{fig:11a}.}
\end{figure}

In general, in our electrostatic simulations, the energy is transferred between kinetic and potential energy of electrons and ions, and
the conservation of the total energy can be written in the form,

\begin{align}
W &= \sum\limits_{j=i}\left( \frac{wt_{j} \, m_{i}\left( v_{x}^{2}+v_{y}^{2}+v_{z}^{2}\right) _{j}}{2} + wt_j \, e\phi \left( x_{j}, y_{j}, z_{j} \right) \right) \notag \\
 & \quad \quad +\sum\limits_{j=e}\left( \frac{wt_{j} \, m_{e}\left( v_{x}^{2}+v_{y}^{2}+v_{z}^{2}\right) _{j}}{2} - wt_j \, e\phi \left( x_{j}, y_{j}, z_{j} \right) \right) = \text{const.},
\end{align}
where the sums are taken over all ions $i$ and electrons $e$, $\phi \left( x_{j},y_{j}, z_{j}\right) $ is the value of the electrostatic potential at the location $\left( x_{j},y_{j}, z_{j}\right) $ of particle $j$, and $wt_j$ is the weight of the particle $j$, defined as the ratio of the number of real particles in the domain to the number of computational particles. In the simulations of Case 1A, when the fluctuations are driven by the $\mathbf{ E\times B}$ drift due to the presence of the external electric field $ \mathbf{E}_{0}=E_{0}\widehat{\mathbf{e}}_{x}$, the electrons shift in the $x$-direction due to the $\widetilde{E}_{z}$ fields. This motion, resulting in the anomalous current $J_{xe}$ defined by Eq. (\ref{eq:J}), also changes the electron energy, which has to be included in the total energy balance. Therefore, the electrostatic potential is written as
\begin{equation}
\phi \left( x,y,z\right) =-E_{0}x+\widetilde{\phi }\left( z\right) .
\end{equation}%
Then, the energy conservation takes the form
\begin{equation}
\frac{dW}{dt}=\frac{d}{dt}\left( E_{i}^{K}+E_{e}^{K}+ \sum_{j=i,e} wt_j \, e_{j}\tilde{%
\phi}(z_{j})\right) + eE_{0}\frac{d}{dt}\sum_{j=e}wt_j \, x_{j}=0,  \label{en1a}
\end{equation}%
where $E^K_{i,e}$ represents the total kinetic energy of ions and electrons respectively, given by
\begin{equation*}
E^K_{i,e} = \sum\limits_{j=i,e} \frac{wt_j \, m_{i,e}\left( v_{x}^{2}+v_{y}^{2}+v_{z}^{2}\right) _{j}}{2}.  
\end{equation*}
In Eq.~\eqref{en1a}, the sum in the first term is taken over all ions and electrons and only over the electrons in the last term. Because we neglect the effects of the magnetic field on ions, ions are not shifted in the $x$-direction and are not included in the last term in Eq.~\eqref{en1a}. This term represents the energy change due to the anomalous current $J_{xe},$ so one can write
\begin{equation}
J_{xe}=-\frac{1}{E_{0}}\frac{d}{dt}\left(
E_{i}^{K}+E_{e}^{K}+ E_s\right),
\end{equation}
where we approximate $E_s$ from Eq.~\eqref{eq:es} as $E_s = \frac{\epsilon_0}{2}\sum E_z^2 \Delta z \equiv \sum_{j=i,e}wt_j \, e_{j}\tilde{\phi}(z_{j})$.
As is shown in Fig.~\ref{fig:20}, the anomalous current obtained from the energy conservation balance is fully consistent with the values obtained from direct calculation in Fig.~\ref{fig:11a} from Eq.~\eqref{eq:J}. 

\begin{figure}[htbp]
  \centering
   \includegraphics[width=0.75\linewidth]{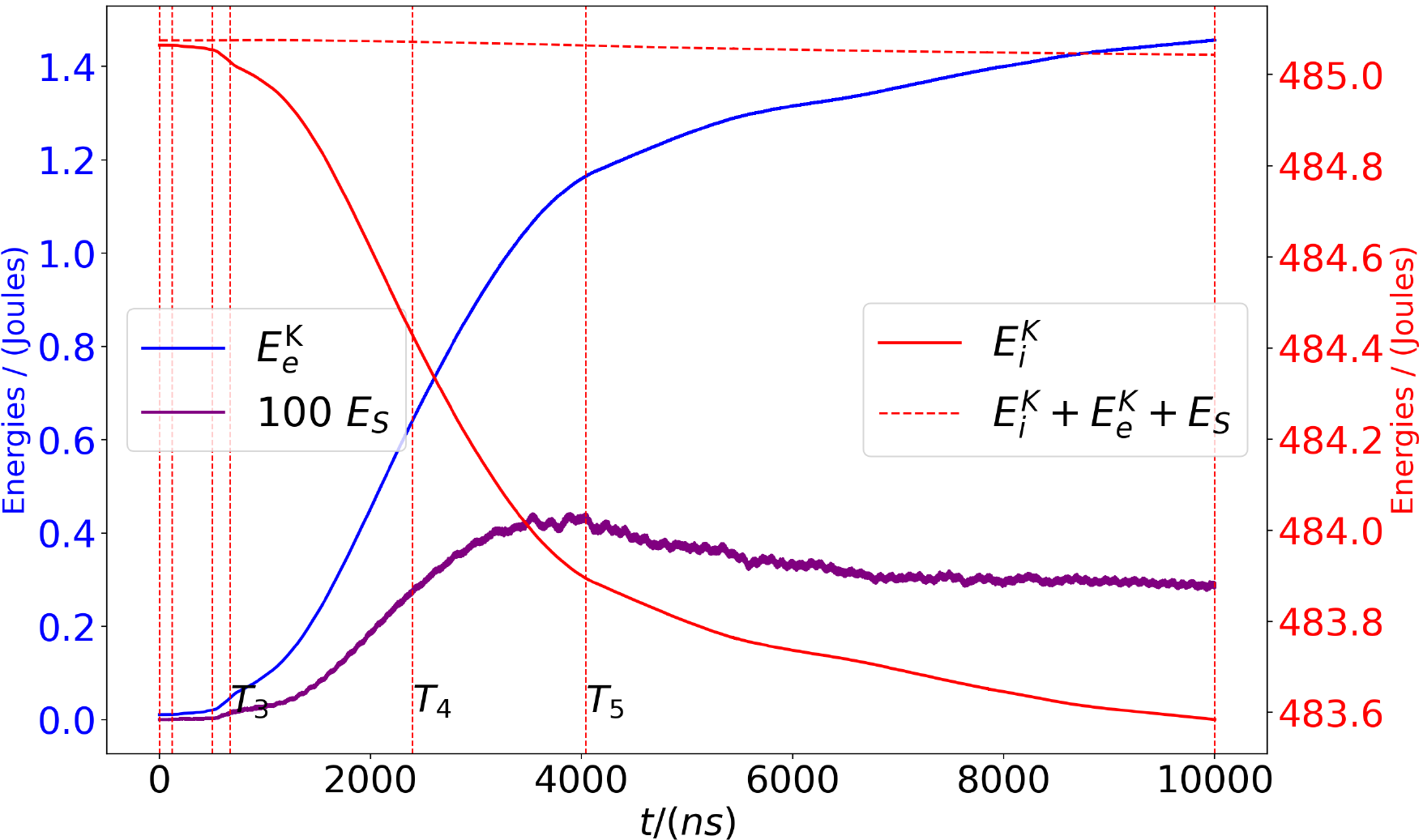}
   \caption{\label{fig:21} Comparison of all total energies in the system ($100 \times$ electrostatic energy, ion energy, and electron energy) for Case 1B. The total energy is shown with a red dotted line.}
\end{figure}

To emphasize the energy exchange between the ions, electrons,  and the electric field, the conservation of energy  can also be written in the moment form as 
\begin{align}
&\frac{\partial }{\partial t}\sum\limits_{i,e}\left( \frac{mn}{2}\mathbf{v}%
^{2}+\frac{3}{2}p\right) +\nabla \cdot \sum\limits_{i,e}\left[ \left( \frac{%
mn}{2}\mathbf{v}^{2}+\frac{5}{2}p\right) \mathbf{v+\pi \cdot v+q}\right]
\nonumber \\
&\hspace{85mm}=\sum\limits_{i,e}e_{\ell }n_{\ell }\mathbf{E\cdot v}_{\ell }=%
\mathbf{E\cdot J}.
\end{align}

On the left-hand side, the sum of the total electron and ion energies is taken, and the right-hand side describes  the energy exchange between plasma and the electromagnetic field due to the electric current
\begin{equation}
\mathbf{J\ =}\sum\limits_{i,e}e_{\ell }n_{\ell }\mathbf{v}_{\ell },
\end{equation}
\begin{equation}
\frac{\partial }{\partial t}\left( \frac{\varepsilon _{0}}{2}E^{2}+\frac{%
B^{2}}{2\mu _{0}}\right) +\nabla \cdot \left( \mathbf{E\times B}\right) =-%
\mathbf{J\cdot E.}
\end{equation}

Summing the particle kinetic energies and field energy and integrating over the periodic boundary, one arrives at the energy conservation in the form 
\begin{equation}
\frac{d}{dt}\left( E_{i}^{K}+E_{e}^{K}+E_{s}\right) =0.  \label{en1b}
\end{equation}%
Here, the kinetic energy for ions  and electrons ($\ell =i,e)$ is defined as
\begin{equation}
E_{\ell }^{K}=\int_{-L_z/2}^{L_z/2}dz\left( \frac{m_{\ell }n_{\ell }}{2}\mathbf{%
v}_{\ell }^{2}+\frac{3}{2}p_{\ell }\right) =\int_{-L_z/2}^{L_z/2}dz\frac{%
m_{\ell }}{2}\int \left( v_{x}^{2}+v_{y}^2 +v_{z}^{2}\right) f_{\ell
}d^{2}v\rightarrow \sum\limits_{j}\frac{wt_j \, m_{\ell }\left(
v_{x}^{2}+v_{y}^2 + v_{z}^{2}\right) _{j}}{2},
\end{equation}%
and the electrostatic energy is
\begin{equation}
E_{s}=\frac{\varepsilon _{0}}{2}\int_{-L_z/2}^{L_z/2}\
E_z^{2}dz \equiv \frac{1}{2} \int_{-L_z/2}^{L_z/2} e \widetilde{\phi}(z) \, dz = \sum\limits_{j=e,i}wt_j \, e_{j}\widetilde{\phi }\left( z_{j}\right) .
\end{equation}%
Energy conservation in the form of  Eq. (\ref{en1b}) is valid for a closed system, as in simulations of Case 1B, shown in Fig.~\ref{fig:21}. In Case 1A, we have the system is not closed: the electric charges maintaining the constant external electric field $\mathbf{E}_{0}$ are excluded resulting in the additional energy exchange term (the last term in Eq. (\ref{en1a})). In both cases, we observe good energy conservation within a deviation of $0.006 \%$.

\section{\label{sec7:level1}Discussion and Conclusions}

In this study, we have performed highly resolved, long-term 1D3V simulations of the ECDI  along the direction of the current due to the relative drift of electrons and ions: Case 1A for the current due to the  $\mathbf{E} \times \mathbf{B}$ electron drift supported by the external electric field, and Case 1B due to the current of the ion beam launched perpendicular to the magnetic  field. Our emphasis is  on the long-term evolution of turbulence, the transition to the ion-acoustic turbulence suggested by early theoretical analysis and numerical simulation, and the nature of the associated anomalous current. 
The transition to the ion-acoustic turbulence, suggested by early theoretical and numerical studies,  was explicitly or implicitly assumed in many studies and served as a basis for interpretation of the anomalous transport.

According to earlier theoretical and numerical studies,~\cite{LampePF1972a,LampePF1972b}  the linear reactive electron cyclotron instability due to the crossing (near the electron cyclotron frequency)  of the electron Bernstein mode with the Doppler-shifted ion-acoustic  mode saturates at a low fluctuation level. Then the band structure of the cyclotron instability disappears, and the growth proceeds at the slower rate of the unmagnetized ion-acoustic instability at the characteristic most-unstable wavelength $k_s=1/\sqrt{2}\lambda_{De} $. The regime of unmagnetized ion-acoustic instability occurs when the non-linear mode broadening $\Delta \omega $ is large, $\Delta \omega >\omega_{ce}/\pi $, and electron motion is demagnetized due to strong perturbation of the electron trajectories.  The above picture of unmagnetized ion-acoustic instability has been widely employed to interpret the anomalous transport observed in numerical simulations. In the 1D geometry, the non-linear mode broadening is the only mechanism for  demagnetization of the electron motion and transition to the ion-acoustic regime. We aimed to investigate this process by following it deeply into the non-linear stage without any ad-hoc mechanisms of turbulence saturation and suppression of electron heating.

Our study reveals that the pumping of turbulence at the cyclotron harmonics persists well into the nonlinear regime. The fluctuations are driven at the characteristic wavelengths of several first cyclotron harmonics,   $k/k_0 = 1,2,3$, which are different from the typical wavelength of the most unstable ion-acoustic mode, $k_s$. 

During the intermediate non-linear stage, we discover an important resonance between the ECDI \(m = 1\) mode and the most dominant mode ($k_s = 1/\sqrt{2}\lambda_{De}$) of the ion-acoustic branch, a phenomenon not previously noted in the literature. The moment $T_3$ for the ion beam case (1B), was earlier interpreted as a point of transition to the purely unmagnetized ion-acoustic turbulence.~\cite{LampePLA1971} In our simulations, at $T_3$, we see enhanced growth of the $m =1$ mode, which subsequently becomes the dominant mode in the instability's later evolution proceeding at the wavelength corresponding to $m = k/k_0=1$ mode, which is much shorter than the ion-acoustic wavelength $~2\pi/k_s$.

Another notable result of our study is the quenching of the anomalous transport in the nonlinear saturated regime that occurs after the fluctuation energy concentrated at the primary cyclotron resonance ($k/k_0 = 1$) starts to spread along the ion-acoustic branch and its non-linear harmonics at \(\omega_{pi} = 1, 2, 3, \ldots\). Thus, the transition to ion-acoustic turbulence occurs at a late stage when the anomalous current present in the intermediate stages is fully quenched. This is consistent with the notion that fully demagnetized electrons should not experience any $\mathbf{E} \times \mathbf{B}$ drift and therefore cannot sustain any transport across the magnetic field.  


Most, if not all, previous simulations of ECDI performed  in 1D or 2D geometries focused on the low-frequency modes and have not resolved higher frequencies of the electron cyclotron harmonics.    In this study, however, we successfully resolve the higher frequencies  observing the linear and non-linear development  of Doppler-shifted electron Bernstein waves. The latter includes non-linear resonance broadening of the cyclotron harmonics, their non-linear spreading, and inverse cascade at lower wavelengths $k\leq k_0$ and lower frequencies $\omega \leq \omega_{pi}$. 

Both high-frequency Doppler-shifted electron Bernstein waves and low-frequency (around $\omega_{pi}$) ion modes show cyclotron resonances in both forward $k/k_0 < 0$ and backward $k/k_0 > 0$ directions.  Interestingly, the backward waves generated in the linear regime at cyclotron resonances $k/k_0 = -1, -2,\dots$, are observed to be present even in the deeply non-linear phases, the signature of which has been observed in experiment. \cite{tsikata2009dispersion}

ECDI results in highly efficient electron  heating at the electron cyclotron resonance.  Resonant wave-electron interactions lead to a strong flattening of the electron distribution function saturating to a final state with a low curtosis.\cite{JanhunenPoP2018a}  Electron heating also saturates at this stage.  

In contrast, the ion phase space dynamics is quite different. 
The early nonlinear stage, which is dominated by the cyclotron mode and its nonlinear resonances $k/k_0 = 1,2,3\dots$ with a typical cnoidal periodic wave structure are changed to a stage with nonlinear wave-ion trapping. Subsequently, however, the individual ion-wave resonances start to overlap and mix and are eventually destroyed,  forming the quasilinear-type high-energy tail in the ion distribution function.\cite{IshiharaPRL1981} 
Strong nonlinear features in the ion dynamics and strong electron heating are the results of large magnitude of the electric field fluctuations, which in our simulations exceed the quasi-static electric fields by an order of magnitude. In this regard, we note Ref.~\onlinecite{wilson2021discrepancy}, which highlights the discrepancies between the large amplitudes electrostatic fluctuations observed experimentally versus the low amplitudes found in the simulations.

We have shown that although the turbulence driven by the ion beam perpendicular to the magnetic field is not identical to that driven by the electron $\mathbf{E}\times \mathbf{B}$ flow, their integral properties remain similar, including the non-linear stage, so that essentially they can be considered as the same turbulent state observed through different Doppler-shifted reference frames. Thus, in a laboratory frame with ECDI driven by an ion beam,  one  observes the dispersion relation for the final saturated state as given in Fig.~\ref{fig:16e}, with the ion-acoustic like dispersion curves Doppler shifted to lie along the Doppler-shifted ion beam mode ($\omega/ k = V_{di}$), alongside a subtle trace of skewed electron Bernstein wave modes.

The effects of the magnetic field on ions have  been omitted in many studies due to the conditions $\omega_{ci} \ll \omega$  and $\rho_{ci} \ll L$, where $\omega$ is the characteristic frequency and $L$ is the characteristics length of the plasma system of interest. These conditions are typical for many devices, e.g., for electric propulsion, where heavy ion species are used, and therefore, ion magnetization can be neglected. For  applications with stronger magnetic fields or in  collisionless shock environments as observed in the Earth's bow shock with lighter species (like hydrogen ions)  and larger length scales, the ion magnetization can be important. Thus, it is of interest to consider effects of the magnetic field on ions. We have performed such simulations with hydrogen ions, where the ions make several Larmor rotations in the plane perpendicular to the applied magnetic field, and observed the generation of periodic growth-saturation cycles of ECDI due to the rotating ion beam.   

In terms of energy conservation throughout the evolution of ECDI, as observed in Case 1B, Fig. \ref{fig:21} shows that the reduction in total ion beam energy ($485.0 \, \text{J} - 483.6 \, \text{J} \sim 1.4$ J) is predominantly transferred to the total electron energy ($\sim 1.4$ J), with only about $1\%$ of total energy going into electrostatic fluctuations. The average electron current in the direction of the ion beam current fluctuates close to zero ($J_{ze}$ from Fig.~\ref{fig:13}), indicating zero mean kinetic energy gained by electrons in the $z$-direction. Thus, the energy gained from the relative drift between ions and electrons primarily leads to turbulent heating of the electrons, also evident from the significant broadening of the electron distribution function, Fig.~\ref{fig:5}. We suggest the interpretation of this turbulent heating of electrons from ECDI as increasing the anomalous resistivity of the plasma to be one of the critical findings of this study.


In many previous 1D simulations, ad-hoc mechanisms such as electron or ion re-injection (or cooling) have been employed to saturate the instability, resulting in a finite value of the  anomalous current. Our simulations here show the inherent suppression of anomalous current in the final non-linear stage of ECDI. It is only in this stage that one observes the transition to stable ion-acoustic modes that do not create any anomalous transport. This finding is rather different from previous views on the transition of ECDI to the ion-acoustic instability. Incidentally, a recent 2D study of ECDI using hydrogen ions\cite{BelloPoP2024} also demonstrates the suppression of the anomalous transport in the final stage of the ECDI. Although not directly related to our study, it is interesting to note that the numerical simulations of 2D ion-acoustic turbulence (IAT) also show\cite{LiuJPP2024} that anomalous resistivity due to IAT is transient and disappears in the final stage.

\acknowledgments{This work is supported in part by the Natural Sciences and Engineering Research Council of Canada (NSERC). The computational resources were provided by Digital Research Alliance of Canada. This research used the open-source particle-in-cell code WarpX https://github.com/ECP-WarpX/WarpX, primarily funded by the US DOE Exascale Computing Project. Primary WarpX contributors are with LBNL, LLNL, CEA-LIDYL, SLAC, DESY, CERN, and TAE Technologies. We acknowledge all WarpX contributors.}


\section*{Data availability}
The data that support the findings of this study are available from the corresponding author upon reasonable request.

\bibliography{aipsamp.bib}

\end{document}